\documentclass[10pt,twocolumn,a4paper]{article}
\usepackage[margin=2cm,columnsep=0.65cm]{geometry}
\usepackage[utf8]{inputenc}
\usepackage[T1]{fontenc}
\usepackage{lmodern}
\usepackage{microtype}
\usepackage{graphicx}
\usepackage{amsmath,amssymb,amsfonts,amsthm,mathrsfs}
\usepackage{booktabs,multirow,tabularx,longtable}
\usepackage{subcaption}
\usepackage{xcolor,textcomp,bm}
\usepackage{algorithm,algorithmicx,algpseudocode,listings}
\usepackage{float,bbm}
\usepackage{pgf,tikz}
\usepackage[numbers,sort&compress]{natbib}
\usepackage[hidelinks]{hyperref}
\newcommand{\num}[1]{\ifcsname num@#1\endcsname\csname num@#1\endcsname\else\PackageError{paper-results}{Missing result key: #1}{Regenerate generated/numbers.tex from the complete final outputs.}\fi}
\expandafter\def\csname num@coverage.10\endcsname{85.831}
\expandafter\def\csname num@coverage.100\endcsname{0.000}
\expandafter\def\csname num@coverage.15\endcsname{0.084}
\expandafter\def\csname num@coverage.20\endcsname{0.000}
\expandafter\def\csname num@coverage.25\endcsname{0.000}
\expandafter\def\csname num@coverage.30\endcsname{0.000}
\expandafter\def\csname num@coverage.35\endcsname{0.000}
\expandafter\def\csname num@coverage.40\endcsname{0.000}
\expandafter\def\csname num@coverage.45\endcsname{0.000}
\expandafter\def\csname num@coverage.50\endcsname{0.000}
\expandafter\def\csname num@coverage.60\endcsname{0.000}
\expandafter\def\csname num@coverage.70\endcsname{0.000}
\expandafter\def\csname num@coverage.80\endcsname{0.000}
\expandafter\def\csname num@coverage.90\endcsname{0.000}
\expandafter\def\csname num@cross.25.dnabert2.variant_unseen_lev\endcsname{0.526}
\expandafter\def\csname num@cross.25.evo2.variant_unseen_lev\endcsname{0.998}
\expandafter\def\csname num@cross.25.ntv2.variant_unseen_lev\endcsname{0.982}
\expandafter\def\csname num@cross.25.reference_n\endcsname{14{,}647}
\expandafter\def\csname num@cross.25.total_n\endcsname{15{,}000}
\expandafter\def\csname num@cross.25.variant_unseen_n\endcsname{338}
\expandafter\def\csname num@emb.dnabert2.spearman.cosine.10\endcsname{0.017}
\expandafter\def\csname num@emb.dnabert2.spearman.cosine.100\endcsname{0.122}
\expandafter\def\csname num@emb.dnabert2.spearman.cosine.15\endcsname{0.028}
\expandafter\def\csname num@emb.dnabert2.spearman.cosine.20\endcsname{0.028}
\expandafter\def\csname num@emb.dnabert2.spearman.cosine.25\endcsname{0.035}
\expandafter\def\csname num@emb.dnabert2.spearman.cosine.30\endcsname{0.042}
\expandafter\def\csname num@emb.dnabert2.spearman.cosine.35\endcsname{0.051}
\expandafter\def\csname num@emb.dnabert2.spearman.cosine.40\endcsname{0.064}
\expandafter\def\csname num@emb.dnabert2.spearman.cosine.45\endcsname{0.054}
\expandafter\def\csname num@emb.dnabert2.spearman.cosine.50\endcsname{0.071}
\expandafter\def\csname num@emb.dnabert2.spearman.cosine.60\endcsname{0.075}
\expandafter\def\csname num@emb.dnabert2.spearman.cosine.70\endcsname{0.080}
\expandafter\def\csname num@emb.dnabert2.spearman.cosine.80\endcsname{0.101}
\expandafter\def\csname num@emb.dnabert2.spearman.cosine.90\endcsname{0.106}
\expandafter\def\csname num@emb.dnabert2.spearman.euclidean.10\endcsname{0.029}
\expandafter\def\csname num@emb.dnabert2.spearman.euclidean.100\endcsname{0.113}
\expandafter\def\csname num@emb.dnabert2.spearman.euclidean.15\endcsname{0.036}
\expandafter\def\csname num@emb.dnabert2.spearman.euclidean.20\endcsname{0.035}
\expandafter\def\csname num@emb.dnabert2.spearman.euclidean.25\endcsname{0.038}
\expandafter\def\csname num@emb.dnabert2.spearman.euclidean.30\endcsname{0.034}
\expandafter\def\csname num@emb.dnabert2.spearman.euclidean.35\endcsname{0.047}
\expandafter\def\csname num@emb.dnabert2.spearman.euclidean.40\endcsname{0.056}
\expandafter\def\csname num@emb.dnabert2.spearman.euclidean.45\endcsname{0.054}
\expandafter\def\csname num@emb.dnabert2.spearman.euclidean.50\endcsname{0.068}
\expandafter\def\csname num@emb.dnabert2.spearman.euclidean.60\endcsname{0.064}
\expandafter\def\csname num@emb.dnabert2.spearman.euclidean.70\endcsname{0.066}
\expandafter\def\csname num@emb.dnabert2.spearman.euclidean.80\endcsname{0.095}
\expandafter\def\csname num@emb.dnabert2.spearman.euclidean.90\endcsname{0.095}
\expandafter\def\csname num@emb.evo2.spearman.cosine.10\endcsname{0.149}
\expandafter\def\csname num@emb.evo2.spearman.cosine.100\endcsname{0.173}
\expandafter\def\csname num@emb.evo2.spearman.cosine.15\endcsname{0.415}
\expandafter\def\csname num@emb.evo2.spearman.cosine.20\endcsname{0.375}
\expandafter\def\csname num@emb.evo2.spearman.cosine.25\endcsname{0.320}
\expandafter\def\csname num@emb.evo2.spearman.cosine.30\endcsname{0.238}
\expandafter\def\csname num@emb.evo2.spearman.cosine.35\endcsname{0.235}
\expandafter\def\csname num@emb.evo2.spearman.cosine.40\endcsname{0.210}
\expandafter\def\csname num@emb.evo2.spearman.cosine.45\endcsname{0.196}
\expandafter\def\csname num@emb.evo2.spearman.cosine.50\endcsname{0.188}
\expandafter\def\csname num@emb.evo2.spearman.cosine.60\endcsname{0.185}
\expandafter\def\csname num@emb.evo2.spearman.cosine.70\endcsname{0.165}
\expandafter\def\csname num@emb.evo2.spearman.cosine.80\endcsname{0.175}
\expandafter\def\csname num@emb.evo2.spearman.cosine.90\endcsname{0.184}
\expandafter\def\csname num@emb.evo2.spearman.euclidean.10\endcsname{0.069}
\expandafter\def\csname num@emb.evo2.spearman.euclidean.100\endcsname{0.147}
\expandafter\def\csname num@emb.evo2.spearman.euclidean.15\endcsname{0.343}
\expandafter\def\csname num@emb.evo2.spearman.euclidean.20\endcsname{0.372}
\expandafter\def\csname num@emb.evo2.spearman.euclidean.25\endcsname{0.314}
\expandafter\def\csname num@emb.evo2.spearman.euclidean.30\endcsname{0.226}
\expandafter\def\csname num@emb.evo2.spearman.euclidean.35\endcsname{0.219}
\expandafter\def\csname num@emb.evo2.spearman.euclidean.40\endcsname{0.190}
\expandafter\def\csname num@emb.evo2.spearman.euclidean.45\endcsname{0.176}
\expandafter\def\csname num@emb.evo2.spearman.euclidean.50\endcsname{0.163}
\expandafter\def\csname num@emb.evo2.spearman.euclidean.60\endcsname{0.161}
\expandafter\def\csname num@emb.evo2.spearman.euclidean.70\endcsname{0.137}
\expandafter\def\csname num@emb.evo2.spearman.euclidean.80\endcsname{0.146}
\expandafter\def\csname num@emb.evo2.spearman.euclidean.90\endcsname{0.153}
\expandafter\def\csname num@emb.ntv2.spearman.cosine.10\endcsname{0.091}
\expandafter\def\csname num@emb.ntv2.spearman.cosine.100\endcsname{0.233}
\expandafter\def\csname num@emb.ntv2.spearman.cosine.15\endcsname{0.094}
\expandafter\def\csname num@emb.ntv2.spearman.cosine.20\endcsname{0.090}
\expandafter\def\csname num@emb.ntv2.spearman.cosine.25\endcsname{0.085}
\expandafter\def\csname num@emb.ntv2.spearman.cosine.30\endcsname{0.084}
\expandafter\def\csname num@emb.ntv2.spearman.cosine.35\endcsname{0.156}
\expandafter\def\csname num@emb.ntv2.spearman.cosine.40\endcsname{0.176}
\expandafter\def\csname num@emb.ntv2.spearman.cosine.45\endcsname{0.155}
\expandafter\def\csname num@emb.ntv2.spearman.cosine.50\endcsname{0.162}
\expandafter\def\csname num@emb.ntv2.spearman.cosine.60\endcsname{0.150}
\expandafter\def\csname num@emb.ntv2.spearman.cosine.70\endcsname{0.194}
\expandafter\def\csname num@emb.ntv2.spearman.cosine.80\endcsname{0.176}
\expandafter\def\csname num@emb.ntv2.spearman.cosine.90\endcsname{0.161}
\expandafter\def\csname num@emb.ntv2.spearman.euclidean.10\endcsname{0.093}
\expandafter\def\csname num@emb.ntv2.spearman.euclidean.100\endcsname{0.240}
\expandafter\def\csname num@emb.ntv2.spearman.euclidean.15\endcsname{0.097}
\expandafter\def\csname num@emb.ntv2.spearman.euclidean.20\endcsname{0.096}
\expandafter\def\csname num@emb.ntv2.spearman.euclidean.25\endcsname{0.095}
\expandafter\def\csname num@emb.ntv2.spearman.euclidean.30\endcsname{0.090}
\expandafter\def\csname num@emb.ntv2.spearman.euclidean.35\endcsname{0.175}
\expandafter\def\csname num@emb.ntv2.spearman.euclidean.40\endcsname{0.194}
\expandafter\def\csname num@emb.ntv2.spearman.euclidean.45\endcsname{0.167}
\expandafter\def\csname num@emb.ntv2.spearman.euclidean.50\endcsname{0.176}
\expandafter\def\csname num@emb.ntv2.spearman.euclidean.60\endcsname{0.168}
\expandafter\def\csname num@emb.ntv2.spearman.euclidean.70\endcsname{0.205}
\expandafter\def\csname num@emb.ntv2.spearman.euclidean.80\endcsname{0.191}
\expandafter\def\csname num@emb.ntv2.spearman.euclidean.90\endcsname{0.180}
\expandafter\def\csname num@fit.dnabert2.random.d\endcsname{0.101}
\expandafter\def\csname num@fit.dnabert2.random.k99\endcsname{3{,}316}
\expandafter\def\csname num@fit.dnabert2.random.k99reachable\endcsname{yes}
\expandafter\def\csname num@fit.dnabert2.random.r\endcsname{1.03}
\expandafter\def\csname num@fit.dnabert2.random.rmse\endcsname{0.005}
\expandafter\def\csname num@fit.dnabert2.snp_enriched.d\endcsname{0.417}
\expandafter\def\csname num@fit.dnabert2.snp_enriched.k99\endcsname{202}
\expandafter\def\csname num@fit.dnabert2.snp_enriched.k99reachable\endcsname{yes}
\expandafter\def\csname num@fit.dnabert2.snp_enriched.r\endcsname{1.08}
\expandafter\def\csname num@fit.dnabert2.snp_enriched.rmse\endcsname{0.006}
\expandafter\def\csname num@fit.evo2.random.d\endcsname{0.309}
\expandafter\def\csname num@fit.evo2.random.k99\endcsname{311}
\expandafter\def\csname num@fit.evo2.random.k99reachable\endcsname{yes}
\expandafter\def\csname num@fit.evo2.random.r\endcsname{0.86}
\expandafter\def\csname num@fit.evo2.random.rmse\endcsname{0.004}
\expandafter\def\csname num@fit.evo2.snp_enriched.d\endcsname{1.110}
\expandafter\def\csname num@fit.evo2.snp_enriched.k99\endcsname{19}
\expandafter\def\csname num@fit.evo2.snp_enriched.k99reachable\endcsname{yes}
\expandafter\def\csname num@fit.evo2.snp_enriched.r\endcsname{0.57}
\expandafter\def\csname num@fit.evo2.snp_enriched.rmse\endcsname{0.003}
\expandafter\def\csname num@fit.ntv2.random.d\endcsname{0.295}
\expandafter\def\csname num@fit.ntv2.random.k99\endcsname{347}
\expandafter\def\csname num@fit.ntv2.random.k99reachable\endcsname{yes}
\expandafter\def\csname num@fit.ntv2.random.r\endcsname{0.88}
\expandafter\def\csname num@fit.ntv2.random.rmse\endcsname{0.004}
\expandafter\def\csname num@fit.ntv2.snp_enriched.d\endcsname{1.070}
\expandafter\def\csname num@fit.ntv2.snp_enriched.k99\endcsname{21}
\expandafter\def\csname num@fit.ntv2.snp_enriched.k99reachable\endcsname{yes}
\expandafter\def\csname num@fit.ntv2.snp_enriched.r\endcsname{0.57}
\expandafter\def\csname num@fit.ntv2.snp_enriched.rmse\endcsname{0.005}
\expandafter\def\csname num@fit.rmse.max\endcsname{0.006}
\expandafter\def\csname num@fit.rmse.min\endcsname{0.003}
\expandafter\def\csname num@id.chance\endcsname{0.1}
\expandafter\def\csname num@id.cohort\endcsname{1{,}000}
\expandafter\def\csname num@id.dnabert2.random.blind.k50\endcsname{--}
\expandafter\def\csname num@id.dnabert2.random.blind.k90\endcsname{--}
\expandafter\def\csname num@id.dnabert2.random.blind.k99\endcsname{--}
\expandafter\def\csname num@id.dnabert2.random.blind.kmax\endcsname{4{,}096}
\expandafter\def\csname num@id.dnabert2.random.blind.topmax\endcsname{0.1}
\expandafter\def\csname num@id.dnabert2.random.blind.topmaxfrac\endcsname{0.001}
\expandafter\def\csname num@id.dnabert2.random.excluded\endcsname{158}
\expandafter\def\csname num@id.dnabert2.random.k50\endcsname{1{,}024}
\expandafter\def\csname num@id.dnabert2.random.k90\endcsname{4{,}096}
\expandafter\def\csname num@id.dnabert2.random.k99\endcsname{4{,}096}
\expandafter\def\csname num@id.dnabert2.random.kmax\endcsname{4{,}096}
\expandafter\def\csname num@id.dnabert2.random.mappedpct\endcsname{76.2}
\expandafter\def\csname num@id.dnabert2.random.nloci\endcsname{4{,}096}
\expandafter\def\csname num@id.dnabert2.random.topmax\endcsname{99.7}
\expandafter\def\csname num@id.dnabert2.random.topmaxfrac\endcsname{0.997}
\expandafter\def\csname num@id.dnabert2.snp_enriched.blind.k50\endcsname{--}
\expandafter\def\csname num@id.dnabert2.snp_enriched.blind.k90\endcsname{--}
\expandafter\def\csname num@id.dnabert2.snp_enriched.blind.k99\endcsname{--}
\expandafter\def\csname num@id.dnabert2.snp_enriched.blind.kmax\endcsname{2{,}048}
\expandafter\def\csname num@id.dnabert2.snp_enriched.blind.topmax\endcsname{0.2}
\expandafter\def\csname num@id.dnabert2.snp_enriched.blind.topmaxfrac\endcsname{0.002}
\expandafter\def\csname num@id.dnabert2.snp_enriched.excluded\endcsname{238}
\expandafter\def\csname num@id.dnabert2.snp_enriched.k50\endcsname{64}
\expandafter\def\csname num@id.dnabert2.snp_enriched.k90\endcsname{256}
\expandafter\def\csname num@id.dnabert2.snp_enriched.k99\endcsname{256}
\expandafter\def\csname num@id.dnabert2.snp_enriched.kmax\endcsname{2{,}048}
\expandafter\def\csname num@id.dnabert2.snp_enriched.mappedpct\endcsname{76.0}
\expandafter\def\csname num@id.dnabert2.snp_enriched.nloci\endcsname{4{,}091}
\expandafter\def\csname num@id.dnabert2.snp_enriched.topmax\endcsname{100.0}
\expandafter\def\csname num@id.dnabert2.snp_enriched.topmaxfrac\endcsname{1.000}
\expandafter\def\csname num@id.evo2.random.blind.k50\endcsname{512}
\expandafter\def\csname num@id.evo2.random.blind.k90\endcsname{1{,}024}
\expandafter\def\csname num@id.evo2.random.blind.k99\endcsname{2{,}048}
\expandafter\def\csname num@id.evo2.random.blind.kmax\endcsname{4{,}096}
\expandafter\def\csname num@id.evo2.random.blind.topmax\endcsname{100.0}
\expandafter\def\csname num@id.evo2.random.blind.topmaxfrac\endcsname{1.000}
\expandafter\def\csname num@id.evo2.random.excluded\endcsname{158}
\expandafter\def\csname num@id.evo2.random.k50\endcsname{128}
\expandafter\def\csname num@id.evo2.random.k90\endcsname{256}
\expandafter\def\csname num@id.evo2.random.k99\endcsname{512}
\expandafter\def\csname num@id.evo2.random.kmax\endcsname{4{,}096}
\expandafter\def\csname num@id.evo2.random.mappedpct\endcsname{99.8}
\expandafter\def\csname num@id.evo2.random.nloci\endcsname{4{,}096}
\expandafter\def\csname num@id.evo2.random.topmax\endcsname{100.0}
\expandafter\def\csname num@id.evo2.random.topmaxfrac\endcsname{1.000}
\expandafter\def\csname num@id.evo2.snp_enriched.blind.k50\endcsname{32}
\expandafter\def\csname num@id.evo2.snp_enriched.blind.k90\endcsname{64}
\expandafter\def\csname num@id.evo2.snp_enriched.blind.k99\endcsname{64}
\expandafter\def\csname num@id.evo2.snp_enriched.blind.kmax\endcsname{2{,}048}
\expandafter\def\csname num@id.evo2.snp_enriched.blind.topmax\endcsname{100.0}
\expandafter\def\csname num@id.evo2.snp_enriched.blind.topmaxfrac\endcsname{1.000}
\expandafter\def\csname num@id.evo2.snp_enriched.excluded\endcsname{238}
\expandafter\def\csname num@id.evo2.snp_enriched.k50\endcsname{16}
\expandafter\def\csname num@id.evo2.snp_enriched.k90\endcsname{16}
\expandafter\def\csname num@id.evo2.snp_enriched.k99\endcsname{32}
\expandafter\def\csname num@id.evo2.snp_enriched.kmax\endcsname{2{,}048}
\expandafter\def\csname num@id.evo2.snp_enriched.mappedpct\endcsname{98.8}
\expandafter\def\csname num@id.evo2.snp_enriched.nloci\endcsname{4{,}091}
\expandafter\def\csname num@id.evo2.snp_enriched.topmax\endcsname{100.0}
\expandafter\def\csname num@id.evo2.snp_enriched.topmaxfrac\endcsname{1.000}
\expandafter\def\csname num@id.ntv2.random.blind.k50\endcsname{512}
\expandafter\def\csname num@id.ntv2.random.blind.k90\endcsname{1{,}024}
\expandafter\def\csname num@id.ntv2.random.blind.k99\endcsname{2{,}048}
\expandafter\def\csname num@id.ntv2.random.blind.kmax\endcsname{4{,}096}
\expandafter\def\csname num@id.ntv2.random.blind.topmax\endcsname{100.0}
\expandafter\def\csname num@id.ntv2.random.blind.topmaxfrac\endcsname{1.000}
\expandafter\def\csname num@id.ntv2.random.excluded\endcsname{158}
\expandafter\def\csname num@id.ntv2.random.k50\endcsname{128}
\expandafter\def\csname num@id.ntv2.random.k90\endcsname{256}
\expandafter\def\csname num@id.ntv2.random.k99\endcsname{512}
\expandafter\def\csname num@id.ntv2.random.kmax\endcsname{4{,}096}
\expandafter\def\csname num@id.ntv2.random.mappedpct\endcsname{98.7}
\expandafter\def\csname num@id.ntv2.random.nloci\endcsname{4{,}096}
\expandafter\def\csname num@id.ntv2.random.topmax\endcsname{100.0}
\expandafter\def\csname num@id.ntv2.random.topmaxfrac\endcsname{1.000}
\expandafter\def\csname num@id.ntv2.snp_enriched.blind.k50\endcsname{32}
\expandafter\def\csname num@id.ntv2.snp_enriched.blind.k90\endcsname{64}
\expandafter\def\csname num@id.ntv2.snp_enriched.blind.k99\endcsname{128}
\expandafter\def\csname num@id.ntv2.snp_enriched.blind.kmax\endcsname{2{,}048}
\expandafter\def\csname num@id.ntv2.snp_enriched.blind.topmax\endcsname{100.0}
\expandafter\def\csname num@id.ntv2.snp_enriched.blind.topmaxfrac\endcsname{1.000}
\expandafter\def\csname num@id.ntv2.snp_enriched.excluded\endcsname{238}
\expandafter\def\csname num@id.ntv2.snp_enriched.k50\endcsname{16}
\expandafter\def\csname num@id.ntv2.snp_enriched.k90\endcsname{16}
\expandafter\def\csname num@id.ntv2.snp_enriched.k99\endcsname{32}
\expandafter\def\csname num@id.ntv2.snp_enriched.kmax\endcsname{2{,}048}
\expandafter\def\csname num@id.ntv2.snp_enriched.mappedpct\endcsname{97.7}
\expandafter\def\csname num@id.ntv2.snp_enriched.nloci\endcsname{4{,}091}
\expandafter\def\csname num@id.ntv2.snp_enriched.topmax\endcsname{100.0}
\expandafter\def\csname num@id.ntv2.snp_enriched.topmaxfrac\endcsname{1.000}
\expandafter\def\csname num@rec.baseline.max\endcsname{44}
\expandafter\def\csname num@rec.baseline.min\endcsname{34}
\expandafter\def\csname num@rec.dnabert2.acc.10\endcsname{0.386}
\expandafter\def\csname num@rec.dnabert2.acc.100\endcsname{0.252}
\expandafter\def\csname num@rec.dnabert2.acc.15\endcsname{0.332}
\expandafter\def\csname num@rec.dnabert2.acc.20\endcsname{0.313}
\expandafter\def\csname num@rec.dnabert2.acc.25\endcsname{0.299}
\expandafter\def\csname num@rec.dnabert2.acc.30\endcsname{0.292}
\expandafter\def\csname num@rec.dnabert2.acc.35\endcsname{0.284}
\expandafter\def\csname num@rec.dnabert2.acc.40\endcsname{0.278}
\expandafter\def\csname num@rec.dnabert2.acc.45\endcsname{0.273}
\expandafter\def\csname num@rec.dnabert2.acc.50\endcsname{0.269}
\expandafter\def\csname num@rec.dnabert2.acc.60\endcsname{0.261}
\expandafter\def\csname num@rec.dnabert2.acc.70\endcsname{0.256}
\expandafter\def\csname num@rec.dnabert2.acc.80\endcsname{0.254}
\expandafter\def\csname num@rec.dnabert2.acc.90\endcsname{0.253}
\expandafter\def\csname num@rec.dnabert2.accsd.10\endcsname{0.266}
\expandafter\def\csname num@rec.dnabert2.accsd.100\endcsname{0.076}
\expandafter\def\csname num@rec.dnabert2.accsd.15\endcsname{0.209}
\expandafter\def\csname num@rec.dnabert2.accsd.20\endcsname{0.177}
\expandafter\def\csname num@rec.dnabert2.accsd.25\endcsname{0.158}
\expandafter\def\csname num@rec.dnabert2.accsd.30\endcsname{0.145}
\expandafter\def\csname num@rec.dnabert2.accsd.35\endcsname{0.136}
\expandafter\def\csname num@rec.dnabert2.accsd.40\endcsname{0.129}
\expandafter\def\csname num@rec.dnabert2.accsd.45\endcsname{0.121}
\expandafter\def\csname num@rec.dnabert2.accsd.50\endcsname{0.116}
\expandafter\def\csname num@rec.dnabert2.accsd.60\endcsname{0.104}
\expandafter\def\csname num@rec.dnabert2.accsd.70\endcsname{0.095}
\expandafter\def\csname num@rec.dnabert2.accsd.80\endcsname{0.089}
\expandafter\def\csname num@rec.dnabert2.accsd.90\endcsname{0.082}
\expandafter\def\csname num@rec.dnabert2.baselev.10\endcsname{0.342}
\expandafter\def\csname num@rec.dnabert2.baselev.100\endcsname{0.438}
\expandafter\def\csname num@rec.dnabert2.baselev.15\endcsname{0.367}
\expandafter\def\csname num@rec.dnabert2.baselev.20\endcsname{0.383}
\expandafter\def\csname num@rec.dnabert2.baselev.25\endcsname{0.392}
\expandafter\def\csname num@rec.dnabert2.baselev.30\endcsname{0.400}
\expandafter\def\csname num@rec.dnabert2.baselev.35\endcsname{0.406}
\expandafter\def\csname num@rec.dnabert2.baselev.40\endcsname{0.411}
\expandafter\def\csname num@rec.dnabert2.baselev.45\endcsname{0.415}
\expandafter\def\csname num@rec.dnabert2.baselev.50\endcsname{0.419}
\expandafter\def\csname num@rec.dnabert2.baselev.60\endcsname{0.425}
\expandafter\def\csname num@rec.dnabert2.baselev.70\endcsname{0.429}
\expandafter\def\csname num@rec.dnabert2.baselev.80\endcsname{0.433}
\expandafter\def\csname num@rec.dnabert2.baselev.90\endcsname{0.436}
\expandafter\def\csname num@rec.dnabert2.exact.10\endcsname{0.055}
\expandafter\def\csname num@rec.dnabert2.exact.100\endcsname{0.000}
\expandafter\def\csname num@rec.dnabert2.exact.15\endcsname{0.015}
\expandafter\def\csname num@rec.dnabert2.exact.20\endcsname{0.004}
\expandafter\def\csname num@rec.dnabert2.exact.25\endcsname{0.002}
\expandafter\def\csname num@rec.dnabert2.exact.30\endcsname{0.002}
\expandafter\def\csname num@rec.dnabert2.exact.35\endcsname{0.002}
\expandafter\def\csname num@rec.dnabert2.exact.40\endcsname{0.002}
\expandafter\def\csname num@rec.dnabert2.exact.45\endcsname{0.001}
\expandafter\def\csname num@rec.dnabert2.exact.50\endcsname{0.001}
\expandafter\def\csname num@rec.dnabert2.exact.60\endcsname{0.001}
\expandafter\def\csname num@rec.dnabert2.exact.70\endcsname{0.000}
\expandafter\def\csname num@rec.dnabert2.exact.80\endcsname{0.000}
\expandafter\def\csname num@rec.dnabert2.exact.90\endcsname{0.000}
\expandafter\def\csname num@rec.dnabert2.exactpct.10\endcsname{5.5}
\expandafter\def\csname num@rec.dnabert2.exactpct.100\endcsname{0.0}
\expandafter\def\csname num@rec.dnabert2.exactpct.15\endcsname{1.5}
\expandafter\def\csname num@rec.dnabert2.exactpct.20\endcsname{0.4}
\expandafter\def\csname num@rec.dnabert2.exactpct.25\endcsname{0.2}
\expandafter\def\csname num@rec.dnabert2.exactpct.30\endcsname{0.2}
\expandafter\def\csname num@rec.dnabert2.exactpct.35\endcsname{0.2}
\expandafter\def\csname num@rec.dnabert2.exactpct.40\endcsname{0.2}
\expandafter\def\csname num@rec.dnabert2.exactpct.45\endcsname{0.1}
\expandafter\def\csname num@rec.dnabert2.exactpct.50\endcsname{0.1}
\expandafter\def\csname num@rec.dnabert2.exactpct.60\endcsname{0.1}
\expandafter\def\csname num@rec.dnabert2.exactpct.70\endcsname{0.0}
\expandafter\def\csname num@rec.dnabert2.exactpct.80\endcsname{0.0}
\expandafter\def\csname num@rec.dnabert2.exactpct.90\endcsname{0.0}
\expandafter\def\csname num@rec.dnabert2.lev.10\endcsname{0.558}
\expandafter\def\csname num@rec.dnabert2.lev.100\endcsname{0.508}
\expandafter\def\csname num@rec.dnabert2.lev.15\endcsname{0.536}
\expandafter\def\csname num@rec.dnabert2.lev.20\endcsname{0.529}
\expandafter\def\csname num@rec.dnabert2.lev.25\endcsname{0.528}
\expandafter\def\csname num@rec.dnabert2.lev.30\endcsname{0.527}
\expandafter\def\csname num@rec.dnabert2.lev.35\endcsname{0.525}
\expandafter\def\csname num@rec.dnabert2.lev.40\endcsname{0.523}
\expandafter\def\csname num@rec.dnabert2.lev.45\endcsname{0.521}
\expandafter\def\csname num@rec.dnabert2.lev.50\endcsname{0.520}
\expandafter\def\csname num@rec.dnabert2.lev.60\endcsname{0.514}
\expandafter\def\csname num@rec.dnabert2.lev.70\endcsname{0.512}
\expandafter\def\csname num@rec.dnabert2.lev.80\endcsname{0.510}
\expandafter\def\csname num@rec.dnabert2.lev.90\endcsname{0.509}
\expandafter\def\csname num@rec.dnabert2.levsd.10\endcsname{0.198}
\expandafter\def\csname num@rec.dnabert2.levsd.100\endcsname{0.089}
\expandafter\def\csname num@rec.dnabert2.levsd.15\endcsname{0.164}
\expandafter\def\csname num@rec.dnabert2.levsd.20\endcsname{0.149}
\expandafter\def\csname num@rec.dnabert2.levsd.25\endcsname{0.142}
\expandafter\def\csname num@rec.dnabert2.levsd.30\endcsname{0.134}
\expandafter\def\csname num@rec.dnabert2.levsd.35\endcsname{0.128}
\expandafter\def\csname num@rec.dnabert2.levsd.40\endcsname{0.124}
\expandafter\def\csname num@rec.dnabert2.levsd.45\endcsname{0.120}
\expandafter\def\csname num@rec.dnabert2.levsd.50\endcsname{0.118}
\expandafter\def\csname num@rec.dnabert2.levsd.60\endcsname{0.111}
\expandafter\def\csname num@rec.dnabert2.levsd.70\endcsname{0.105}
\expandafter\def\csname num@rec.dnabert2.levsd.80\endcsname{0.103}
\expandafter\def\csname num@rec.dnabert2.levsd.90\endcsname{0.095}
\expandafter\def\csname num@rec.dnabert2.strongmax\endcsname{--}
\expandafter\def\csname num@rec.dnabert2.strongmin\endcsname{--}
\expandafter\def\csname num@rec.evo2.acc.10\endcsname{0.655}
\expandafter\def\csname num@rec.evo2.acc.100\endcsname{0.597}
\expandafter\def\csname num@rec.evo2.acc.15\endcsname{1.000}
\expandafter\def\csname num@rec.evo2.acc.20\endcsname{0.999}
\expandafter\def\csname num@rec.evo2.acc.25\endcsname{0.997}
\expandafter\def\csname num@rec.evo2.acc.30\endcsname{0.992}
\expandafter\def\csname num@rec.evo2.acc.35\endcsname{0.983}
\expandafter\def\csname num@rec.evo2.acc.40\endcsname{0.966}
\expandafter\def\csname num@rec.evo2.acc.45\endcsname{0.942}
\expandafter\def\csname num@rec.evo2.acc.50\endcsname{0.908}
\expandafter\def\csname num@rec.evo2.acc.60\endcsname{0.822}
\expandafter\def\csname num@rec.evo2.acc.70\endcsname{0.741}
\expandafter\def\csname num@rec.evo2.acc.80\endcsname{0.682}
\expandafter\def\csname num@rec.evo2.acc.90\endcsname{0.633}
\expandafter\def\csname num@rec.evo2.accsd.10\endcsname{0.128}
\expandafter\def\csname num@rec.evo2.accsd.100\endcsname{0.166}
\expandafter\def\csname num@rec.evo2.accsd.15\endcsname{0.003}
\expandafter\def\csname num@rec.evo2.accsd.20\endcsname{0.012}
\expandafter\def\csname num@rec.evo2.accsd.25\endcsname{0.019}
\expandafter\def\csname num@rec.evo2.accsd.30\endcsname{0.033}
\expandafter\def\csname num@rec.evo2.accsd.35\endcsname{0.050}
\expandafter\def\csname num@rec.evo2.accsd.40\endcsname{0.075}
\expandafter\def\csname num@rec.evo2.accsd.45\endcsname{0.099}
\expandafter\def\csname num@rec.evo2.accsd.50\endcsname{0.119}
\expandafter\def\csname num@rec.evo2.accsd.60\endcsname{0.143}
\expandafter\def\csname num@rec.evo2.accsd.70\endcsname{0.154}
\expandafter\def\csname num@rec.evo2.accsd.80\endcsname{0.161}
\expandafter\def\csname num@rec.evo2.accsd.90\endcsname{0.164}
\expandafter\def\csname num@rec.evo2.baselev.10\endcsname{0.342}
\expandafter\def\csname num@rec.evo2.baselev.100\endcsname{0.438}
\expandafter\def\csname num@rec.evo2.baselev.15\endcsname{0.366}
\expandafter\def\csname num@rec.evo2.baselev.20\endcsname{0.382}
\expandafter\def\csname num@rec.evo2.baselev.25\endcsname{0.392}
\expandafter\def\csname num@rec.evo2.baselev.30\endcsname{0.401}
\expandafter\def\csname num@rec.evo2.baselev.35\endcsname{0.406}
\expandafter\def\csname num@rec.evo2.baselev.40\endcsname{0.411}
\expandafter\def\csname num@rec.evo2.baselev.45\endcsname{0.415}
\expandafter\def\csname num@rec.evo2.baselev.50\endcsname{0.419}
\expandafter\def\csname num@rec.evo2.baselev.60\endcsname{0.425}
\expandafter\def\csname num@rec.evo2.baselev.70\endcsname{0.429}
\expandafter\def\csname num@rec.evo2.baselev.80\endcsname{0.433}
\expandafter\def\csname num@rec.evo2.baselev.90\endcsname{0.436}
\expandafter\def\csname num@rec.evo2.exact.10\endcsname{0.006}
\expandafter\def\csname num@rec.evo2.exact.100\endcsname{0.016}
\expandafter\def\csname num@rec.evo2.exact.15\endcsname{0.999}
\expandafter\def\csname num@rec.evo2.exact.20\endcsname{0.987}
\expandafter\def\csname num@rec.evo2.exact.25\endcsname{0.962}
\expandafter\def\csname num@rec.evo2.exact.30\endcsname{0.899}
\expandafter\def\csname num@rec.evo2.exact.35\endcsname{0.805}
\expandafter\def\csname num@rec.evo2.exact.40\endcsname{0.667}
\expandafter\def\csname num@rec.evo2.exact.45\endcsname{0.498}
\expandafter\def\csname num@rec.evo2.exact.50\endcsname{0.318}
\expandafter\def\csname num@rec.evo2.exact.60\endcsname{0.085}
\expandafter\def\csname num@rec.evo2.exact.70\endcsname{0.046}
\expandafter\def\csname num@rec.evo2.exact.80\endcsname{0.036}
\expandafter\def\csname num@rec.evo2.exact.90\endcsname{0.024}
\expandafter\def\csname num@rec.evo2.exactpct.10\endcsname{0.6}
\expandafter\def\csname num@rec.evo2.exactpct.100\endcsname{1.6}
\expandafter\def\csname num@rec.evo2.exactpct.15\endcsname{99.9}
\expandafter\def\csname num@rec.evo2.exactpct.20\endcsname{98.7}
\expandafter\def\csname num@rec.evo2.exactpct.25\endcsname{96.2}
\expandafter\def\csname num@rec.evo2.exactpct.30\endcsname{89.9}
\expandafter\def\csname num@rec.evo2.exactpct.35\endcsname{80.5}
\expandafter\def\csname num@rec.evo2.exactpct.40\endcsname{66.7}
\expandafter\def\csname num@rec.evo2.exactpct.45\endcsname{49.8}
\expandafter\def\csname num@rec.evo2.exactpct.50\endcsname{31.8}
\expandafter\def\csname num@rec.evo2.exactpct.60\endcsname{8.5}
\expandafter\def\csname num@rec.evo2.exactpct.70\endcsname{4.6}
\expandafter\def\csname num@rec.evo2.exactpct.80\endcsname{3.6}
\expandafter\def\csname num@rec.evo2.exactpct.90\endcsname{2.4}
\expandafter\def\csname num@rec.evo2.lev.10\endcsname{0.669}
\expandafter\def\csname num@rec.evo2.lev.100\endcsname{0.665}
\expandafter\def\csname num@rec.evo2.lev.15\endcsname{1.000}
\expandafter\def\csname num@rec.evo2.lev.20\endcsname{0.999}
\expandafter\def\csname num@rec.evo2.lev.25\endcsname{0.997}
\expandafter\def\csname num@rec.evo2.lev.30\endcsname{0.993}
\expandafter\def\csname num@rec.evo2.lev.35\endcsname{0.986}
\expandafter\def\csname num@rec.evo2.lev.40\endcsname{0.973}
\expandafter\def\csname num@rec.evo2.lev.45\endcsname{0.954}
\expandafter\def\csname num@rec.evo2.lev.50\endcsname{0.929}
\expandafter\def\csname num@rec.evo2.lev.60\endcsname{0.864}
\expandafter\def\csname num@rec.evo2.lev.70\endcsname{0.799}
\expandafter\def\csname num@rec.evo2.lev.80\endcsname{0.748}
\expandafter\def\csname num@rec.evo2.lev.90\endcsname{0.701}
\expandafter\def\csname num@rec.evo2.levsd.10\endcsname{0.118}
\expandafter\def\csname num@rec.evo2.levsd.100\endcsname{0.140}
\expandafter\def\csname num@rec.evo2.levsd.15\endcsname{0.003}
\expandafter\def\csname num@rec.evo2.levsd.20\endcsname{0.010}
\expandafter\def\csname num@rec.evo2.levsd.25\endcsname{0.016}
\expandafter\def\csname num@rec.evo2.levsd.30\endcsname{0.026}
\expandafter\def\csname num@rec.evo2.levsd.35\endcsname{0.039}
\expandafter\def\csname num@rec.evo2.levsd.40\endcsname{0.056}
\expandafter\def\csname num@rec.evo2.levsd.45\endcsname{0.072}
\expandafter\def\csname num@rec.evo2.levsd.50\endcsname{0.087}
\expandafter\def\csname num@rec.evo2.levsd.60\endcsname{0.107}
\expandafter\def\csname num@rec.evo2.levsd.70\endcsname{0.119}
\expandafter\def\csname num@rec.evo2.levsd.80\endcsname{0.128}
\expandafter\def\csname num@rec.evo2.levsd.90\endcsname{0.135}
\expandafter\def\csname num@rec.evo2.strongmax\endcsname{50}
\expandafter\def\csname num@rec.evo2.strongmin\endcsname{15}
\expandafter\def\csname num@rec.ntest\endcsname{50{,}000}
\expandafter\def\csname num@rec.ntv2.acc.10\endcsname{0.983}
\expandafter\def\csname num@rec.ntv2.acc.100\endcsname{0.585}
\expandafter\def\csname num@rec.ntv2.acc.15\endcsname{0.981}
\expandafter\def\csname num@rec.ntv2.acc.20\endcsname{0.980}
\expandafter\def\csname num@rec.ntv2.acc.25\endcsname{0.979}
\expandafter\def\csname num@rec.ntv2.acc.30\endcsname{0.970}
\expandafter\def\csname num@rec.ntv2.acc.35\endcsname{0.929}
\expandafter\def\csname num@rec.ntv2.acc.40\endcsname{0.903}
\expandafter\def\csname num@rec.ntv2.acc.45\endcsname{0.886}
\expandafter\def\csname num@rec.ntv2.acc.50\endcsname{0.856}
\expandafter\def\csname num@rec.ntv2.acc.60\endcsname{0.743}
\expandafter\def\csname num@rec.ntv2.acc.70\endcsname{0.697}
\expandafter\def\csname num@rec.ntv2.acc.80\endcsname{0.663}
\expandafter\def\csname num@rec.ntv2.acc.90\endcsname{0.590}
\expandafter\def\csname num@rec.ntv2.accsd.10\endcsname{0.055}
\expandafter\def\csname num@rec.ntv2.accsd.100\endcsname{0.152}
\expandafter\def\csname num@rec.ntv2.accsd.15\endcsname{0.054}
\expandafter\def\csname num@rec.ntv2.accsd.20\endcsname{0.049}
\expandafter\def\csname num@rec.ntv2.accsd.25\endcsname{0.046}
\expandafter\def\csname num@rec.ntv2.accsd.30\endcsname{0.055}
\expandafter\def\csname num@rec.ntv2.accsd.35\endcsname{0.101}
\expandafter\def\csname num@rec.ntv2.accsd.40\endcsname{0.107}
\expandafter\def\csname num@rec.ntv2.accsd.45\endcsname{0.108}
\expandafter\def\csname num@rec.ntv2.accsd.50\endcsname{0.114}
\expandafter\def\csname num@rec.ntv2.accsd.60\endcsname{0.145}
\expandafter\def\csname num@rec.ntv2.accsd.70\endcsname{0.140}
\expandafter\def\csname num@rec.ntv2.accsd.80\endcsname{0.142}
\expandafter\def\csname num@rec.ntv2.accsd.90\endcsname{0.155}
\expandafter\def\csname num@rec.ntv2.baselev.10\endcsname{0.342}
\expandafter\def\csname num@rec.ntv2.baselev.100\endcsname{0.438}
\expandafter\def\csname num@rec.ntv2.baselev.15\endcsname{0.367}
\expandafter\def\csname num@rec.ntv2.baselev.20\endcsname{0.382}
\expandafter\def\csname num@rec.ntv2.baselev.25\endcsname{0.393}
\expandafter\def\csname num@rec.ntv2.baselev.30\endcsname{0.400}
\expandafter\def\csname num@rec.ntv2.baselev.35\endcsname{0.406}
\expandafter\def\csname num@rec.ntv2.baselev.40\endcsname{0.412}
\expandafter\def\csname num@rec.ntv2.baselev.45\endcsname{0.416}
\expandafter\def\csname num@rec.ntv2.baselev.50\endcsname{0.419}
\expandafter\def\csname num@rec.ntv2.baselev.60\endcsname{0.425}
\expandafter\def\csname num@rec.ntv2.baselev.70\endcsname{0.429}
\expandafter\def\csname num@rec.ntv2.baselev.80\endcsname{0.433}
\expandafter\def\csname num@rec.ntv2.baselev.90\endcsname{0.436}
\expandafter\def\csname num@rec.ntv2.exact.10\endcsname{0.880}
\expandafter\def\csname num@rec.ntv2.exact.100\endcsname{0.042}
\expandafter\def\csname num@rec.ntv2.exact.15\endcsname{0.826}
\expandafter\def\csname num@rec.ntv2.exact.20\endcsname{0.768}
\expandafter\def\csname num@rec.ntv2.exact.25\endcsname{0.711}
\expandafter\def\csname num@rec.ntv2.exact.30\endcsname{0.580}
\expandafter\def\csname num@rec.ntv2.exact.35\endcsname{0.477}
\expandafter\def\csname num@rec.ntv2.exact.40\endcsname{0.347}
\expandafter\def\csname num@rec.ntv2.exact.45\endcsname{0.269}
\expandafter\def\csname num@rec.ntv2.exact.50\endcsname{0.175}
\expandafter\def\csname num@rec.ntv2.exact.60\endcsname{0.081}
\expandafter\def\csname num@rec.ntv2.exact.70\endcsname{0.070}
\expandafter\def\csname num@rec.ntv2.exact.80\endcsname{0.071}
\expandafter\def\csname num@rec.ntv2.exact.90\endcsname{0.050}
\expandafter\def\csname num@rec.ntv2.exactpct.10\endcsname{88.0}
\expandafter\def\csname num@rec.ntv2.exactpct.100\endcsname{4.2}
\expandafter\def\csname num@rec.ntv2.exactpct.15\endcsname{82.6}
\expandafter\def\csname num@rec.ntv2.exactpct.20\endcsname{76.8}
\expandafter\def\csname num@rec.ntv2.exactpct.25\endcsname{71.1}
\expandafter\def\csname num@rec.ntv2.exactpct.30\endcsname{58.0}
\expandafter\def\csname num@rec.ntv2.exactpct.35\endcsname{47.7}
\expandafter\def\csname num@rec.ntv2.exactpct.40\endcsname{34.7}
\expandafter\def\csname num@rec.ntv2.exactpct.45\endcsname{26.9}
\expandafter\def\csname num@rec.ntv2.exactpct.50\endcsname{17.5}
\expandafter\def\csname num@rec.ntv2.exactpct.60\endcsname{8.1}
\expandafter\def\csname num@rec.ntv2.exactpct.70\endcsname{7.0}
\expandafter\def\csname num@rec.ntv2.exactpct.80\endcsname{7.1}
\expandafter\def\csname num@rec.ntv2.exactpct.90\endcsname{5.0}
\expandafter\def\csname num@rec.ntv2.lev.10\endcsname{0.983}
\expandafter\def\csname num@rec.ntv2.lev.100\endcsname{0.675}
\expandafter\def\csname num@rec.ntv2.lev.15\endcsname{0.982}
\expandafter\def\csname num@rec.ntv2.lev.20\endcsname{0.981}
\expandafter\def\csname num@rec.ntv2.lev.25\endcsname{0.979}
\expandafter\def\csname num@rec.ntv2.lev.30\endcsname{0.971}
\expandafter\def\csname num@rec.ntv2.lev.35\endcsname{0.936}
\expandafter\def\csname num@rec.ntv2.lev.40\endcsname{0.914}
\expandafter\def\csname num@rec.ntv2.lev.45\endcsname{0.900}
\expandafter\def\csname num@rec.ntv2.lev.50\endcsname{0.875}
\expandafter\def\csname num@rec.ntv2.lev.60\endcsname{0.789}
\expandafter\def\csname num@rec.ntv2.lev.70\endcsname{0.755}
\expandafter\def\csname num@rec.ntv2.lev.80\endcsname{0.729}
\expandafter\def\csname num@rec.ntv2.lev.90\endcsname{0.677}
\expandafter\def\csname num@rec.ntv2.levsd.10\endcsname{0.051}
\expandafter\def\csname num@rec.ntv2.levsd.100\endcsname{0.116}
\expandafter\def\csname num@rec.ntv2.levsd.15\endcsname{0.048}
\expandafter\def\csname num@rec.ntv2.levsd.20\endcsname{0.045}
\expandafter\def\csname num@rec.ntv2.levsd.25\endcsname{0.043}
\expandafter\def\csname num@rec.ntv2.levsd.30\endcsname{0.051}
\expandafter\def\csname num@rec.ntv2.levsd.35\endcsname{0.086}
\expandafter\def\csname num@rec.ntv2.levsd.40\endcsname{0.090}
\expandafter\def\csname num@rec.ntv2.levsd.45\endcsname{0.089}
\expandafter\def\csname num@rec.ntv2.levsd.50\endcsname{0.092}
\expandafter\def\csname num@rec.ntv2.levsd.60\endcsname{0.105}
\expandafter\def\csname num@rec.ntv2.levsd.70\endcsname{0.104}
\expandafter\def\csname num@rec.ntv2.levsd.80\endcsname{0.109}
\expandafter\def\csname num@rec.ntv2.levsd.90\endcsname{0.119}
\expandafter\def\csname num@rec.ntv2.strongmax\endcsname{40}
\expandafter\def\csname num@rec.ntv2.strongmin\endcsname{10}

\title{How Private Are DNA Embeddings? Inverting Foundation Model Representations of Genomic Sequences}
\author{\parbox{0.95\textwidth}{\centering
Jules Kreuer$^{1,2,*}$, Sofiane Ouaari$^{1,2}$, Nico Pfeifer$^{1,2}$\\[0.5em]
\small $^1$Methods in Medical Informatics, Department of Computer Science, University of Tuebingen, Germany\\
$^2$Institute for Bioinformatics and Medical Informatics (IBMI), University of Tuebingen, Germany\\
$^*$\texttt{jules.kreuer@uni-tuebingen.de}}}
\date{}
\begin{document}
\maketitle
\begin{abstract}
Genomic foundation models are increasingly used to share dense representations rather than raw DNA sequences. This is particularly pursued for mean-pooled representations that are later used for downstream prediction tasks. Whether this practice protects the privacy of the underlying sequence is largely unknown. We evaluated model-inversions against zero-shot embeddings from \textit{DNABERT-2}, \textit{Evo~2}, and Nucleotide Transformer v2 (\textit{NTv2}). Substantial reconstruction remains possible: a single multi-length Length-Conditioned Query Decoder reconstructed \textit{NTv2} sequences with mean Levenshtein similarity from $0.98$ at 10 nucleotides to $0.68$ at 100, and \textit{Evo~2} sequences from nearly $1.00$ at 15 nucleotides to $0.67$ at 100. \textit{DNABERT-2} showed higher resistance ($0.56$ to $0.51$), although it remained above the random baseline at short lengths. Therefore, we further evaluated, whether a linkage attack similar to Homer's attack is possible based on the mean-pooled representations of human DNA sequences. At SNP-enriched length-25 loci and known coordinates, \textit{Evo~2} and \textit{NTv2} reached 99\% identification at $\num{id.ntv2.snp_enriched.k99}$ loci and \textit{DNABERT-2} at $\num{id.dnabert2.snp_enriched.k99}$ loci. With alignment-based coordinate recovery, the corresponding 99\% budgets were $\num{id.evo2.snp_enriched.blind.k99}$ and $\num{id.ntv2.snp_enriched.blind.k99}$ loci for \textit{Evo~2} and \textit{NTv2}, while \textit{DNABERT-2} did not reach the reported identification thresholds. Random loci required larger budgets. These findings show that mean-pooled embeddings can leak sequence and identity, with risk depending on sequence length, representation, locus informativeness, and coordinate information.
\end{abstract}


\section{Introduction}

Foundation models are large, broadly pre-trained models whose representations can be adapted to many downstream tasks \citep{bommasani2021opportunities, awais2025foundation, zhou2025comprehensive}. This approach is especially attractive in genomics, where unlabelled sequence data are abundant but task-specific labels are scarce.

Genomic foundation models are trained on reference, population, and multi-species genomes and support downstream tasks such as promoter prediction, functional-variant annotation, and splice-site prediction \citep{guo2025foundation}. Embeddings-as-a-Service (EaaS) and Representation-as-a-Service (RaaS) expose such representations to a second party without transferring the original model or the raw sequence \citep{adilazero, ouaari2025robust}.

Concrete genomic workflows already materialise short, locus-level pooled representations. Feng et al. \citep{feng2025benchmarking} benchmarked frozen DNA foundation models by mean-pooling non-padding token states into one vector per input sequence and fitting lightweight downstream classifiers; their suite includes a 70-nt promoter-identification task \citep{feng2025benchmarking}. Independently, Benegas et al. \citep{benegas2023dna} average contextual nucleotide embeddings over non-overlapping 100-bp genomic windows and use the resulting locus-level vectors for visualisation and logistic-regression classification of genomic regions. These studies describe research pipelines rather than documented services releasing patient embeddings, but they establish that short, fixed-dimensional genomic representations are operational downstream artefacts, not merely a hypothetical interface.

More generally, existing work benchmarks DNA foundation models by the downstream utility of their embeddings \citep{marin2024bend,feng2025benchmarking}: an embedding is judged good when it separates classes well and transfers across tasks. That criterion rewards representations that retain as much sequence information as possible, which is precisely what makes the privacy question urgent. The same density that makes an embedding useful may also make it invertible.

This matters more for genomic data than for most other modalities. Learned representations can inadvertently encode sensitive information \citep{naveed2015privacy, bonomi2020privacy}, and a genome is immutable and uniquely identifying: a leak cannot be revoked, reissued, or outgrown, and it implicates biological relatives who never consented to the release.

\begin{figure*}
    \centering
    \includegraphics[width=0.9\linewidth,trim={0 0cm 0 0cm},clip]{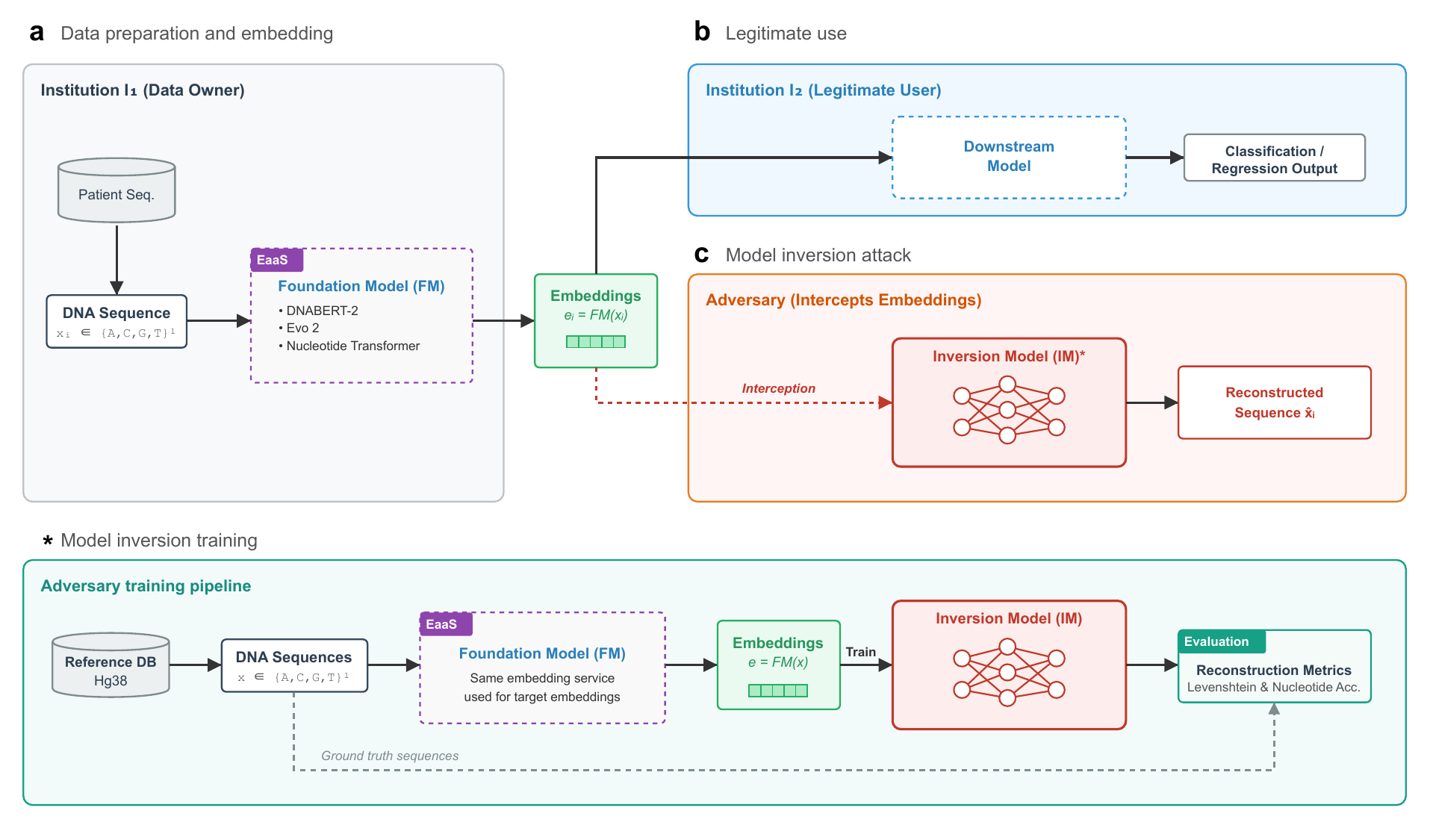}
    \caption{Threat model and training pipeline. An institution shares mean-pooled genomic foundation-model embeddings. An attacker trains a compact decoder on auxiliary sequence--embedding pairs, reconstructs intercepted sequences, and can subsequently link reconstructions to individuals in a candidate panel.}
    \label{fig:overall_attack_scenario}
\end{figure*}

We therefore benchmark the privacy of shared genomic embeddings under model inversion, covering \textit{DNABERT-2} \citep{zhou2024dnabert}, \textit{Evo~2} \citep{brixi2026genome}, and Nucleotide Transformer v2 (\textit{NTv2}) \citep{nucleotidetransf}. Our primary setting is fixed-dimensional mean pooling, which removes the explicit sequence axis but may retain order-sensitive information inside contextual token states. Our contributions are an end-to-end reconstruction benchmark across 14 sequence lengths, a single variable-length decoder trained once per foundation model, and a cohort-linkage evaluation that examines reconstruction and two distinct ways of attaching reconstructed sequences to genomic coordinates. Figure~\ref{fig:overall_attack_scenario} illustrates the threat model and training pipeline.

\section{Background}
\subsection{Model Inversion Attack}
Model inversion attacks, first introduced by Fredrikson et al.~\citep{fredrikson2015model}, are a type of privacy attack that aims to reconstruct the features of input data based on either the classification output or the representation provided by a machine learning (ML) model. To achieve this, an adversary can utilise either white-box or black-box access to the model. These attacks can be classified as \textbf{optimisation-based}~\citep{zhang2020secret, nguyen2023re, wu2023learning} or \textbf{training-based}~\citep{zhou2023boosting}. We use the latter in our study.

Consider a model $\mathcal{M}$ trained on a private dataset $D_{\text{priv}} = \{(x_i, y_i)\}_{i=1}^{n}$. Training-based inversion attacks aim to recover sensitive input data by learning an inversion model $\mathcal{A}$ parameterised as a decoder network. The inversion model is optimised to minimise the reconstruction loss $L = R(x, \mathcal{A}(\mathcal{M}(x)))$, where $R$ denotes a reconstruction metric that quantifies the fidelity between the original input $x$ and its reconstruction $\mathcal{A}(\mathcal{M}(x))$.

\subsection{DNA Foundation Models}
In this work, we consider embeddings from three popular DNA foundation models: \textit{DNABERT-2}, \textit{Evo~2}, and \textit{NTv2}.

\textbf{DNABERT-2} is a transformer-based model for DNA sequence analysis that advances its predecessor by replacing k-mer tokenisation with Byte Pair Encoding (BPE), which is a popular encoding technique for language models \citep{radford2019language}. This approach creates variable-length tokens by iteratively merging frequent nucleotide pairs, enabling more efficient genome representation and improved sample efficiency. Unlike fixed-length k-mer tokenisation, BPE's variable-length tokens present a more challenging prediction task during training, as the model must simultaneously predict both the number and identity of masked nucleotides, ultimately enhancing its understanding of genomic semantic structure. Its tokeniser has a vocabulary of 4{,}096 tokens.

\textbf{Evo~2} is a large-scale foundation model trained on a wide collection of genomes that captures the breadth of observed evolutionary diversity. Rather than focusing on task-specific optimisation, \textit{Evo~2} prioritises broad generalist capabilities, demonstrating strong performance in both prediction and generation tasks that span from molecular-level analyses to genome-scale applications across all domains of life. Two model variants were developed with 7 billion and 40 billion parameters, respectively, utilising a training corpus exceeding 9.3 trillion tokens at single-nucleotide resolution. Its character-level tokeniser yields one content token per nucleotide; the separate attack decoder predicts A/C/G/T plus EOS. Both models support an extended context window of up to 1 million tokens and exhibit effective information retrieval capabilities throughout the entire contextual range.

\textbf{Nucleotide Transformer v2} (\textit{NTv2}) is a BERT-based model pre-trained via masked language modelling on a multi-species genomic corpus; we use the 500M multi-species checkpoint. It uses 6-mer tokenisation with single-nucleotide tokens for remaining positions when the sequence length is not divisible by six or is interrupted by \textit{N}, giving a vocabulary of 4{,}107 tokens. \textit{NTv2} improves upon standard BERT by incorporating rotary positional embeddings \citep{su2024roformer}.

\section{Methods}
\subsection{General Pipeline}
We consider a dataset of DNA sequences $\mathcal{D} = \{x_1, x_2, \ldots, x_N\}$, where each sequence $x_i \in \{A, C, G, T\}^{l}$ has length $l$. Given a DNA foundation model $\mathcal{F}$, we obtain a corresponding set of embeddings $\mathcal{E} = \{e_1, e_2, \ldots, e_N\}$, with  $e_i = \mathcal{F}(x_i)$.

For the main benchmark, we chose extraction points that follow common embedding-extraction conventions: final-layer hidden states for \textit{DNABERT-2} and \textit{NTv2}, and an intermediate MLP representation for \textit{Evo~2} (\texttt{blocks.26.mlp.l3}). Supplementary Note~\ref{subsec:layer_sensitivity} evaluates sensitivity to this choice.

\textbf{Mean-pooled embeddings}: We obtain a fixed-size representation by averaging the contextual hidden states of the content tokens. For a sequence $x_i$, let $n$ denote the number of content tokens produced by the foundation model's tokeniser and let $h_i^{(j)} \in \mathbb{R}^{d}$ denote the hidden state of token $j$. The mean-pooled embedding is
\[
e_i = \frac{1}{n} \sum_{j=1}^{n} h_i^{(j)} \in \mathbb{R}^{d}.
\]
This aggregation removes the explicit token-position axis and produces a single fixed-dimensional vector. It does not make the representation order-invariant in a strict sense: the contextual states being averaged can themselves encode sequence order and long-range context.

We define our privacy threat scenario as follows: we consider two institutions, $\mathcal{I}_1$ and $\mathcal{I}_2$, that have agreed to collaborate in an Embedding-as-a-Service (EaaS) framework for a downstream task involving genomic data. Institution $\mathcal{I}_1$ possesses a labelled dataset $\mathcal{D} = \{(x_i, y_i)\}_{i=1}^{N}$, where $x_i \in \{A, C, G, T\}^{l}$ represents a DNA sequence, $y_i$ denotes its corresponding label and $N$ is the number of sequences. To preserve privacy while enabling collaboration, $\mathcal{I}_1$ transforms the original dataset into an embedding-based variant $\mathcal{D}^{(\text{emb})} = \{(e_i, y_i)\}_{i=1}^{N}$, where $e_i = \mathcal{F}(x_i)$ is the embedding generated by a DNA foundation model $\mathcal{F}$. This transformed dataset $\mathcal{D}^{(\text{emb})}$ is then shared with institution $\mathcal{I}_2$ for training downstream models.

We assume the presence of an adversary $\mathcal{A}$ who intercepts the shared embedding dataset $\mathcal{D}^{(\text{emb})}$. The adversary's objective is to perform a model inversion attack by training a reconstruction model $\mathcal{M}: \mathbb{R}^{d} \to \{A, C, G, T\}^{*}$ that aims to recover the original DNA sequences from their embeddings. Formally, given an embedding $e_i$, the adversary seeks to reconstruct the corresponding sequence $\hat{x}_i = \mathcal{M}(e_i)$,
where $\hat{x}_i$ represents the reconstructed approximation of the original sequence $x_i$. The success of this attack would compromise the privacy guarantees that the embedding transformation was intended to provide.

We assume that the adversary can infer the originating foundation model from the distinct shape and distribution of its embeddings. The attack is nevertheless representation-specific: the adversary must also identify the extracted layer and pooling method, although these properties should be straightforward to infer in practice from metadata, embedding dimensionality, or a small set of model queries.
\subsection{Metrics}
Our reconstruction metric is normalized Levenshtein similarity. Levenshtein distance \citep{levenshtein1966binary, berger2020levenshtein} is the minimum number of single-nucleotide substitutions, insertions, and deletions required to transform one sequence into another. We report $\texttt{sim}_{\texttt{lev}}(x_1, x_2) = 1 - \texttt{lev}(x_1, x_2) / \max(|x_1|, |x_2|) \in [0, 1]$, where 1 indicates identical sequences. The dynamic-programming definition is provided in Supplementary Note~\ref{subsec:levenshtein}; the secondary positionwise match rate is defined in Supplementary Note~\ref{sec:detailed_metrics}.
\subsection{Inversion Model}
For the mean-pooled scenario, we train one Length-Conditioned Query Decoder (LCQD) per foundation model on all sequence lengths jointly. The LCQD maps a fixed-dimensional mean embedding to a variable-length token sequence and is evaluated at every length used during training. A preliminary internal architecture sweep over transformer encoder-only, transformer decoder-only, ResNet, and nearest-neighbour alternatives motivated this choice; configurations and baseline results are reported in Supplementary Notes~\ref{sec:model_architecture} and~\ref{sec:per_length_sweep}, respectively.

\textbf{Length-Conditioned Query Decoder (LCQD)}: The LCQD works by projecting the mean embedding to a context of $n_{\text{context}}$ learnable memory tokens (context projection). Simultaneously, a length head predicts a distribution over the number of output content tokens directly from the mean embedding, which is mixed with a length embedding to condition the position queries. The position queries, together with the length conditioning, are passed to a stack of transformer decoder layers that cross-attend to the context tokens and self-attend to the queries. The output layer predicts per-position token logits over the vocabulary, including the end-of-sequence (EOS) token. At inference, decoding stops at the first predicted EOS. The complete decoded string is compared with the true sequence without truncating either string to the other's length, so premature or delayed EOS predictions are penalised by Levenshtein similarity.

\subsection{Cohort Identification (Linkage Attack)}
Homer's attack \citep{homer2008resolving} shows that an individual's presence in a genomic mixture can be inferred by accumulating weak signals across many variants, with implications for sharing aggregate cohort statistics. Our linkage attack instead identifies an individual within a candidate panel using reconstructed haplotypes from shared mean-pooled embeddings.

The ultimate threat of reconstruction attacks in genomic EaaS frameworks \citep{marin2024bend,vilov2025investigating,awasthi2025unsupervised,feng2025benchmarking, tang2025evaluating} is the de-anonymisation of individuals by linking their reconstructed sequences to database profiles (linkage attacks). We consider a closed-set linkage scenario where an adversary intercepts embeddings of both phased haplotypes of a target individual $t$ across a set of $k$ coordinates $\mathcal{K}$, also known as loci, reconstructs the sequences using the trained inversion model, and attempts to identify $t$ within a cohort of $N$ candidates that contains the target.

Two independent factors govern how much identifying evidence such an attack collects, and conflating them makes the risk appear either uniformly high or uniformly negligible. The first is \emph{which loci} are released; the second is \emph{how much the attacker knows} about where those loci sit in the genome. We therefore vary both.

\textbf{\textit{Which loci} (random versus SNP-enriched)}: Reconstruction similarity describes how faithfully a window is recovered, but only positions that differ between candidates can carry identifying information. Random sampling draws from eligible held-out autosomal reference windows. SNP-enriched sampling requires a source-VCF SNP with minor allele frequency at least 5\% (using the VCF allele frequency, not an estimate from the selected cohort). Both modes enforce at least 50 kb between selected windows. The enriched mode is an informative common-variant panel; it is not a generic model of exome, disease-region, or clinical embedding releases.

We evaluate embeddings generated from sequences in the reference orientation, which is assumed known to the attacker. This provides a consistent orientation for reconstruction and linkage. Our results therefore do not establish attack performance when sequence orientation is unknown or randomised.

\textbf{\textit{Positional knowledge of the attacker}:} We evaluate two scenarios and describe a third attack route:
\textbf{Locus-blind adversary} (evaluated): the attacker does not know which loci the intercepted embeddings represent, but knows their nominal window length, forward orientation, autosomal scope, and target grouping. The coordinates are sought by aligning each complete decoded reconstruction with the reference genome (BWA-MEM~\citep{li2013bwamem} with high gap penalties, enforcing ungapped alignments without soft clipping); up to five retained hits can contribute to candidate scoring. The decoded sequence is neither truncated nor padded before alignment or candidate scoring, so premature and delayed EOS predictions remain penalised.

\textbf{Oracle adversary} (partially-informed, evaluated): the attacker knows the exact genomic coordinates of every query and maps each reconstruction directly onto its locus in the candidate panel. Reconstruction is still required, because the panel holds genotypes rather than embeddings. This provides a known-coordinate reference attack for the evaluated reconstruction and scoring procedure.

\textbf{Direct embedding matching} (not evaluated): an attacker who knows the coordinates could embed candidate haplotypes and compare their embeddings directly with the intercepted representations. The number of additional foundation-model evaluations depends on the number of unique candidate sequences, caching, and batching; this study measures neither its cost nor its accuracy.

Coordinate knowledge is a continuum rather than a binary choice. Provenance metadata or assay information may narrow a query to a region without revealing its exact coordinate. The two evaluated configurations also differ in scoring and candidate-panel construction, so their accuracy gap does not isolate the effect of coordinate knowledge. A controlled coordinate comparison would hold those components fixed.

For the oracle evaluation, each intercepted locus contributes two embeddings, one for each phased chromosome copy of target $t$. The inversion model decodes these embeddings separately. Let $\hat{x}_{t,\ell,h}$ denote the reconstructed haplotype sequence for chromosome copy $h\in\{1,2\}$ of target $t$ at locus $\ell$, and let $x_{c,\ell,h}$ denote the true haplotype sequence of candidate $c$. Here, $h$ labels the two homologous chromosome copies, not strand orientation.

Because haplotype labels are arbitrary across individuals, we evaluate both possible one-to-one pairings at each locus: matching copies 1 and 2 directly, or swapping their assignment. Let $\mathcal{P}=\{(1,2),(2,1)\}$ denote these pairings. The locus score is the higher mean similarity:
\begin{equation}
\begin{aligned}
q_{t,\ell}(c) &= \frac{1}{2}\max_{(u,v)\in\mathcal{P}}\Bigl[
\operatorname{sim}_{\mathrm{lev}}\!\left(\hat{x}_{t,\ell,1},x_{c,\ell,u}\right)\\
&\qquad +\operatorname{sim}_{\mathrm{lev}}\!\left(\hat{x}_{t,\ell,2},x_{c,\ell,v}\right)\Bigr].
\end{aligned}
\end{equation}
The pairing is selected independently at each locus. For a set $\mathcal{K}$ of $k$ intercepted loci, we sum the locus scores and identify the candidate with the highest total:
\begin{equation}
S_t(c;k)=\sum_{\ell\in\mathcal{K}}q_{t,\ell}(c),
\qquad
\hat{c}_t=\arg\max_c S_t(c;k).
\end{equation}
For the locus-blind evaluation, each reconstructed haplotype is compared independently with the closer candidate haplotype at retained alignment hits, and the two distances are summed across loci. Unlike the known-coordinate score, this rule need not pair the two reconstructions with distinct candidate copies. The locus-blind implementation also drops all multiply represented VCF positions when constructing candidate haplotypes, whereas the known-coordinate panel retains valid split multiallelic variants and drops only conflicting assignments. The observed difference between arms therefore includes scoring and panel differences as well as coordinate recovery. Shared, monomorphic windows cannot distinguish candidates under either score; polymorphic windows can. We report the fraction of targets for which the correct individual is ranked first (top-1) as a function of $k$.

\section{Datasets}
We use the human reference genome (\texttt{GRCh38}/\texttt{hg38}) as our primary dataset. Rather than drawing uniformly from the combinatorial space of $4^l$ DNA strings, hg38 restricts both the attacker's auxiliary data and the evaluation targets to a biologically plausible, human-reference-supported subset. This represents a realistic public prior for an attacker and makes the learning problem tractable; it also means that the benchmark estimates inversion for reference-like human windows rather than arbitrary DNA strings. Although hg38 is a publicly available composite assembly from multiple anonymous donors, it provides a controlled and reproducible test bed for evaluating inversion attacks. Individual genomes additionally contain person-specific variants and ancestry-informative markers \citep{bollas2024snvstory, spiliopoulou2015genomic, lippert2017identification}. The percentage of the combinatorial $4^l$ space represented by the benchmark is reported in Supplementary Note~\ref{subsec:data_embedding}.

To evaluate beyond the reference genome, we additionally use individual-specific sequences derived from the 1000 Genomes Project~\citep{1000genomes}. We take autosomal test loci from the coordinate-disjoint \texttt{hg38} corpus and apply biallelic SNPs from a randomly selected phased individual and haplotype. As shown in Supplementary Note~\ref{sec:1000g_cross_dataset}, aggregate reconstruction performance resembles the \texttt{hg38} results. Most evaluated windows are reference-identical, so aggregate agreement alone is limited evidence for recovery of person-specific variation. Supplementary Note~\ref{sec:1000g_cross_dataset} separately reports whole-window similarity and exact-match counts for variant-bearing windows absent as exact strings from decoder training; a site-level alternate-allele recovery rate was not measured.

\section{Experiments and Results}
\subsection{Empirical Collision Audit}
Prior to model inversion, we inspected whether distinct sequences produced exact or near-colliding mean-pooled embeddings. Here, a near collision refers only to a pair of embeddings separated by a small normalised Euclidean distance. Exact collisions would preclude unique recovery for the affected sequences, while near collisions may make reconstruction less stable; however, this finite diagnostic neither establishes injectivity nor determines whether an embedding is practically invertible. Formally, a function $f: A \to B$ is injective if $\forall x_1, x_2 \in A, f(x_1) = f(x_2) \Rightarrow x_1 = x_2$; recent work analyses conditions under which transformer representations may be injective \citep{injective-nikolaou2025}.

\begin{figure}[h]
    \centering
    \includegraphics[width=\linewidth]{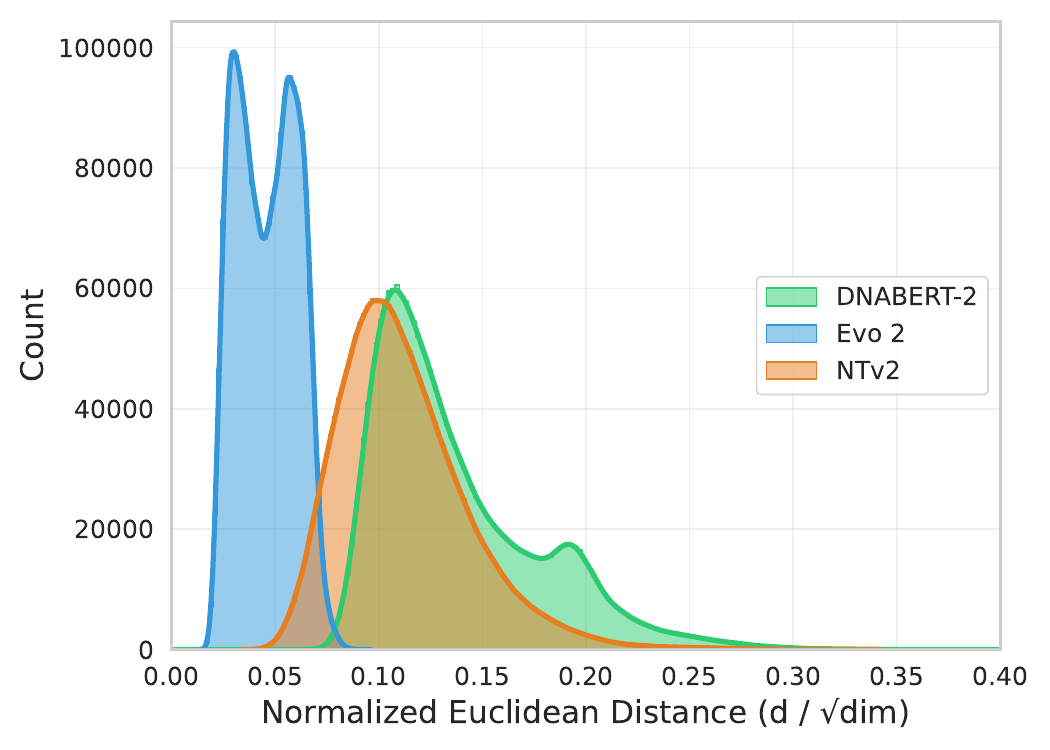}
    \caption{Empirical collision audit for mean-pooled embeddings of sequences of length $l=100$. The figure shows normalised Euclidean distances for all pairs in a random subsample of $2{,}000$ sequences (Supplementary Note~\ref{subsec:collision_all_seqlens} reports all lengths).}
    \label{fig:collision_audit}
\end{figure}

Given a set of mean-pooled embeddings $\mathcal{E} = \{e_1, e_2, \ldots, e_N\}$ where $e_i \in \mathbb{R}^d$ for all $i \in \{1, 2, \ldots, N\}$, the pairwise normalised Euclidean distance matrix $\mathbf{D} \in \mathbb{R}^{N \times N}$ is defined element-wise as:
\[
D_{ij} = \frac{\|e_i - e_j\|_2}{\sqrt{d}} = \frac{\sqrt{\sum_{k=1}^{d} (e_i^{(k)} - e_j^{(k)})^2}}{\sqrt{d}}
\]
where $e_i^{(k)}$ denotes the $k$-th component of embedding $e_i$. \autoref{fig:collision_audit} shows the distribution at $l=100$. In this finite empirical audit, which is not a test of injectivity over the complete sequence domain, we observe no zero-distance pairs among the sampled points. The same sampled diagnostic is reported across all lengths in Supplementary Note~\ref{subsec:collision_all_seqlens}.

\begin{figure}[h]
    \centering
    \includegraphics[width=\linewidth]{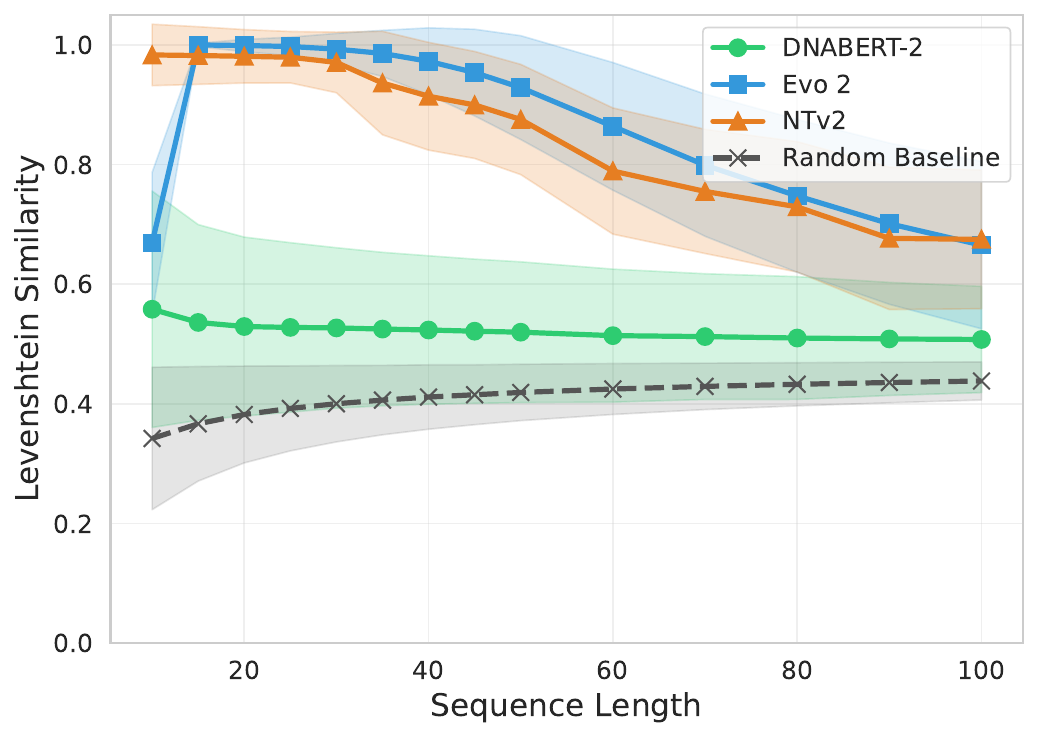}
    \caption{Mean-pooled reconstruction performance across sequence lengths for the multi-length LCQD architecture, measured by normalised Levenshtein similarity.}
    \label{fig:levenshtein_reconstruction}
\end{figure}
\subsection{Reconstruction Evaluation}

In our analysis, we evaluate DNA sequences sampled from the default chromosomes (chr1--chr22, chrX, chrY, chrM) of the human reference genome \texttt{hg38} across sequence lengths $l \in \{10, 15, \dots, 50, 60, \dots, 100\}$. The theoretical search space is $4^l$ (e.g., $4^{100} \approx 1.6 \times 10^{60}$), whereas the benchmark deliberately uses the much smaller reference-supported subset quantified in Supplementary Figure~\ref{fig:sequence_space_coverage}. All sampled intervals are coordinate-disjoint, including across sequence lengths and train, validation, and test splits. We train a single multi-length model per foundation model on the union of all lengths. The corpus size scales from $1{,}000{,}000$ distinct sampled windows per length for $l \leq 20$ to $4{,}000{,}000$ for $l \geq 90$; a fixed $\num{rec.ntest}$ sequences per length are held out for validation and $\num{rec.ntest}$ for testing. As a random baseline, we sample nucleotides according to the empirical target distribution, yielding $\num{rec.baseline.min}$--$\num{rec.baseline.max}$\% mean Levenshtein similarity depending on length.

\textbf{Mean-Pooled Reconstruction.}
Pooling compresses the position-specific hidden states into a single fixed-dimensional vector, creating a more demanding inversion problem. \autoref{fig:levenshtein_reconstruction} compares the three multi-length LCQDs on $\num{rec.ntest}$ held-out sequences per length. The attack remains strong for \textit{Evo~2} and \textit{NTv2}, but its behaviour varies by sequence length and foundation model.

\textit{DNABERT-2} is the most resilient model, with mean Levenshtein similarity decreasing from $\num{rec.dnabert2.lev.10} \pm \num{rec.dnabert2.levsd.10}$ at $l=10$ to $\num{rec.dnabert2.lev.100} \pm \num{rec.dnabert2.levsd.100}$ at $l=100$. Edit-based similarity remains above its length-matched random baseline, but the present comparison does not isolate which architectural, training, layer-selection, or tokenisation differences cause the gap between foundation models. \textit{Evo~2} exhibits a distinctive non-monotonic pattern. Similarity is $\num{rec.evo2.lev.10} \pm \num{rec.evo2.levsd.10}$ at $l=10$, rises to $\num{rec.evo2.lev.15} \pm \num{rec.evo2.levsd.15}$ at $l=15$ and $\num{rec.evo2.lev.20} \pm \num{rec.evo2.levsd.20}$ at $l=20$, and then decreases steadily to $\num{rec.evo2.lev.100} \pm \num{rec.evo2.levsd.100}$ at $l=100$. The anomalously weaker performance of \textit{Evo~2} at $l=10$ may reflect a short-context boundary effect, as causal convolution boundaries dominate a larger proportion of very short sequences and may be amplified by mean pooling. Consistent with this hypothesis, cosine and Euclidean Spearman correlations increase from $\num{emb.evo2.spearman.cosine.10}$ and $\num{emb.evo2.spearman.euclidean.10}$ at $l=10$ to $\num{emb.evo2.spearman.cosine.20}$ and $\num{emb.evo2.spearman.euclidean.20}$ at $l=20$, although confirming the mechanism would require operator-specific ablations. For \textit{NTv2}, Levenshtein similarity decreases from $\num{rec.ntv2.lev.10} \pm \num{rec.ntv2.levsd.10}$ at $l=10$ to $\num{rec.ntv2.lev.100} \pm \num{rec.ntv2.levsd.100}$ at $l=100$. Full numerical results, including secondary diagnostics, are given in Supplementary Table~\ref{tab:query_decoder_results}, and the preliminary per-length architecture sweep that motivated this decoder is reported in Supplementary Note~\ref{sec:per_length_sweep}.

\textbf{Embedding Geometry Audit.}
Supplementary Note~\ref{subsec:embedding_analysis} reports Spearman associations between pairwise embedding similarity and sequence similarity. These statistics describe whether nearby sequences tend to occupy nearby regions of embedding space, but they do not consistently rank reconstruction performance across models and lengths. We therefore treat them as descriptive geometry diagnostics rather than predictors of attack success.

\subsection{Cohort Identification Performance}
\begin{figure*}[!t]
    \centering
    \includegraphics[width=\textwidth]{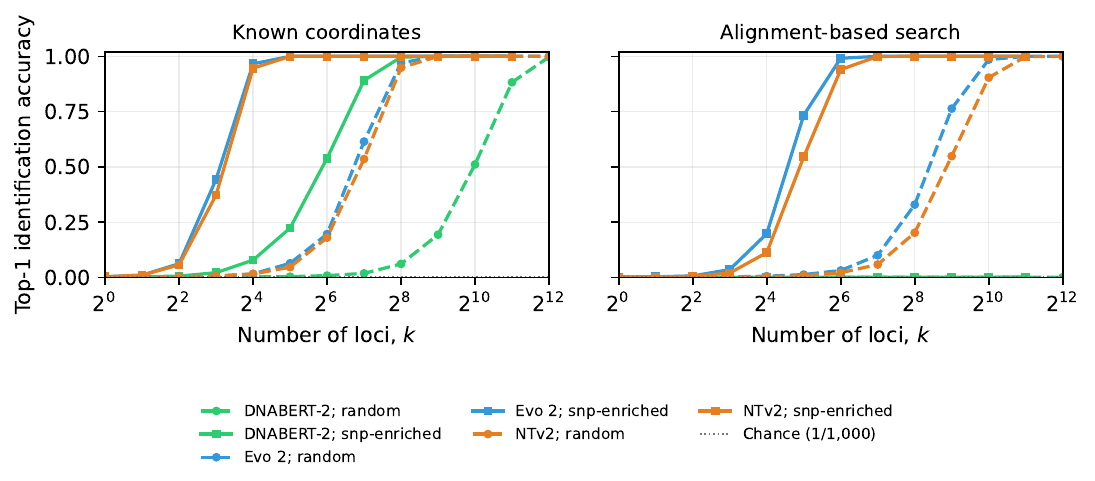}
    \caption{Top-1 cohort identification accuracy for $\num{id.cohort}$ individuals using length-25 reconstructions, averaged over 50 seeded locus permutations of the same cohort and locus pools. Colours identify foundation models consistently with Figure~\ref{fig:levenshtein_reconstruction}; solid lines with square markers use SNP-enriched loci and dashed lines with circular markers use random loci. The left panel scores reconstructed haplotypes at their known coordinates with one-to-one diploid pairing. The right panel searches for coordinates by aligning complete EOS-decoded reconstructions and independently matches each reconstructed haplotype to the closer candidate copy across retained hits. Candidate-panel handling of split multiallelic variants also differs between panels. The separation between panels therefore does not isolate the effect of coordinate knowledge.}
    \label{fig:identification_performance}
\end{figure*}

We evaluate linkage at sequence length 25 on $N=\num{id.cohort}$ individuals from the 1000 Genomes Project, sweeping the locus budget from $k=1$ to $\num{id.ntv2.random.kmax}$ for random loci and to $\num{id.ntv2.snp_enriched.kmax}$ for SNP-enriched loci. The two pools contain $\num{id.ntv2.random.nloci}$ and $\num{id.ntv2.snp_enriched.nloci}$ eligible loci, respectively. Each point is an empirical mean over 50 seeded locus permutations of the same cohort and locus pool. The cohort haplotypes contain real phased 1000 Genomes variants at held-out loci that are coordinate-disjoint from the auxiliary training windows. Results are shown in \autoref{fig:identification_performance} and Supplementary Table~\ref{tab:identification_results}.

\textbf{Known coordinates and informative loci make linkage practical.} With SNP-enriched windows and known coordinates, the smallest evaluated locus budgets at which both \textit{Evo~2} and \textit{NTv2} exceed 90\% and 99\% top-1 accuracy are $k=\num{id.ntv2.snp_enriched.k90}$ and $k=\num{id.ntv2.snp_enriched.k99}$, respectively. \textit{DNABERT-2} first crosses 50\%, 90\% and 99\% at the evaluated budgets $k=\num{id.dnabert2.snp_enriched.k50}$, $\num{id.dnabert2.snp_enriched.k90}$ and $\num{id.dnabert2.snp_enriched.k99}$, respectively. Thus between tens and a few hundred short windows suffice in this SNP-enriched setting. The \textit{DNABERT-2} result also shows that high mean sequence fidelity is not necessary for ranking: smaller genotype-consistent differences can accumulate across many known loci even when the full reconstruction is imperfect.

\textbf{Locus informativeness changes the required budget.} Under known coordinates, the smallest evaluated random-window budgets reaching 99\% are $k=\num{id.evo2.random.k99}$ for \textit{Evo~2}, $\num{id.ntv2.random.k99}$ for \textit{NTv2}, and $\num{id.dnabert2.random.k99}$ for \textit{DNABERT-2}, compared with $\num{id.ntv2.snp_enriched.k99}$--$\num{id.dnabert2.snp_enriched.k99}$ for SNP-enriched windows. Positions shared by every candidate do not change their ordering under this score, whereas segregating variants can contribute identity signal.

\textbf{Coordinate recovery is model-dependent rather than uniformly prohibitive.} In the evaluated alignment-based configuration, the smallest SNP-enriched budget reaching 99\% is $k=\num{id.evo2.snp_enriched.blind.k99}$ for \textit{Evo~2} and $k=\num{id.ntv2.snp_enriched.blind.k99}$ for \textit{NTv2}. On random windows the corresponding budgets are $\num{id.evo2.random.blind.k99}$ and $\num{id.ntv2.random.blind.k99}$. In contrast, neither \textit{DNABERT-2} alignment curve reaches 50\% top-1 accuracy within the evaluated locus budget. DNABERT-2 nevertheless has at least one usable alignment for $\num{id.dnabert2.random.mappedpct}$\% of random-window queries and $\num{id.dnabert2.snp_enriched.mappedpct}$\% of SNP-enriched queries. Its near-chance linkage therefore cannot be attributed to every query failing to align; usable hits alone do not establish correct or genotype-informative placement. These configurations do not isolate the effect of coordinate suppression because their scoring rules and candidate panels also differ. Longer windows or provenance metadata may change alignment performance; direct embedding matching is a separate, unevaluated attack route.

\textbf{Exact-string training-unseen sensitivity.} Excluding loci containing any cohort haplotype sequence observed in decoder training left the SNP-enriched 99\% identification thresholds unchanged. This control addresses exact-string overlap with decoder training, but not overlap with foundation-model pretraining.
\section{Discussion}

In this work, we systematically evaluated the privacy of DNA foundation-model embeddings against model inversion attacks in an Embeddings-as-a-Service (EaaS) setting. The results show that sharing embeddings instead of raw sequences can reduce direct exposure, but does not by itself prevent recovery of sensitive sequence information.

\textbf{Foundation models differ substantially in reconstruction vulnerability.} At the extraction points used in the main benchmark, mean-pooled \textit{Evo~2} and \textit{NTv2} embeddings retain enough information for high-fidelity reconstruction over a substantial part of the evaluated length range, whereas \textit{DNABERT-2} is consistently more difficult to invert. The comparison does not isolate a single cause: architecture, pretraining data, model scale, tokenization, and the selected representation layer all differ between models. In the layer-sensitivity study, earlier \textit{DNABERT-2} and intermediate \textit{NTv2} layers were easier to invert than their final layers, while \textit{Evo~2} block 26 gave the highest reconstruction similarity among its tested extraction points (Supplementary Table~\ref{tab:layer_sensitivity}). Comparisons of reconstruction vulnerability therefore depend on the extraction point, sequence length and attack budget.

\textbf{Substantial reconstruction remains possible after mean pooling.} Averaging removes the explicit token-position axis of the embeddings. Even so, \textit{NTv2} sequences up to $\num{rec.ntv2.strongmax}$ nucleotides and \textit{Evo~2} sequences from $\num{rec.evo2.strongmin}$ to $\num{rec.evo2.strongmax}$ nucleotides exceed 90\% mean Levenshtein similarity. Moreover, one decoder per foundation model covers all evaluated lengths, showing that an attacker does not need to train a separate inversion model for every sequence length exposed by a service.

\textbf{Sequence length remains an important privacy parameter.} Shorter sequences are generally easier to reconstruct from the fixed-dimensional pooled representation, while longer windows contain more genomic content and have a greater opportunity to include identifying variants. These effects act in opposite directions. Consequently, there is no single sequence length that can be considered intrinsically safe; release decisions must account for both reconstruction fidelity and the biological information contained in the shared windows.

\textbf{Reconstruction can translate into individual identification.} The linkage experiments show that imperfectly reconstructed windows can still identify an individual when small genotype-consistent differences accumulate across informative loci. SNP-enriched windows require far fewer observations than randomly sampled windows in this selected common-variant panel. Individual-derived clinical sequences can contain variants, but that alone does not imply the same enrichment or spacing rule. The known-coordinate configuration needs fewer loci than the alignment-based configuration; their difference cannot be assigned solely to coordinate access. Alignment-based coordinate recovery remains sufficient for near-perfect \textit{Evo~2} and \textit{NTv2} linkage, while the weaker \textit{DNABERT-2} reconstructions do not yield effective identification in this setting. Thus, reconstruction quality is central but not sufficient to characterise privacy risk: locus variation and adversarial side information also determine whether sequence leakage becomes identifying.

\textbf{Practical implications.} Mean pooling should be regarded as an obfuscation step rather than a privacy guarantee, particularly for short sequence windows. Embedding releases should be evaluated end to end, including reconstruction, the variation present in the released loci, and the coordinate information available to an attacker. A monomorphic window contributes no ranking signal within the evaluated cohort, but this does not make its sequence generally safe: it may vary in another population, carry information through linkage, or itself reveal sensitive biology.

Our study has several limitations. The inversion models use \texttt{hg38} as their auxiliary reference rather than a diverse population cohort. All sampled windows are coordinate-disjoint across sequence lengths and data splits, but they originate from the same reference genome; a chromosome- or region-level holdout would provide a stronger test of generalisation to unseen genomic contexts. All three decoders reached the fixed 20-epoch horizon without early stopping, and \textit{Evo~2} and \textit{NTv2} validation loss was still decreasing at the final epoch, so their results are fixed-compute estimates rather than converged optima. The linkage benchmark is restricted to length-25 windows and a $\num{id.cohort}$-person cohort. We also do not evaluate direct matching of candidate haplotype embeddings, or defences such as differential privacy, representation perturbation, access control, and secure computation. Broader auxiliary cohorts, region-level holdouts, continued decoder training, additional reconstruction architectures, longer alignment queries, and explicit defence--utility evaluations are important directions for future work.
\section{Conclusion}
We presented a benchmark for model inversion and downstream identification from mean-pooled genomic foundation-model embeddings. Compact multi-length decoders recover substantial sequence information after mean pooling from \textit{Evo~2} and \textit{NTv2}; \textit{DNABERT-2} is more resistant but not uniformly safe. Reconstruction also translates into cohort identification when informative coordinates are available. The evaluated alignment-based configuration needs more loci for high-accuracy linkage and is ineffective for the weaker \textit{DNABERT-2} reconstructions in this setting. Embedding releases should therefore be treated as disclosures of sensitive genomic information and evaluated jointly for reconstruction, locus informativeness, and adversarial side information before deployment.

\subsection*{Supplementary information}

\subsection*{Acknowledgements}
The authors thank the International Max Planck Research School for Intelligent Systems (IMPRS-IS) for supporting Sofiane Ouaari in his research.

\section*{Declarations}

\subsection*{Funding}
This work was supported by the Carl Zeiss Stiftung Research Project ``Certification and Foundations of Safe Machine Learning Systems in Healthcare'', by the German Research Foundation (DFG) under Germany's Excellence Strategy (EXC 2064/1, project 390727645), and by the German Federal Ministry of Research, Technology and Space (BMFTR) within the PrivateAIM project (funding number 01ZZ2316D).

\subsection*{Competing interests}
The authors declare no competing interests.

\subsection*{Ethics approval and consent to participate}
Not applicable

\subsection*{Consent for publication}
Not applicable

\subsection*{Data availability}
Result files are available at \url{https://github.com/not-a-feature/DNA-Embedding-Inversion}.\\
Model weights are available at \url{https://huggingface.co/not-a-feature/DNA-Embedding-Inversion}.\\
The human reference genome UCSC hg38 / GRCh38, initial December 2013 assembly is available at \url{ https://hgdownload.soe.ucsc.edu/goldenPath/hg38/bigZips/hg38.fa.gz}\\
The 1000 Genomes high-coverage phased SNV/indel/SV panel, 2022-04-22 release is available at \url{https://ftp.1000genomes.ebi.ac.uk/vol1/ftp/data_collections/1000G_2504_high_coverage/working/20220422_3202_phased_SNV_INDEL_SV/}

\subsection*{Materials availability}
Not applicable

\subsection*{Code availability}
The training, evaluation and exploration code is available at \url{https://github.com/not-a-feature/DNA-Embedding-Inversion}.

\subsection*{Author contributions}
J.K. and S.O. conceived the project idea and the experimental setup. S.O. established the threat analysis scenario. J.K. conducted the experiments including the identification attack. S.O. and N.P. supervised the experiments and provided additional ideas. All authors analysed the results, wrote and reviewed the manuscript.

\noindent

\bigskip

\bibliographystyle{plainnat}
\bibliography{sample}%

@inproceedings{zhou2024dnabert,
  title={DNABERT-2: Efficient foundation model and benchmark for multi-species genomes},
  author={Zhou, Zhihan and Ji, Yanrong and Li, Weijian and Dutta, Pratik and Davuluri, Ramana and Liu, Han},
  booktitle={International Conference on Learning Representations},
  volume={2024},
  pages={41642--41665},
  year={2024}
}

@inproceedings{fredrikson2015model,
  title={Model inversion attacks that exploit confidence information and basic countermeasures},
  author={Fredrikson, Matt and Jha, Somesh and Ristenpart, Thomas},
  booktitle={Proc. 22nd ACM SIGSAC Conf. Comput. Commun. Security},
  pages={1322--1333},
  year={2015}
}

@article{bommasani2021opportunities,
  title={On the opportunities and risks of foundation models},
  author={Bommasani, Rishi},
  journal={arXiv preprint arXiv:2108.07258},
  year={2021}
}

@article{awais2025foundation,
  title={Foundation models defining a new era in vision: a survey and outlook},
  author={Awais, Muhammad and Naseer, Muzammal and Khan, Salman and others},
  journal={IEEE Trans. Pattern Anal. Mach. Intell.},
  year={2025}
}

@article{zhou2025comprehensive,
  title={A comprehensive survey on pretrained foundation models: A history from bert to chatgpt},
  author={Zhou, Ce and Li, Qian and Li, Chen and others},
  journal={Int. J. Mach. Learn. Cybern.},
  volume={16},
  number={12},
  pages={9851--9915},
  year={2025}
}

@article{homer2008resolving,
  title={Resolving individuals contributing trace amounts of DNA to highly complex mixtures using high-density SNP genotyping microarrays},
  author={Homer, Nils and Szelinger, Szabolcs and Redman, Margot and Duggan, David and Tembe, Waibhav and Muehling, Jill and Pearson, John V and Stephan, Dietrich A and Nelson, Stanley F and Craig, David W},
  journal={PLoS genetics},
  volume={4},
  number={8},
  pages={e1000167},
  year={2008},
  publisher={Public Library of Science, San Francisco, USA}
}

@article{guo2025foundation,
  title={Foundation models in bioinformatics},
  author={Guo, Fei and Guan, Renchu and Li, Yaohang and others},
  journal={Natl. Sci. Rev.},
  volume={12},
  number={4},
  pages={nwaf028},
  year={2025}
}

@article{feng2025benchmarking,
  title={Benchmarking DNA foundation models for genomic and genetic tasks},
  author={Feng, Haonan and Wu, Lang and Zhao, Bingxin and others},
  journal={Nat. Commun.},
  volume={16},
  number={1},
  pages={10780},
  year={2025},
  doi={10.1038/s41467-025-65823-8}
}

@article{benegas2023dna,
  title={DNA language models are powerful predictors of genome-wide variant effects},
  author={Benegas, Gonzalo and Batra, Sanjit Singh and Song, Yun S.},
  journal={Proc. Natl. Acad. Sci. U.S.A.},
  volume={120},
  number={44},
  pages={e2311219120},
  year={2023},
  doi={10.1073/pnas.2311219120}
}

@inproceedings{adilazero,
  title={Zero-Shot Robustification of Zero-Shot Models},
  author={Adila, Dyah and Shin, Changho and Cai, Linrong and others},
  booktitle={Int. Conf. Learn. Represent. (ICLR)},
  year={2024}
}

@article{ouaari2025robust,
  title={Robust Representation Learning for Privacy-Preserving Machine Learning: A Multi-Objective Autoencoder Approach},
  author={Ouaari, Sofiane and {\"U}nal, Ali Burak and Akg{\"u}n, Mete and others},
  journal={IEEE Access},
  year={2025}
}

@inproceedings{marin2024bend,
  title={Bend: Benchmarking dna language models on biologically meaningful tasks},
  author={Marin, Frederikke and Teufel, Felix and Horlacher, Marc and Madsen, Dennis and Pultz, Dennis and Winther, Ole and Boomsma, Wouter},
  booktitle={International Conference on Learning Representations},
  volume={2024},
  pages={15246--15281},
  year={2024}
}

@article{naveed2015privacy,
  title={Privacy in the genomic era},
  author={Naveed, Muhammad and Ayday, Erman and Clayton, Ellen W and others},
  journal={ACM Comput. Surv.},
  volume={48},
  number={1},
  pages={1--44},
  year={2015}
}

@article{bonomi2020privacy,
  title={Privacy challenges and research opportunities for genomic data sharing},
  author={Bonomi, Luca and Huang, Yingxiang and Ohno-Machado, Lucila},
  journal={Nat. Genet.},
  volume={52},
  number={7},
  pages={646--654},
  year={2020}
}

@inproceedings{zhang2020secret,
  title={The secret revealer: Generative model-inversion attacks against deep neural networks},
  author={Zhang, Yuheng and Jia, Ruoxi and Pei, Hengzhi and others},
  booktitle={Proc. IEEE/CVF Conf. Comput. Vis. Pattern Recognit. (CVPR)},
  pages={253--261},
  year={2020}
}

@inproceedings{nguyen2023re,
  title={Re-thinking model inversion attacks against deep neural networks},
  author={Nguyen, Ngoc-Bao and Chandrasegaran, Keshigeyan and Abdollahzadeh, Milad and others},
  booktitle={Proc. IEEE/CVF Conf. Comput. Vis. Pattern Recognit. (CVPR)},
  pages={16384--16393},
  year={2023}
}

@inproceedings{wu2023learning,
  title={Learning to invert: Simple adaptive attacks for gradient inversion in federated learning},
  author={Wu, Ruihan and Chen, Xiangyu and Guo, Chuan and others},
  booktitle={Uncertainty Artif. Intell.},
  pages={2293--2303},
  year={2023},
  organization={PMLR}
}

@article{zhou2023boosting,
  title={Boosting model inversion attacks with adversarial examples},
  author={Zhou, Shuai and Zhu, Tianqing and Ye, Dayong and others},
  journal={IEEE Trans. Dependable Secure Comput.},
  year={2023}
}

@article{brixi2026genome,
  title={Genome modelling and design across all domains of life with Evo 2},
  author={Brixi, Garyk and Durrant, Matthew G and Ku, Jerome and Naghipourfar, Mohsen and Poli, Michael and Sun, Gwanggyu and Brockman, Greg and Chang, Daniel and Fanton, Alison and Gonzalez, Gabriel A and others},
  journal={Nature},
  volume={652},
  number={8112},
  pages={1349--1361},
  year={2026},
  publisher={Nature Publishing Group UK London}
}

@article{radford2019language,
  title={Language models are unsupervised multitask learners},
  author={Radford, Alec and Wu, Jeffrey and Child, Rewon and others},
  journal={OpenAI blog},
  volume={1},
  number={8},
  pages={9},
  year={2019}
}

@inproceedings{bostrom2020byte,
  title={Byte Pair Encoding is Suboptimal for Language Model Pretraining},
  author={Bostrom, Kaj and Durrett, Greg},
  booktitle={Findings Assoc. Comput. Linguist.: EMNLP 2020},
  pages={4617--4624},
  year={2020}
}

@article{su2024roformer,
  title={Roformer: Enhanced transformer with rotary position embedding},
  author={Su, Jianlin and Ahmed, Murtadha and Lu, Yu and others},
  journal={Neurocomputing},
  volume={568},
  pages={127063},
  year={2024}
}

@article{nucleotidetransf,
  title={Nucleotide Transformer: building and evaluating robust foundation models for human genomics},
  author={Dalla-Torre, Hugo and Gonzalez, Liam and Mendoza-Revilla, Javier and others},
  journal={Nat. Methods},
  volume={22},
  number={2},
  pages={287--297},
  year={2025}
}

@article{berger2020levenshtein,
  title={Levenshtein distance, sequence comparison and biological database search},
  author={Berger, Bonnie and Waterman, Michael S and Yu, Yun William},
  journal={IEEE Trans. Inf. Theory},
  volume={67},
  number={6},
  pages={3287--3294},
  year={2020}
}

@inproceedings{levenshtein1966binary,
  title={Binary codes capable of correcting deletions, insertions, and reversals},
  author={Levenshtein, Vladimir I and others},
  booktitle={Soviet physics doklady},
  volume={10},
  number={8},
  pages={707--710},
  year={1966}
}

@article{bollas2024snvstory,
  title={SNVstory: inferring genetic ancestry from genome sequencing data},
  author={Bollas, Audrey E and Rajkovic, Andrei and Ceyhan, Defne and others},
  journal={BMC Bioinformatics},
  volume={25},
  number={1},
  pages={76},
  year={2024}
}

@article{injective-nikolaou2025,
  title={Language models are injective and hence invertible},
  author={Nikolaou, Giorgos and Mencattini, Tommaso and Crisostomi, Donato and others},
  journal={arXiv preprint arXiv:2510.15511},
  year={2025}
}

@article{lippert2017identification,
  title={Identification of individuals by trait prediction using whole-genome sequencing data},
  author={Lippert, Christoph and Sabatini, Riccardo and Maher, M Cyrus and others},
  journal={Proc. Natl. Acad. Sci. USA},
  volume={114},
  number={38},
  pages={10166--10171},
  year={2017}
}

@article{spiliopoulou2015genomic,
  title={Genomic prediction of complex human traits: relatedness, trait architecture and predictive meta-models},
  author={Spiliopoulou, Athina and Nagy, Reka and Bermingham, Mairead L and others},
  journal={Hum. Mol. Genet.},
  volume={24},
  number={14},
  pages={4167--4182},
  year={2015}
}

@article{1000genomes,
  title={The International Genome Sample Resource (IGSR) collection of open human genomic variation resources},
  author={Fairley, Susan and Lowy-Gallego, Ernesto and Perry, Emily and others},
  journal={Nucleic Acids Res.},
  volume={48},
  number={D1},
  pages={D941--D947},
  year={2019}
}

@article{li2013bwamem,
  title={Aligning sequence reads, clone sequences and assembly contigs with {BWA-MEM}},
  author={Li, Heng},
  journal={arXiv preprint arXiv:1303.3997},
  year={2013},
  doi={10.48550/arXiv.1303.3997}
}

@article{vilov2025investigating,
  title={Investigating the performance of foundation models on human 3'UTR sequences},
  author={Vilov, Sergey and Heinig, Matthias},
  journal={Nucleic Acids Research},
  volume={53},
  number={17},
  pages={gkaf871},
  year={2025},
  doi={10.1093/nar/gkaf871}
}

@article{awasthi2025unsupervised,
  title={Unsupervised evaluation of pre-trained DNA language model embeddings},
  author={Awasthi, Raghav and Mend Mend Arachchige, Gayan Samuditha and Zhu, Xiaofeng},
  journal={BMC Genomics},
  volume={26},
  pages={710},
  year={2025},
  doi={10.1186/s12864-025-11913-2}
}

@article{tang2025evaluating,
  title={Evaluating the representational power of pre-trained DNA language models for regulatory genomics},
  author={Tang, Ziqi and Somia, Nirali and Yu, Yiyang and Koo, Peter K.},
  journal={Genome Biology},
  volume={26},
  pages={203},
  year={2025},
  doi={10.1186/s13059-025-03674-8}
}

\clearpage
\onecolumn
\appendix
\setcounter{figure}{0}
\setcounter{table}{0}
\renewcommand{\thefigure}{S\arabic{figure}}
\renewcommand{\thetable}{S\arabic{table}}
\section*{Appendix}
\label{sec:appendix}
\section{Data Preparation and Embedding Extraction}
\label{subsec:data_embedding}
We extract non-ambiguous subsequences (no \texttt{N} characters) from the regular chromosomes (chr1--22, chrX, chrY, chrM) of the \texttt{hg38} reference genome, using a fixed random seed of 42 for reproducibility. Sequence lengths are scheduled jointly over shuffled reference blocks, and every accepted interval advances the genomic cursor. Consequently, no two sampled windows overlap, including windows of different lengths and windows assigned to different train, validation, and test splits.

\begin{figure}[H]
    \centering
    \includegraphics[width=\linewidth]{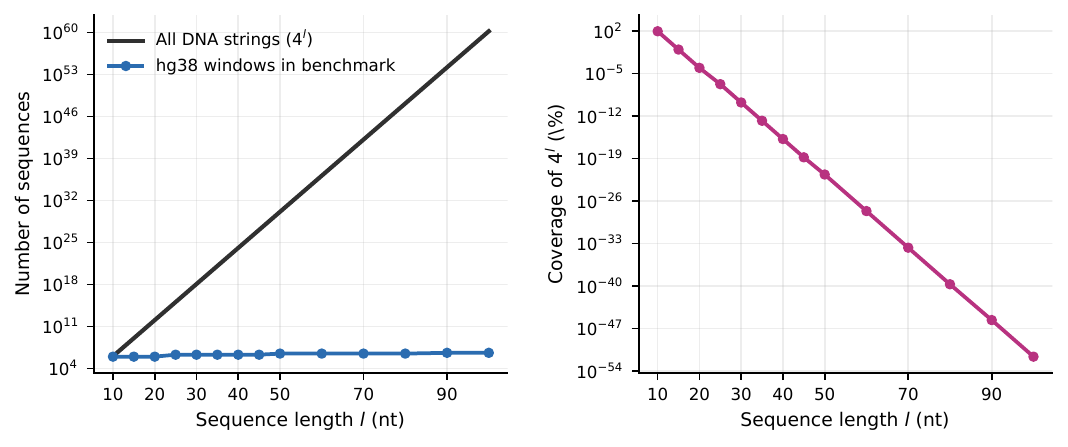}
    \caption{Percentage of the combinatorial DNA sequence space represented by the coordinate-disjoint \texttt{hg38} training windows at each benchmark length. Training-set coverage is \num{coverage.10}\% at $l=10$, \num{coverage.15}\% at $l=15$, and then decreases exponentially with length. The $4^l$ denominator is a combinatorial reference, not an estimate of the number of biologically plausible human sequences.}
    \label{fig:sequence_space_coverage}
\end{figure}

For the mean-pooled experiments, a single multi-length model is trained per foundation model, so we generate data for every sequence length and scale the amount of training data with length: the per-length totals range from $1{,}000{,}000$ sequences ($l \leq 20$) through $2{,}000{,}000$ ($25 \leq l \leq 45$) and $3{,}000{,}000$ ($50 \leq l \leq 80$) up to $4{,}000{,}000$ ($l \geq 90$). We use a fixed-count split that routes exactly $50{,}000$ sequences per length to validation and $50{,}000$ to testing, with all remaining sequences used for training; this keeps the validation and test sets constant across lengths while only the training-set size grows. Within each length, DNA strings are unique and no exact sequence is shared between training, validation, and test splits. Supplementary Figure~\ref{fig:sequence_space_coverage} reports the resulting training-set percentage coverage of the combinatorial $4^l$ sequence space.

\newpage
We extract embeddings without fine-tuning the foundation models, using the following layers:
\begin{itemize}
    \item \textbf{DNABERT-2} (checkpoint \texttt{zhihan1996/DNABERT-2-117M}): We use the last hidden-layer output of the transformer encoder. We remove the special \texttt{[CLS]} and \texttt{[SEP]} tokens and average the remaining hidden states into a 768-dimensional embedding.
    \item \textbf{NTv2} (checkpoint \texttt{InstaDeepAI/nucleotide-transformer-v2-500m-multi-species}): We extract the last hidden state via \texttt{output\_hidden\_states}. We remove the leading \texttt{[CLS]} token and average the remaining hidden states into a 1{,}024-dimensional embedding.
    \item \textbf{Evo~2} (checkpoint \texttt{evo2\_7b}, 7B parameters): We extract embeddings from the intermediate MLP projection \texttt{blocks.26.mlp.l3}. This extraction point gives the highest reconstruction similarity among the four Evo~2 points tested in Supplementary Note~\ref{subsec:layer_sensitivity}. Each nucleotide corresponds to exactly one token, and averaging its hidden states yields an embedding of dimension 4{,}096.
\end{itemize}
Averaging the contextual hidden states across content-token positions produces a single fixed-size vector per sequence. All embeddings are stored in HDF5 format with SHA-256 checksums for integrity verification. Training-set statistics (mean, standard deviation) are computed and used for z-score normalisation across all splits, ensuring no information leakage from the validation or test sets.

\clearpage
\section{Sensitivity to the Embedding Extraction Layer}
\label{subsec:layer_sensitivity}
We tested four extraction points per foundation model, including the point used in the main benchmark (Supplementary Table~\ref{tab:layer_sensitivity}). For each point, we trained a fresh LCQD with the same architecture and optimiser settings for 20 epochs, using seed 42 and frozen foundation-model weights. Each decoder was trained jointly on lengths 25, 50 and 100, with $100{,}000$ training, $5{,}000$ validation and $10{,}000$ test sequences per length.

All conditions used the same subsets of the reference corpus and retained its split assignments. We verified exact-sequence uniqueness and coordinate disjointness across lengths and splits. Embeddings were mean-pooled, and normalisation statistics were fitted separately for each extraction point using training data only. The checkpoint with the lowest validation loss was evaluated on each test set.

\begin{table}[H]
    \centering
    \caption{Mean normalised Levenshtein similarity for the layer-sensitivity study, using $10{,}000$ test sequences per cell. Bold values identify the highest similarity within each foundation model and length; $\dagger$ marks the extraction point used in the main benchmark, retrained here with the study's training budget. DNABERT-2 and NTv2 indices count completed layers from 1. Evo~2 uses zero-based block indices, with \texttt{blocks.\{index\}.mlp.l3} at every point.}
    \label{tab:layer_sensitivity}
    \begin{tabular}{llrrr}
        \toprule
        Foundation model & Layer/block & 25 nt & 50 nt & 100 nt \\
        \midrule
        DNABERT-2 & 3 & \textbf{0.718} & \textbf{0.597} & \textbf{0.530} \\
                   & 6 & 0.596 & 0.538 & 0.508 \\
                   & 9 & 0.540 & 0.501 & 0.488 \\
                   & $12^{\dagger}$ & 0.496 & 0.475 & 0.480 \\
        \midrule
        NTv2       & 7 & 0.931 & 0.705 & 0.588 \\
                   & 15 & \textbf{0.957} & \textbf{0.805} & 0.655 \\
                   & 22 & 0.946 & 0.786 & \textbf{0.659} \\
                   & $29^{\dagger}$ & 0.908 & 0.697 & 0.572 \\
        \midrule
        Evo~2      & 7 & 0.570 & 0.475 & 0.418 \\
                   & 15 & 0.608 & 0.483 & 0.419 \\
                   & $26^{\dagger}$ & \textbf{0.729} & \textbf{0.534} & \textbf{0.437} \\
                   & 31 & 0.371 & 0.365 & 0.312 \\
        \bottomrule
    \end{tabular}
\end{table}

DNABERT-2 reconstruction decreased with depth at all three lengths. NTv2 performed best at layers 15--22, while Evo~2 block 26 gave the highest similarity at every length. These results support block 26 as a high-leakage extraction point among those tested, without establishing an upper bound on leakage.

Evo~2 block 31 performed below the random baseline; its mean predicted lengths were 31.3 and 42.1 nt for 50- and 100-nt inputs, respectively. Its low reconstruction scores therefore do not establish that sequence information is absent. This single-seed study measures recovery under a fixed training budget, rather than converged attack performance or variability across training seeds.

\newpage
\section{Tokenisation Analysis}
\label{subsec:tokenisation}
The three foundation models employ different tokenisation strategies. \textbf{Evo~2} uses a single-nucleotide (character-level) tokeniser, producing exactly $l$ tokens for a sequence of length $l$, while the attack decoder uses four nucleotide symbols plus EOS. The foundation-model tokenizer has additional reserved symbols. \textbf{NTv2} uses non-overlapping 6-mers followed by single-nucleotide tokens for the residual suffix, producing $\lfloor l/6 \rfloor + (l \bmod 6)$ content tokens and using a vocabulary of 4{,}107 tokens. \textbf{DNABERT-2} uses Byte Pair Encoding (BPE) \citep{bostrom2020byte}, which produces a variable number of tokens depending on sequence content. For a sequence of length 100, DNABERT-2 typically generates 20 tokens, mostly spanning 1--8 nucleotides, drawn from a vocabulary of 4{,}096 tokens. The NTv2 and DNABERT-2 vocabulary counts characterise the decoder target spaces; because the foundation models also differ in architecture, scale, training corpus, and extracted layer, the present benchmark does not identify tokenisation as the cause of performance differences.

\begin{figure}[H]
    \centering
    \includegraphics[width=0.6\linewidth]{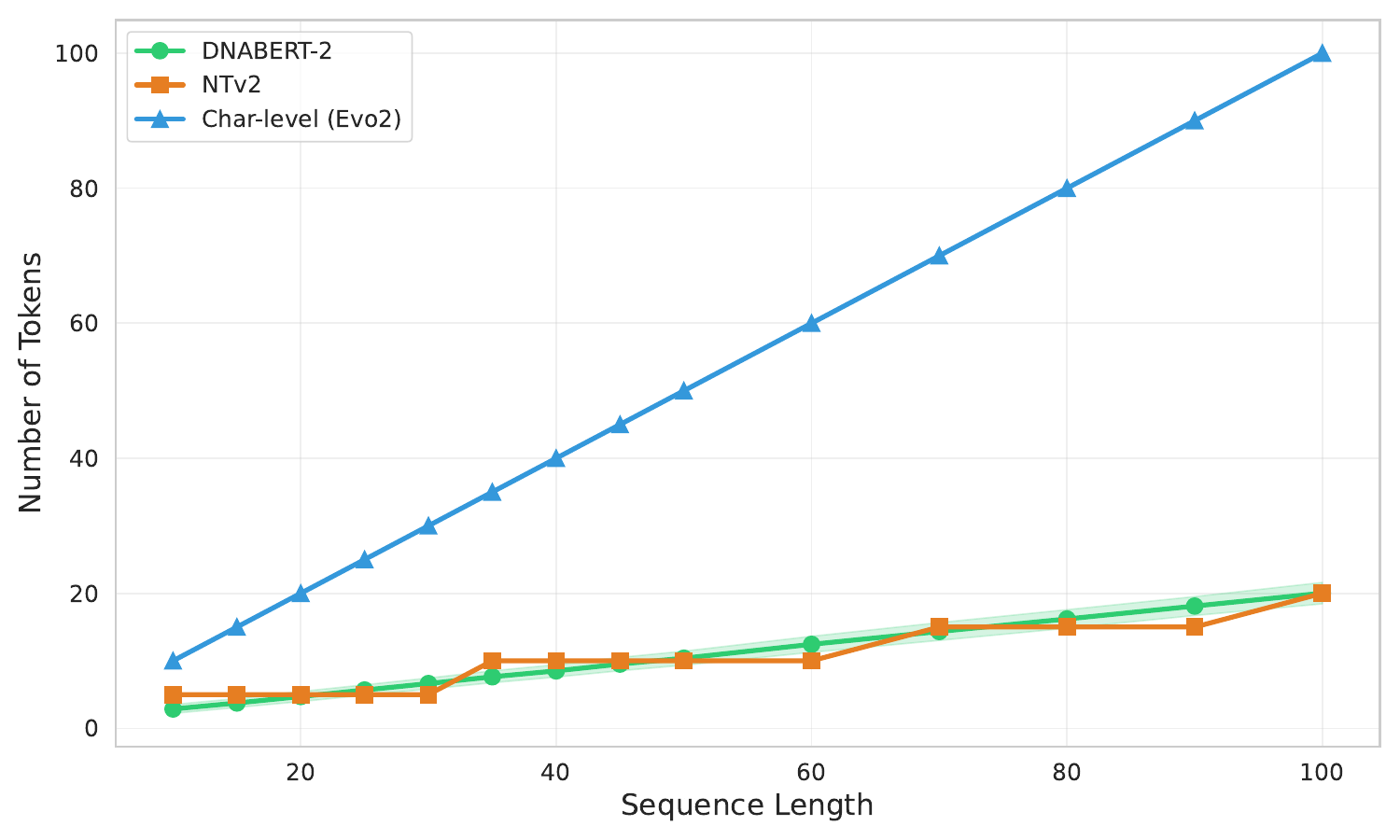}
    \caption{Token count vs.\ sequence length for the three foundation models. Evo~2 produces exactly $l$ tokens, NTv2 uses full non-overlapping 6-mers plus one token per residual nucleotide, and DNABERT-2 exhibits variable, content-dependent BPE tokenisation. Shaded regions indicate $\pm 1$ standard deviation.}
    \label{fig:token_counts_vs_length}
\end{figure}

\autoref{fig:token_counts_vs_length} illustrates how the number of tokens produced by each tokeniser scales with sequence length. Evo~2 maintains a strict 1:1 ratio, NTv2 follows the deterministic 6-mer-plus-residual rule above, and DNABERT-2's BPE tokeniser exhibits content-dependent growth with higher variance.

\newpage
\section{Levenshtein Distance Definition}
\label{subsec:levenshtein}
Let $a=a_1\cdots a_m$ and $b=b_1\cdots b_n$ be nucleotide sequences. Define $D(i,j)$ as the minimum number of single-nucleotide insertions, deletions and substitutions needed to transform the first $i$ nucleotides of $a$ into the first $j$ nucleotides of $b$, with unit cost for each edit. The boundary conditions are
\[
D(i,0)=i,\qquad D(0,j)=j.
\]
For $i>0$ and $j>0$, the dynamic-programming recurrence is
\begin{equation}
D(i,j)=\min\begin{cases}
    D(i-1,j)+1,\\
    D(i,j-1)+1,\\
    D(i-1,j-1)+\delta(a_i,b_j),
\end{cases}
\qquad
\delta(a_i,b_j)=\begin{cases}
    0,&a_i=b_j,\\
    1,&a_i\ne b_j.
\end{cases}
\end{equation}
The Levenshtein distance is $\operatorname{lev}(a,b)=D(m,n)$. We report normalised similarity,
\[
\operatorname{sim}_{\mathrm{lev}}(a,b)=1-\frac{D(m,n)}{\max(m,n)},
\]
which ranges from 0 to 1, with 1 indicating identical sequences. If both sequences are empty, similarity is defined as 1; if only one is empty, it is 0. Evaluation uses the complete decoded nucleotide strings, so insertions, deletions and differences in predicted length contribute to the distance.

\vfill
\clearpage

\section{Collision Analysis Across Sequence Lengths}
\label{subsec:collision_all_seqlens}

We extend the empirical collision audit from the main text by showing pairwise normalised Euclidean distance distributions for mean-pooled embeddings across all evaluated sequence lengths. For each length, we compute pairwise distances in a random subsample of $2{,}000$ sequences. The sampled distributions contain no exact zero distances, but small distances indicate only that two embeddings differ little under this metric. This finite audit neither establishes injectivity over the full input domain nor excludes exact or near collisions outside the sample.

\begin{figure}[H]
    \centering
        \begin{minipage}[t]{0.24\linewidth}%
            \centering%
            \includegraphics[width=\linewidth]{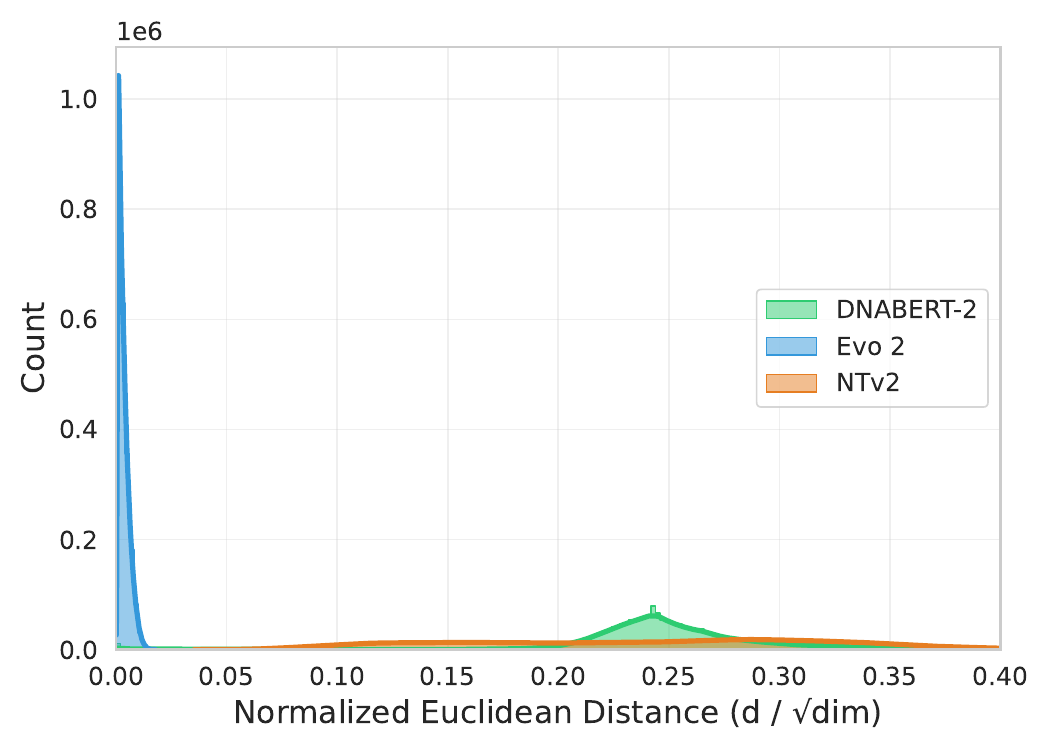}%
            \par\smallskip{\small $l = 10$}%
        \end{minipage}%
        \hspace{0.01\linewidth}%
        \begin{minipage}[t]{0.24\linewidth}%
            \centering%
            \includegraphics[width=\linewidth]{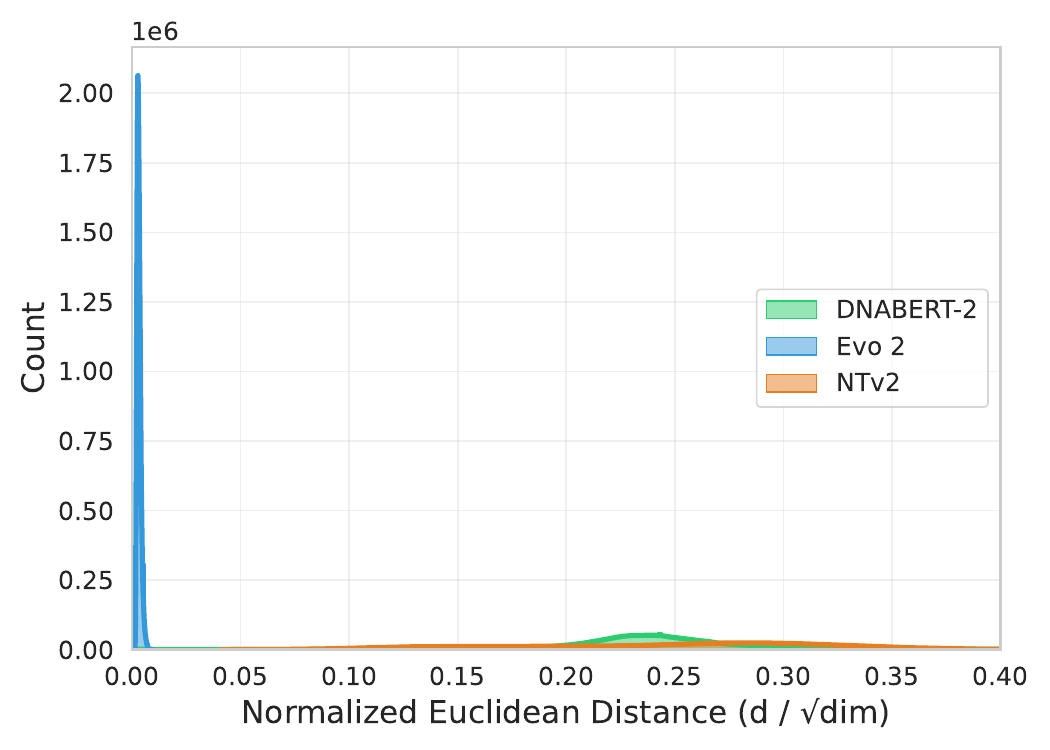}%
            \par\smallskip{\small $l = 15$}%
        \end{minipage}%
        \hspace{0.01\linewidth}%
        \begin{minipage}[t]{0.24\linewidth}%
            \centering%
            \includegraphics[width=\linewidth]{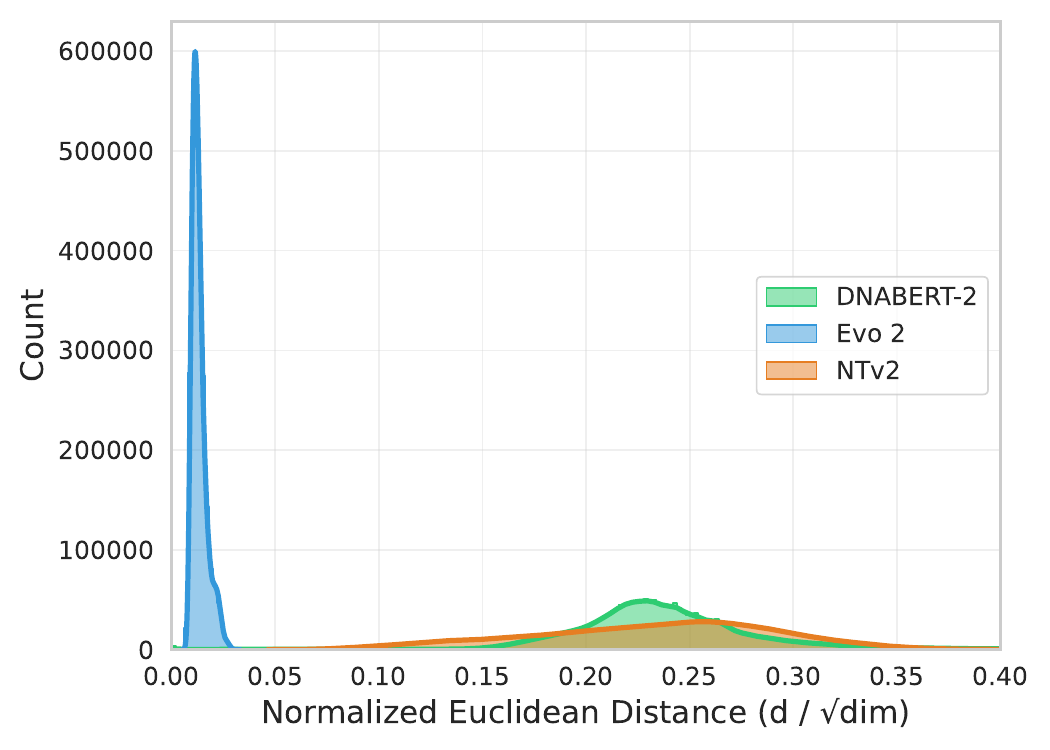}%
            \par\smallskip{\small $l = 20$}%
        \end{minipage}%
        \hspace{0.01\linewidth}%
        \begin{minipage}[t]{0.24\linewidth}%
            \centering%
            \includegraphics[width=\linewidth]{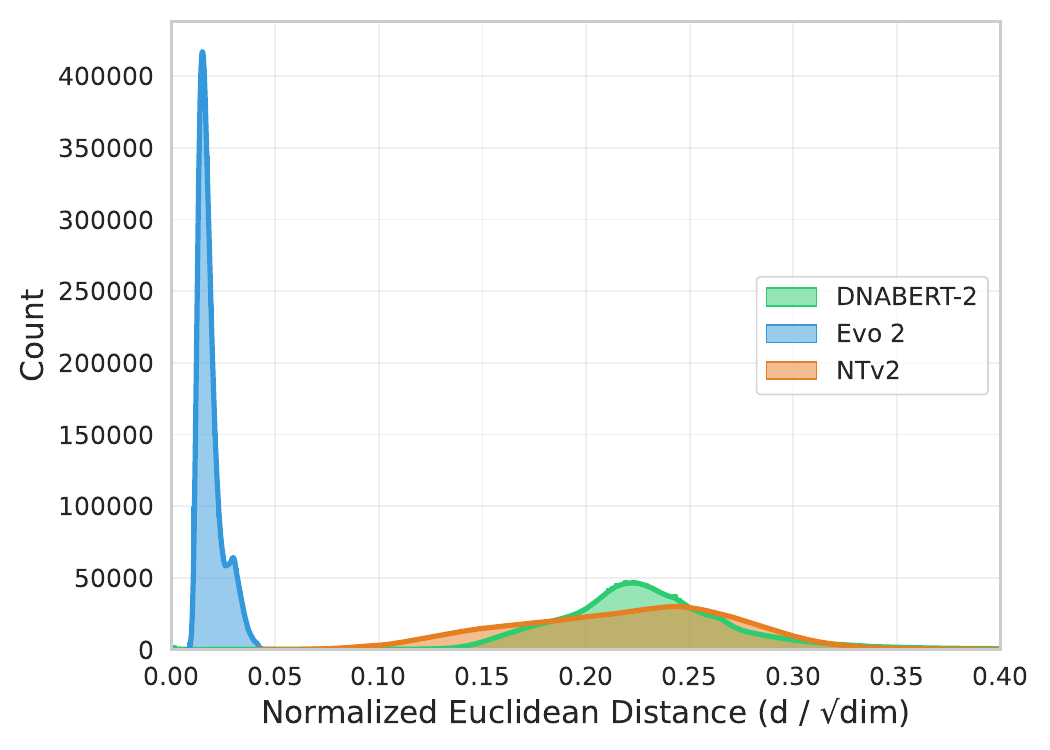}%
            \par\smallskip{\small $l = 25$}%
        \end{minipage}%
        \\[1em]%
        \begin{minipage}[t]{0.24\linewidth}%
            \centering%
            \includegraphics[width=\linewidth]{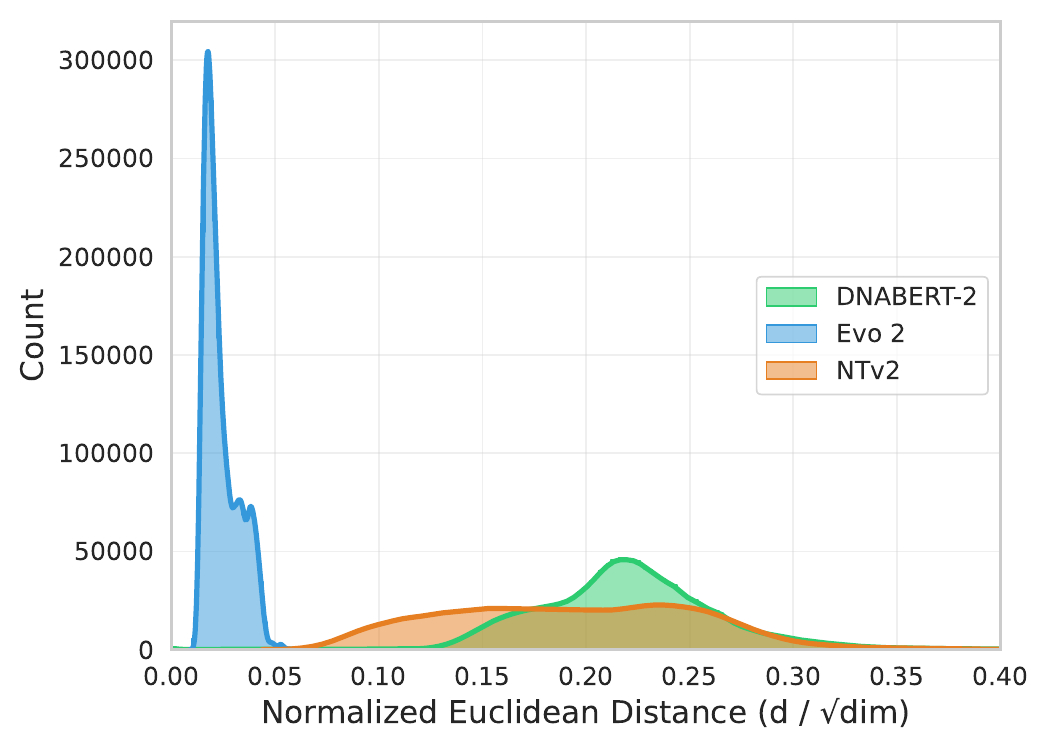}%
            \par\smallskip{\small $l = 30$}%
        \end{minipage}%
        \hspace{0.01\linewidth}%
        \begin{minipage}[t]{0.24\linewidth}%
            \centering%
            \includegraphics[width=\linewidth]{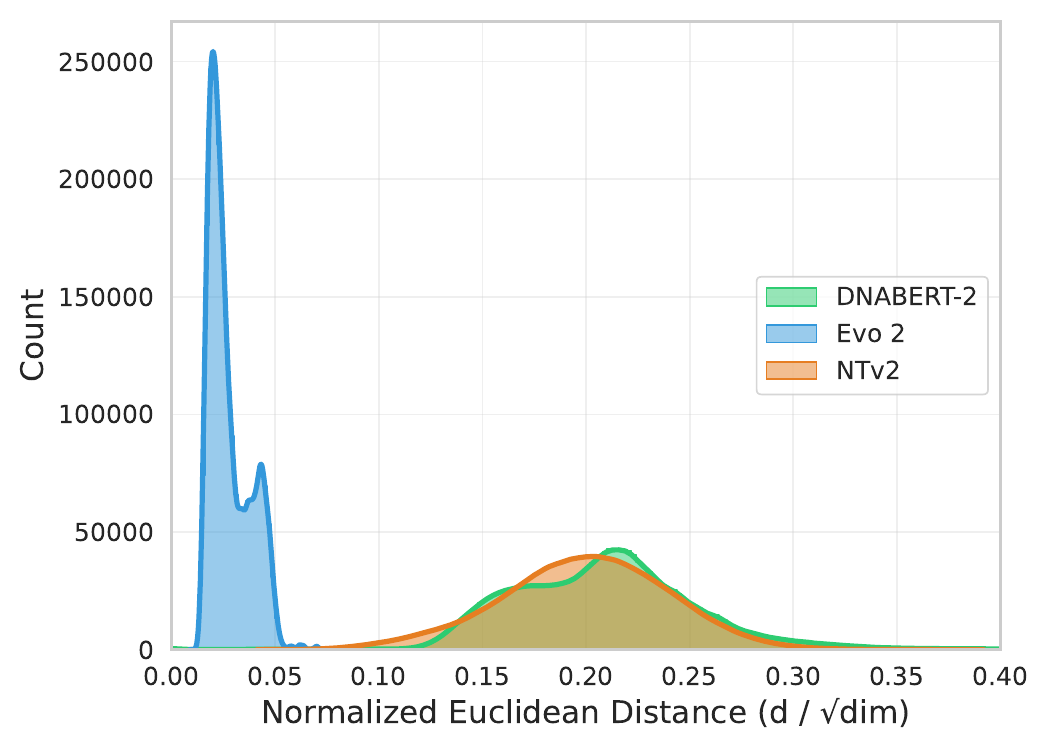}%
            \par\smallskip{\small $l = 35$}%
        \end{minipage}%
        \hspace{0.01\linewidth}%
        \begin{minipage}[t]{0.24\linewidth}%
            \centering%
            \includegraphics[width=\linewidth]{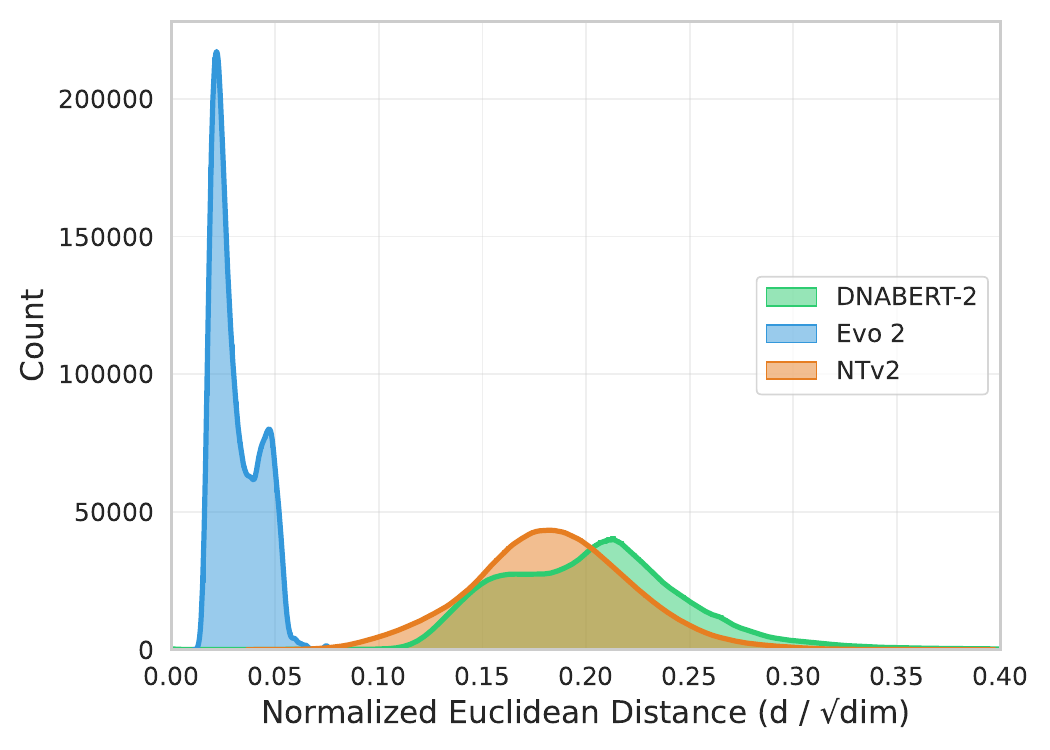}%
            \par\smallskip{\small $l = 40$}%
        \end{minipage}%
        \hspace{0.01\linewidth}%
        \begin{minipage}[t]{0.24\linewidth}%
            \centering%
            \includegraphics[width=\linewidth]{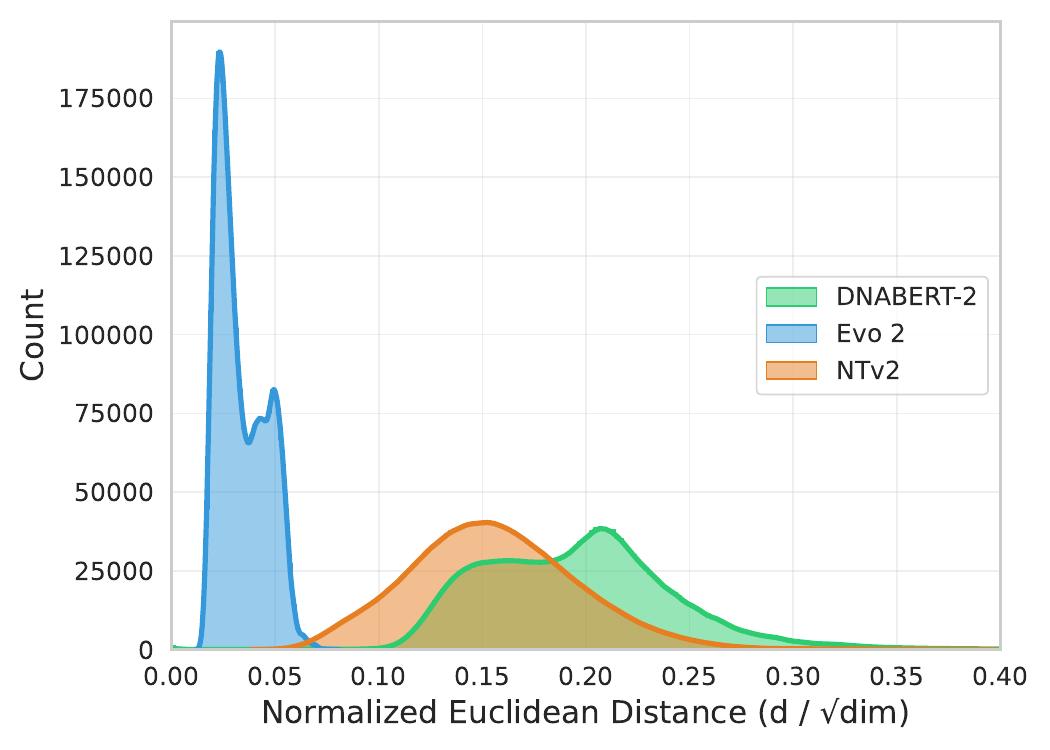}%
            \par\smallskip{\small $l = 45$}%
        \end{minipage}%
        \\[1em]%
        \begin{minipage}[t]{0.24\linewidth}%
            \centering%
            \includegraphics[width=\linewidth]{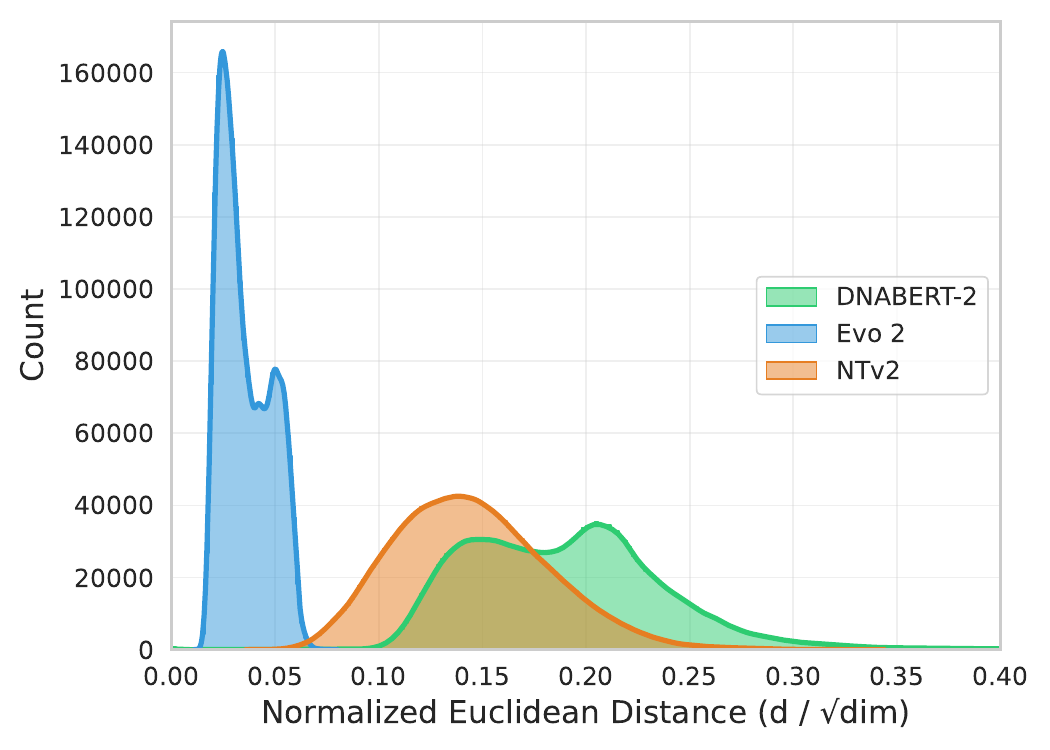}%
            \par\smallskip{\small $l = 50$}%
        \end{minipage}%
        \hspace{0.01\linewidth}%
        \begin{minipage}[t]{0.24\linewidth}%
            \centering%
            \includegraphics[width=\linewidth]{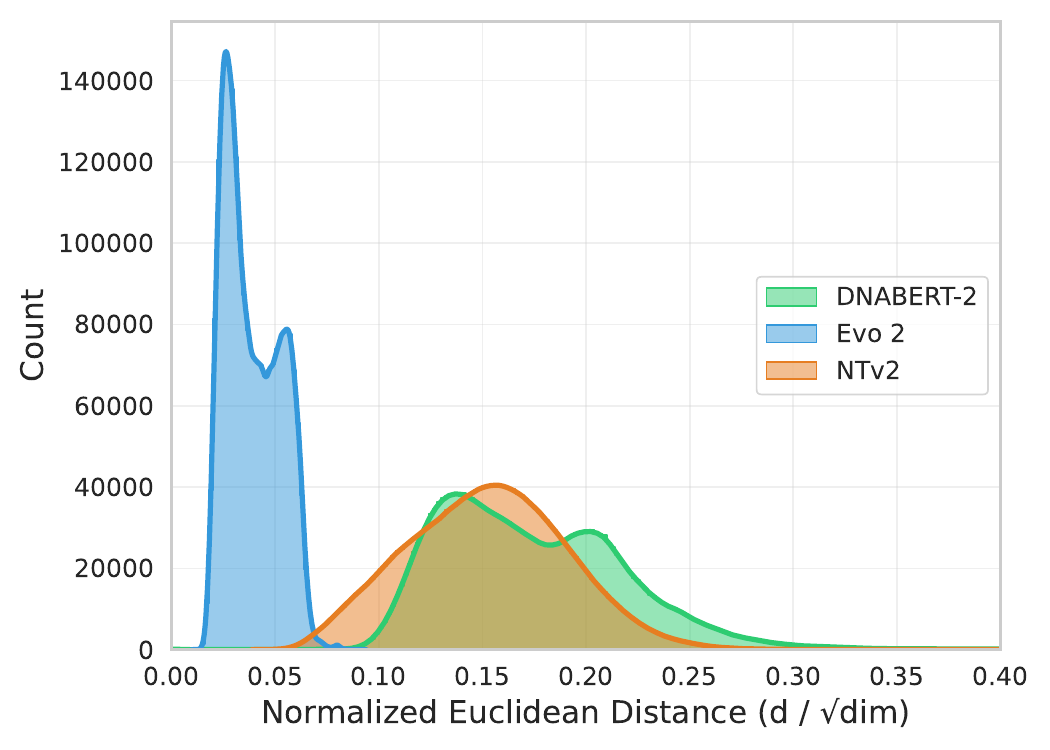}%
            \par\smallskip{\small $l = 60$}%
        \end{minipage}%
        \hspace{0.01\linewidth}%
        \begin{minipage}[t]{0.24\linewidth}%
            \centering%
            \includegraphics[width=\linewidth]{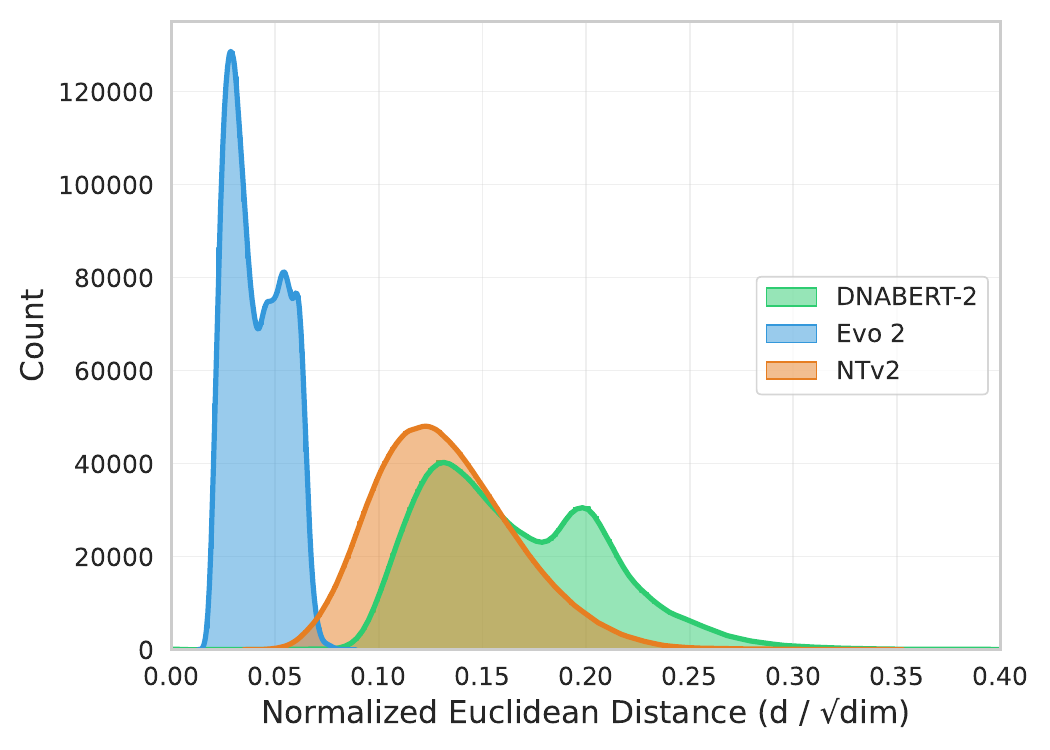}%
            \par\smallskip{\small $l = 70$}%
        \end{minipage}%
        \hspace{0.01\linewidth}%
        \begin{minipage}[t]{0.24\linewidth}%
            \centering%
            \includegraphics[width=\linewidth]{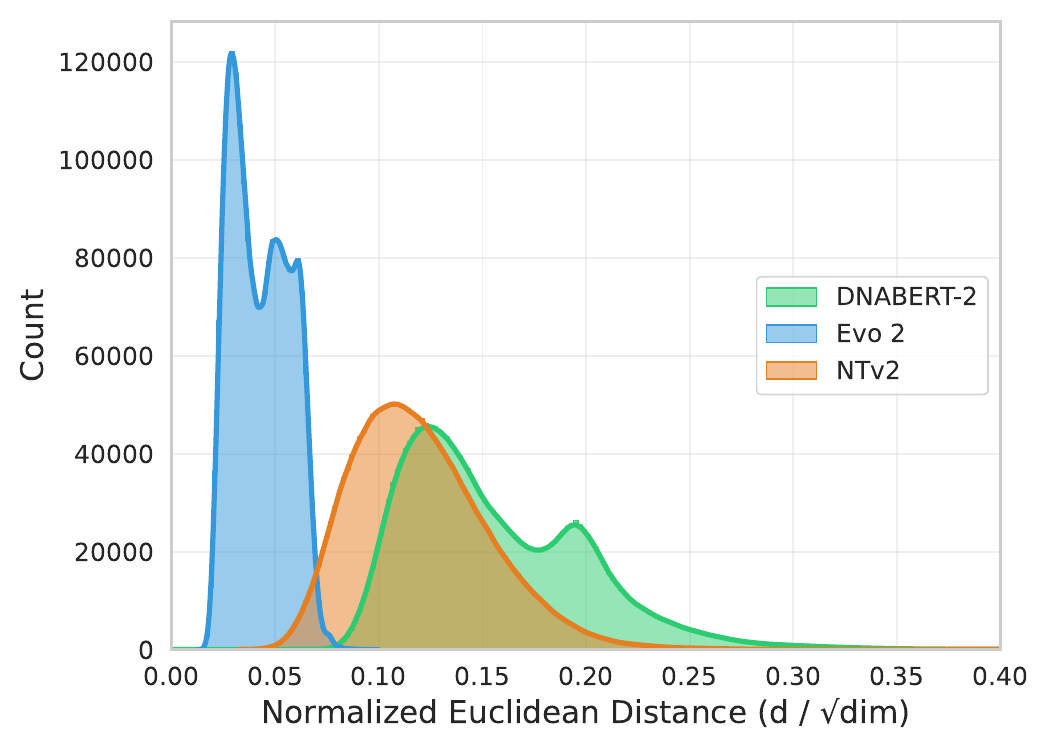}%
            \par\smallskip{\small $l = 80$}%
        \end{minipage}%
        \\[1em]%
        \begin{minipage}[t]{0.24\linewidth}%
            \centering%
            \includegraphics[width=\linewidth]{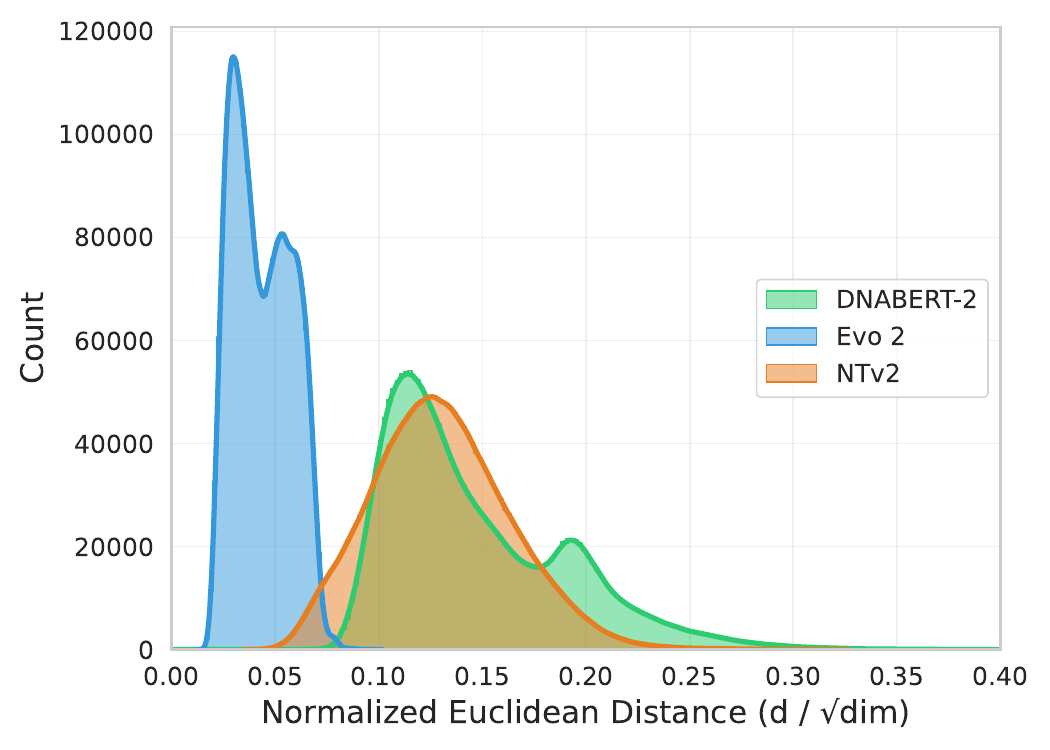}%
            \par\smallskip{\small $l = 90$}%
        \end{minipage}%
        \hspace{0.01\linewidth}%
        \begin{minipage}[t]{0.24\linewidth}%
            \centering%
            \includegraphics[width=\linewidth]{figures/cross_dataset_comparison/ned_merged/seqlen_100.pdf}%
            \par\smallskip{\small $l = 100$}%
        \end{minipage}%
    \caption{Collision analysis: pairwise normalised Euclidean distance distributions for mean-pooled embeddings across all evaluated sequence lengths.}
    \label{fig:collision_all_seqlens}
\end{figure}

\clearpage

\section{Embedding Analysis}
\label{subsec:embedding_analysis}

\begin{figure}[H]
    \centering
        \begin{minipage}[t]{0.20\linewidth}%
            \centering%
            \includegraphics[width=\linewidth]{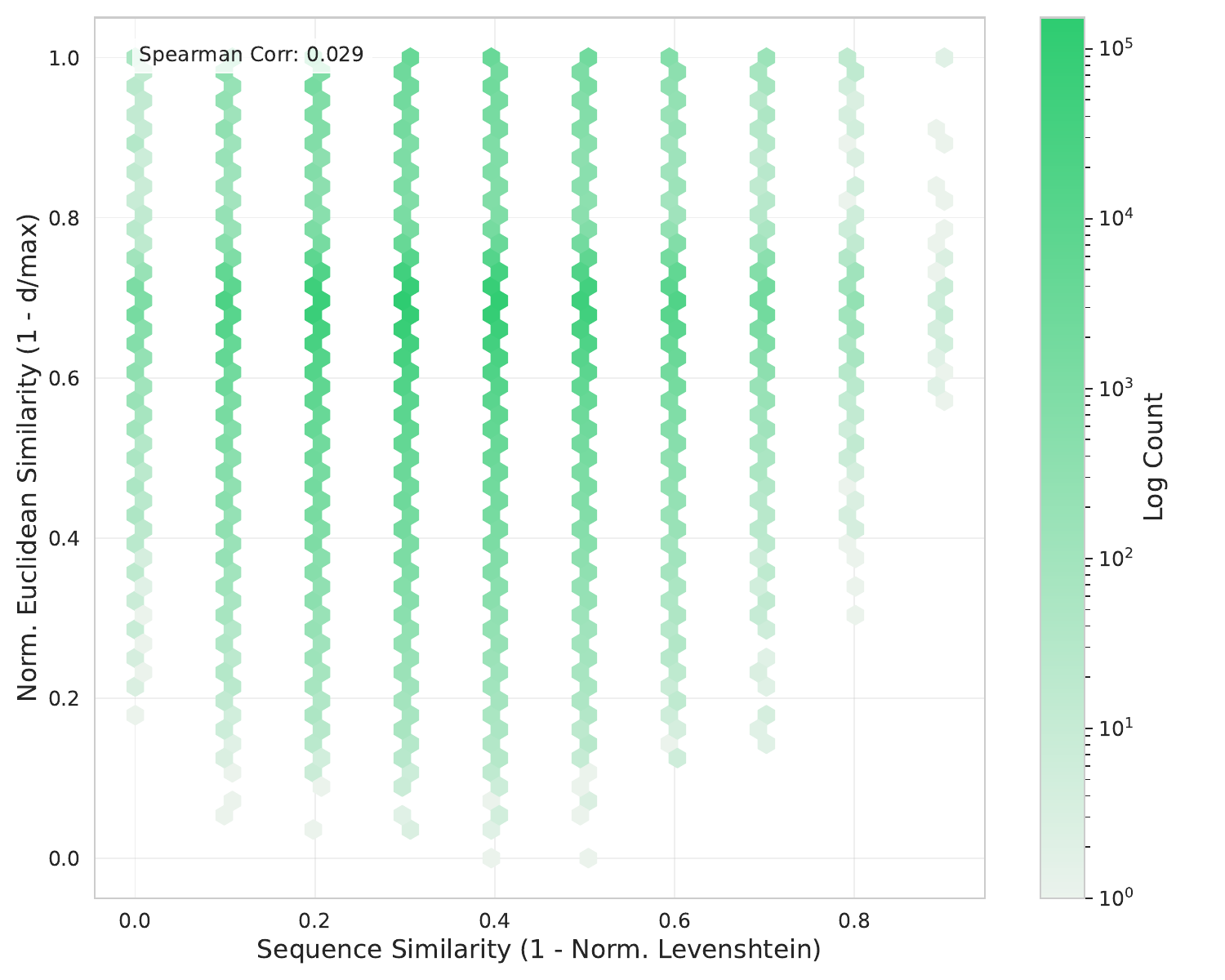}%
            \par\smallskip{\small $l = 10$}%
        \end{minipage}%
        \hspace{0.01\linewidth}%
        \begin{minipage}[t]{0.20\linewidth}%
            \centering%
            \includegraphics[width=\linewidth]{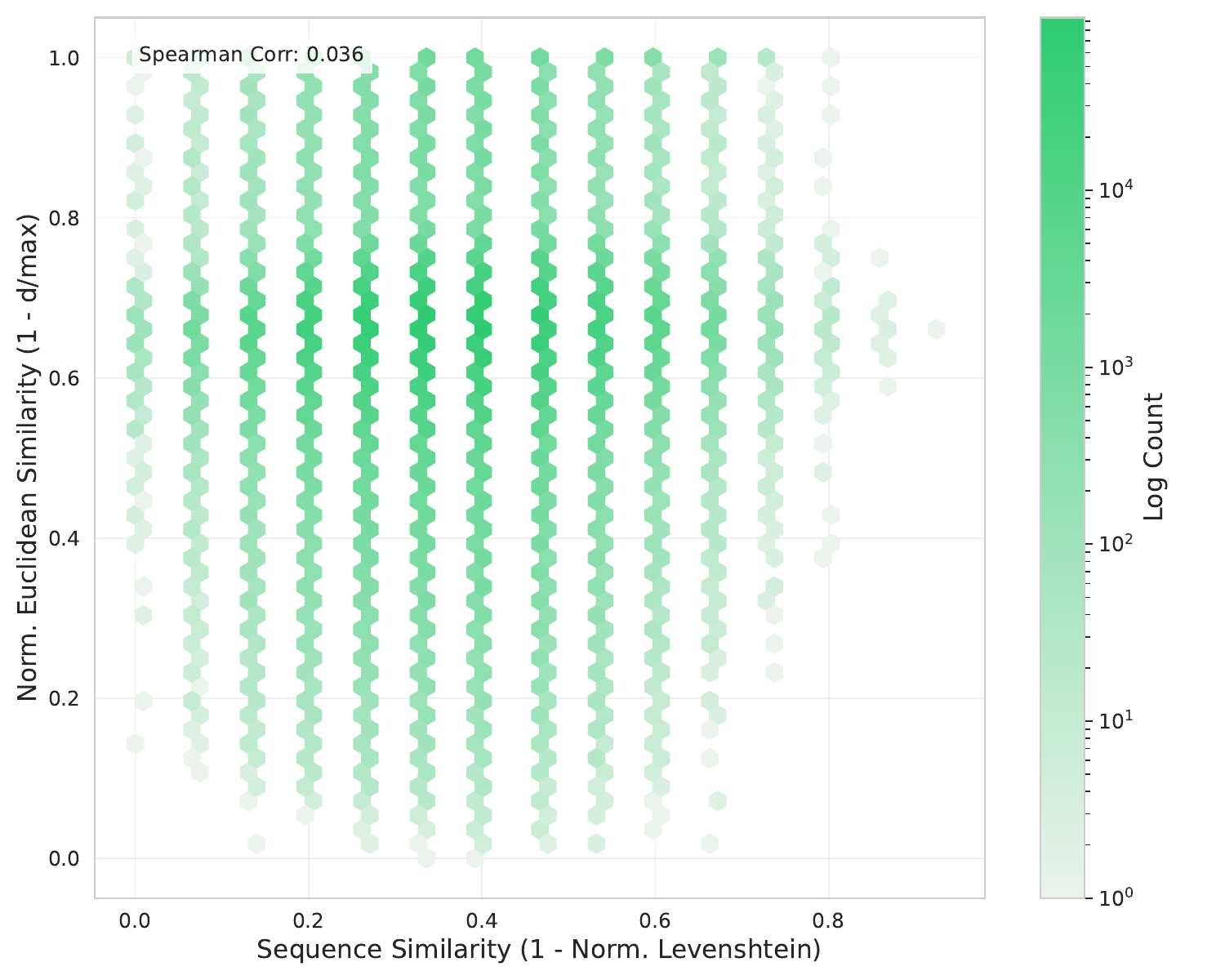}%
            \par\smallskip{\small $l = 15$}%
        \end{minipage}%
        \hspace{0.01\linewidth}%
        \begin{minipage}[t]{0.20\linewidth}%
            \centering%
            \includegraphics[width=\linewidth]{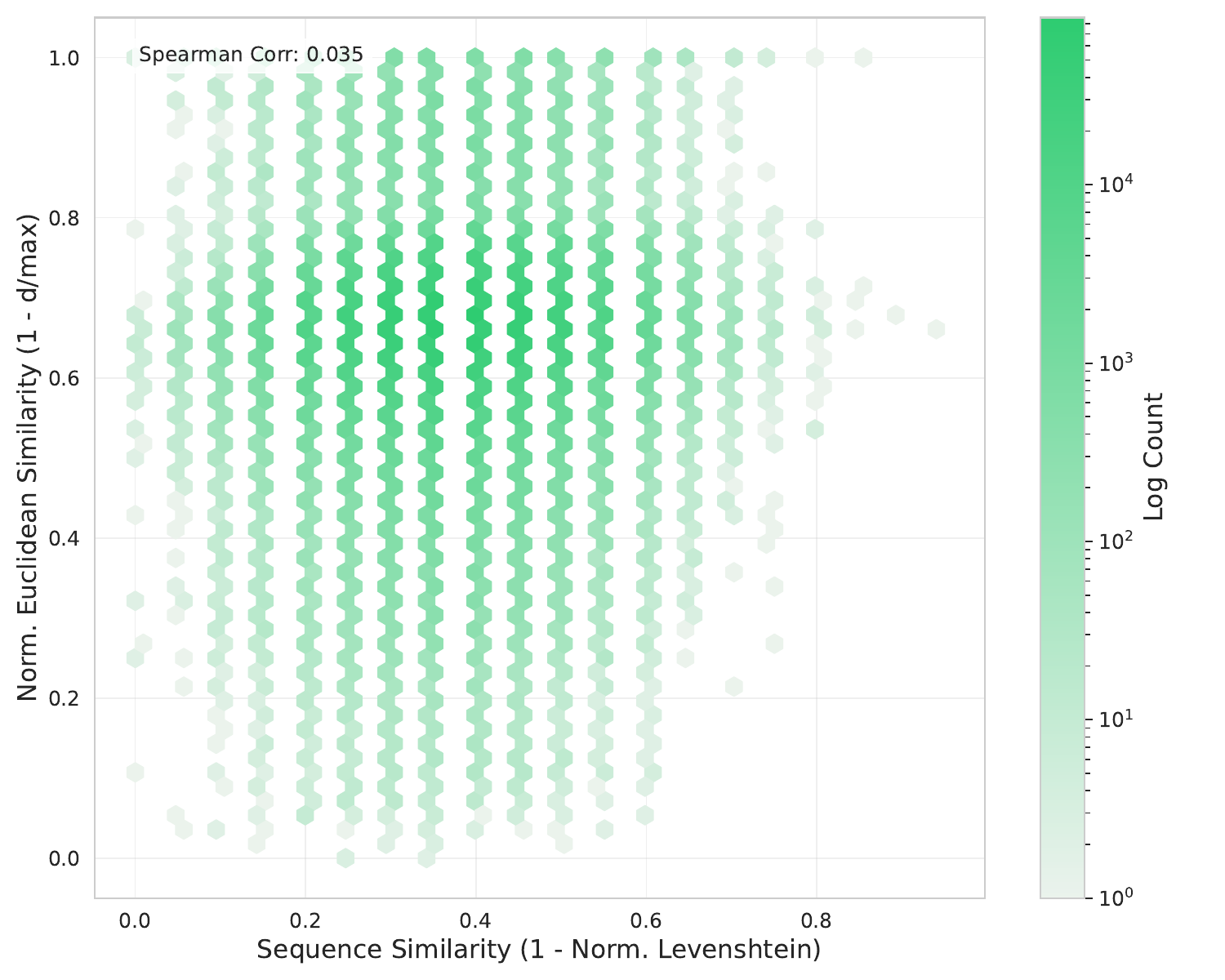}%
            \par\smallskip{\small $l = 20$}%
        \end{minipage}%
        \hspace{0.01\linewidth}%
        \begin{minipage}[t]{0.20\linewidth}%
            \centering%
            \includegraphics[width=\linewidth]{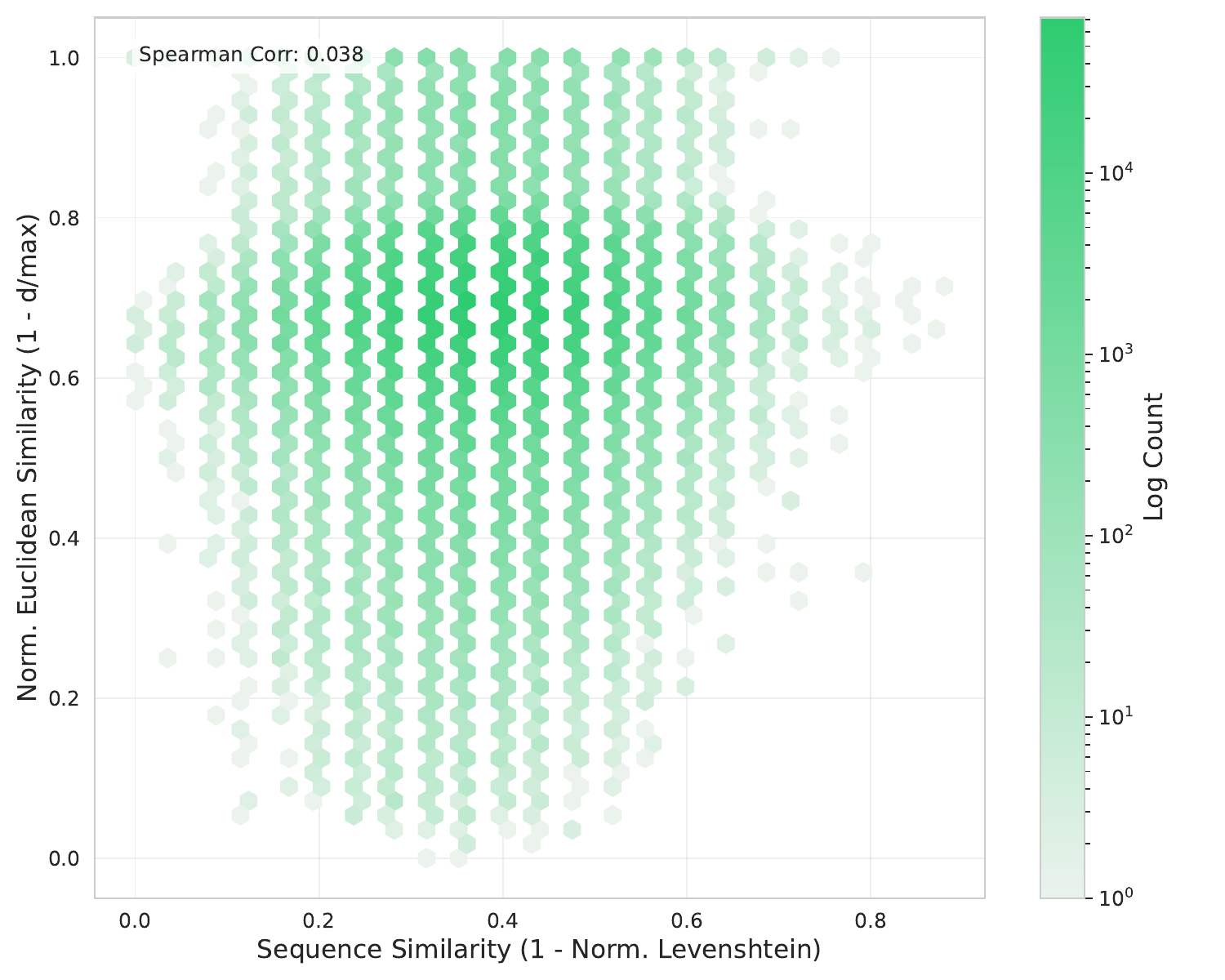}%
            \par\smallskip{\small $l = 25$}%
        \end{minipage}%
        \\[1em]%
        \begin{minipage}[t]{0.20\linewidth}%
            \centering%
            \includegraphics[width=\linewidth]{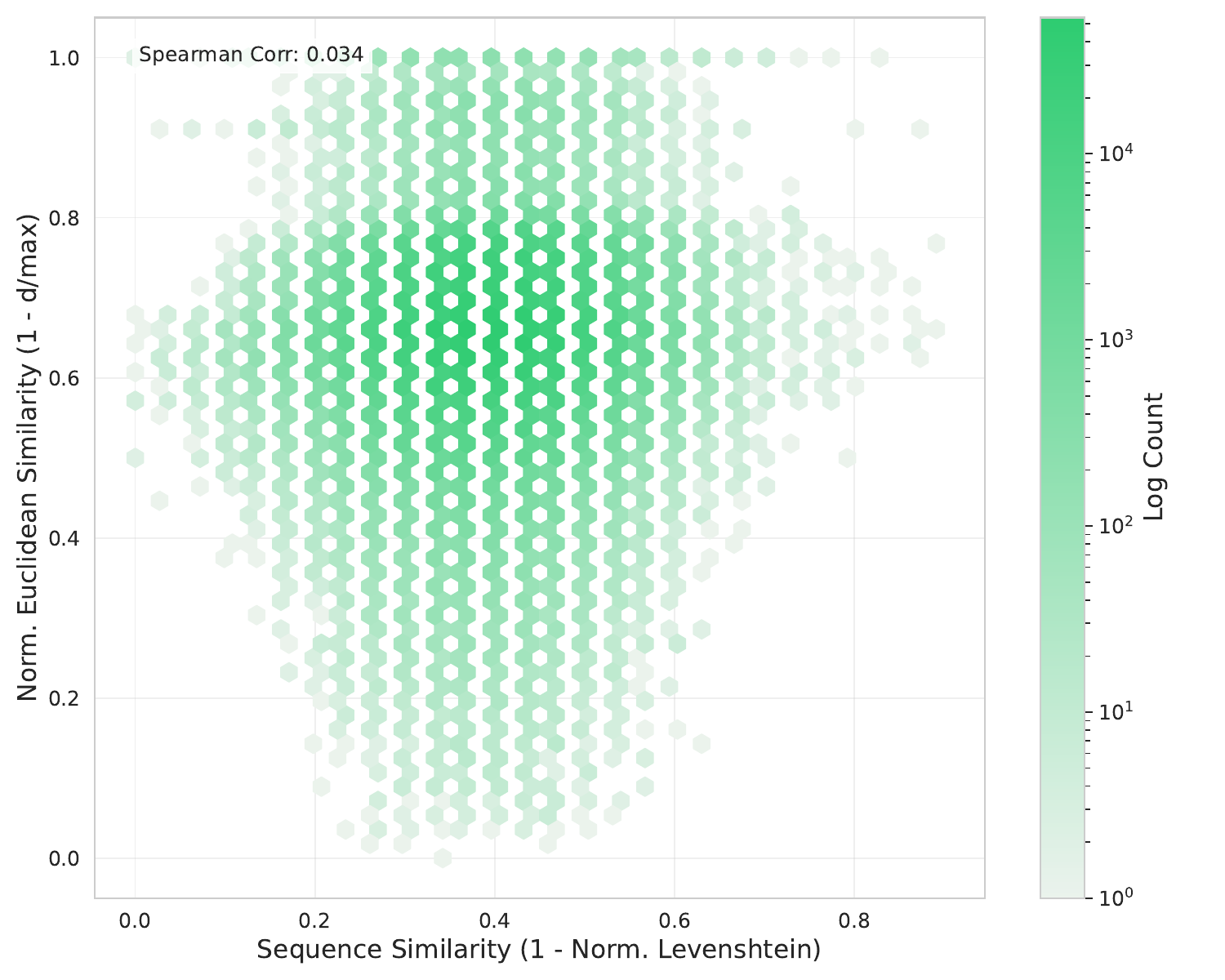}%
            \par\smallskip{\small $l = 30$}%
        \end{minipage}%
        \hspace{0.01\linewidth}%
        \begin{minipage}[t]{0.20\linewidth}%
            \centering%
            \includegraphics[width=\linewidth]{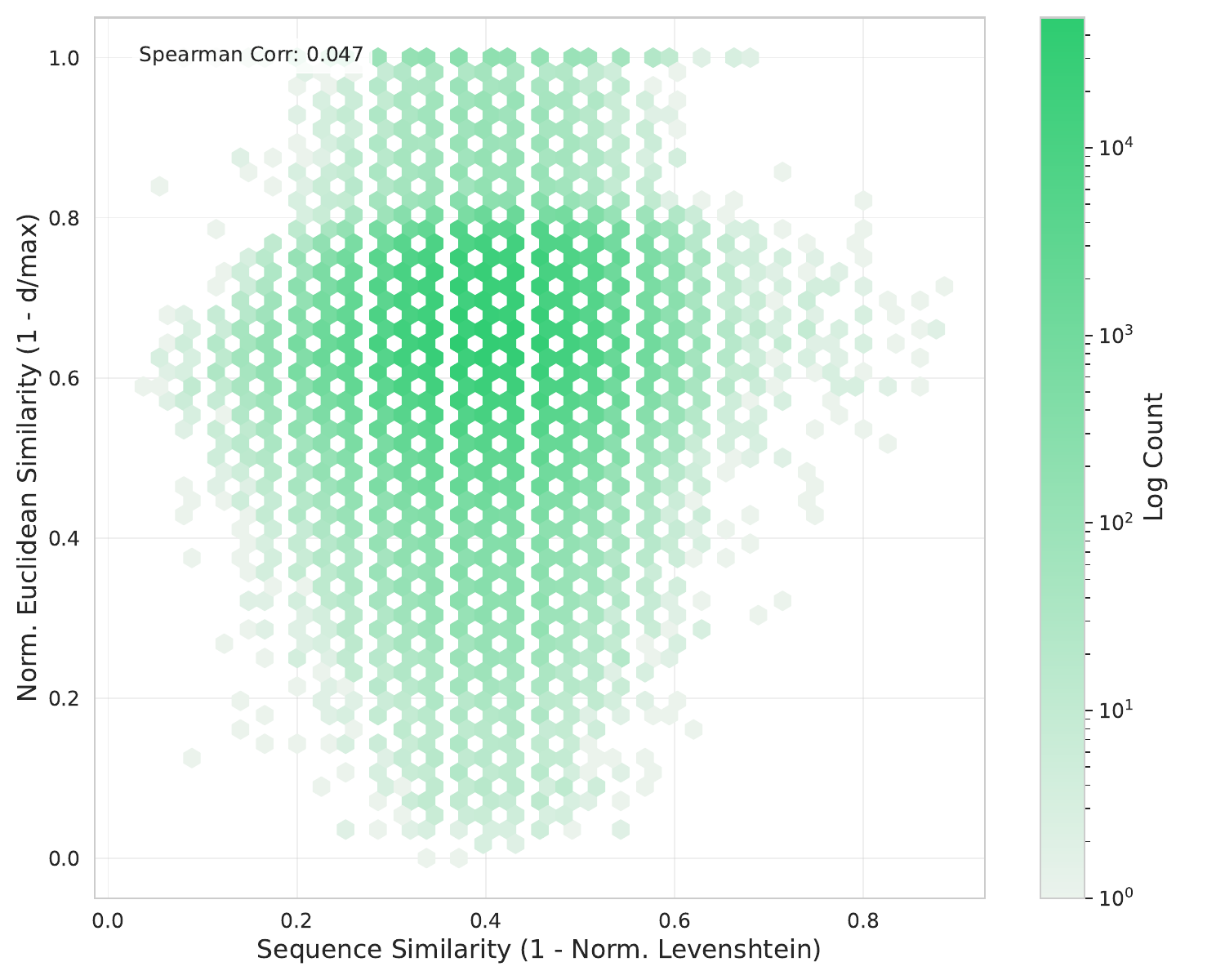}%
            \par\smallskip{\small $l = 35$}%
        \end{minipage}%
        \hspace{0.01\linewidth}%
        \begin{minipage}[t]{0.20\linewidth}%
            \centering%
            \includegraphics[width=\linewidth]{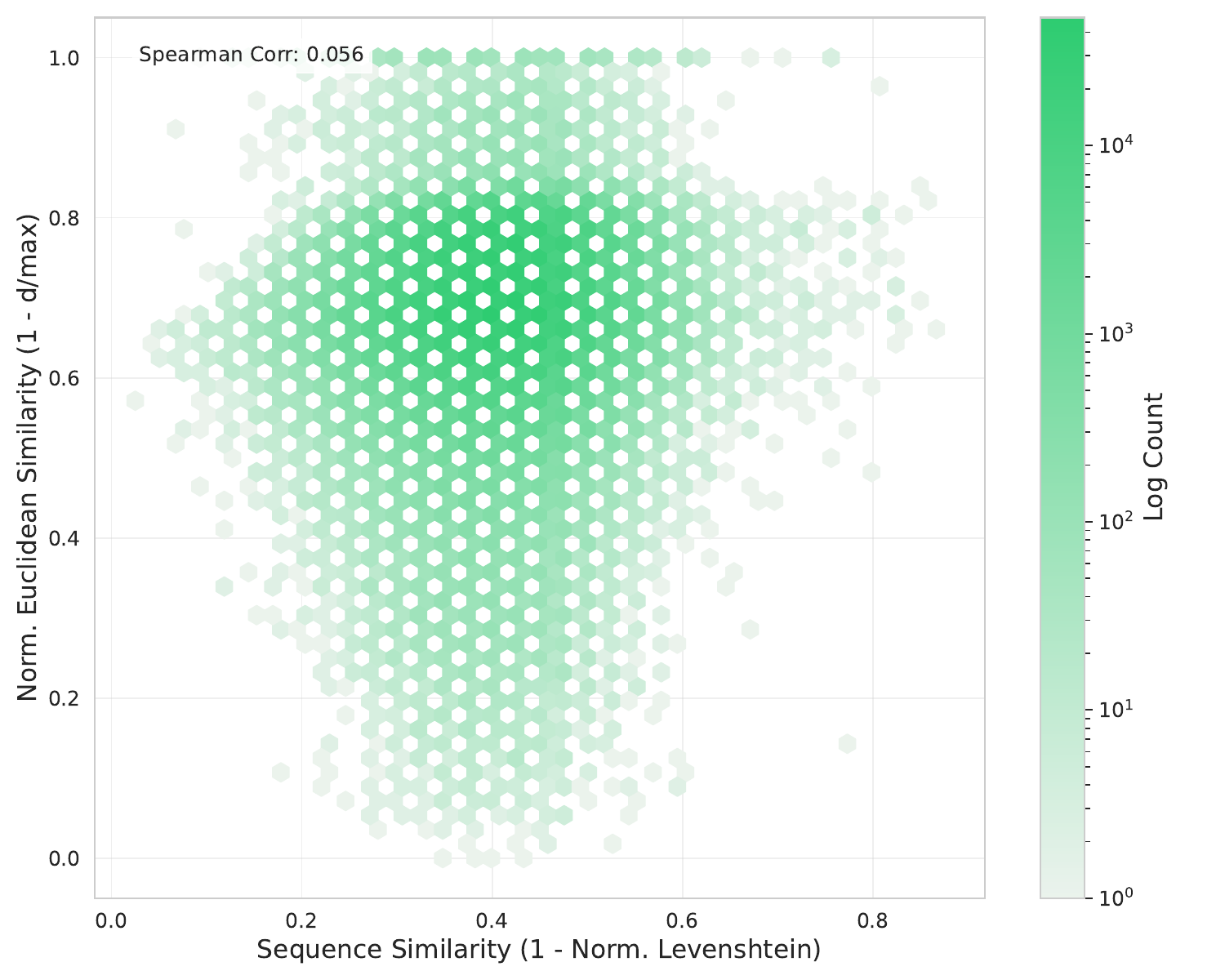}%
            \par\smallskip{\small $l = 40$}%
        \end{minipage}%
        \hspace{0.01\linewidth}%
        \begin{minipage}[t]{0.20\linewidth}%
            \centering%
            \includegraphics[width=\linewidth]{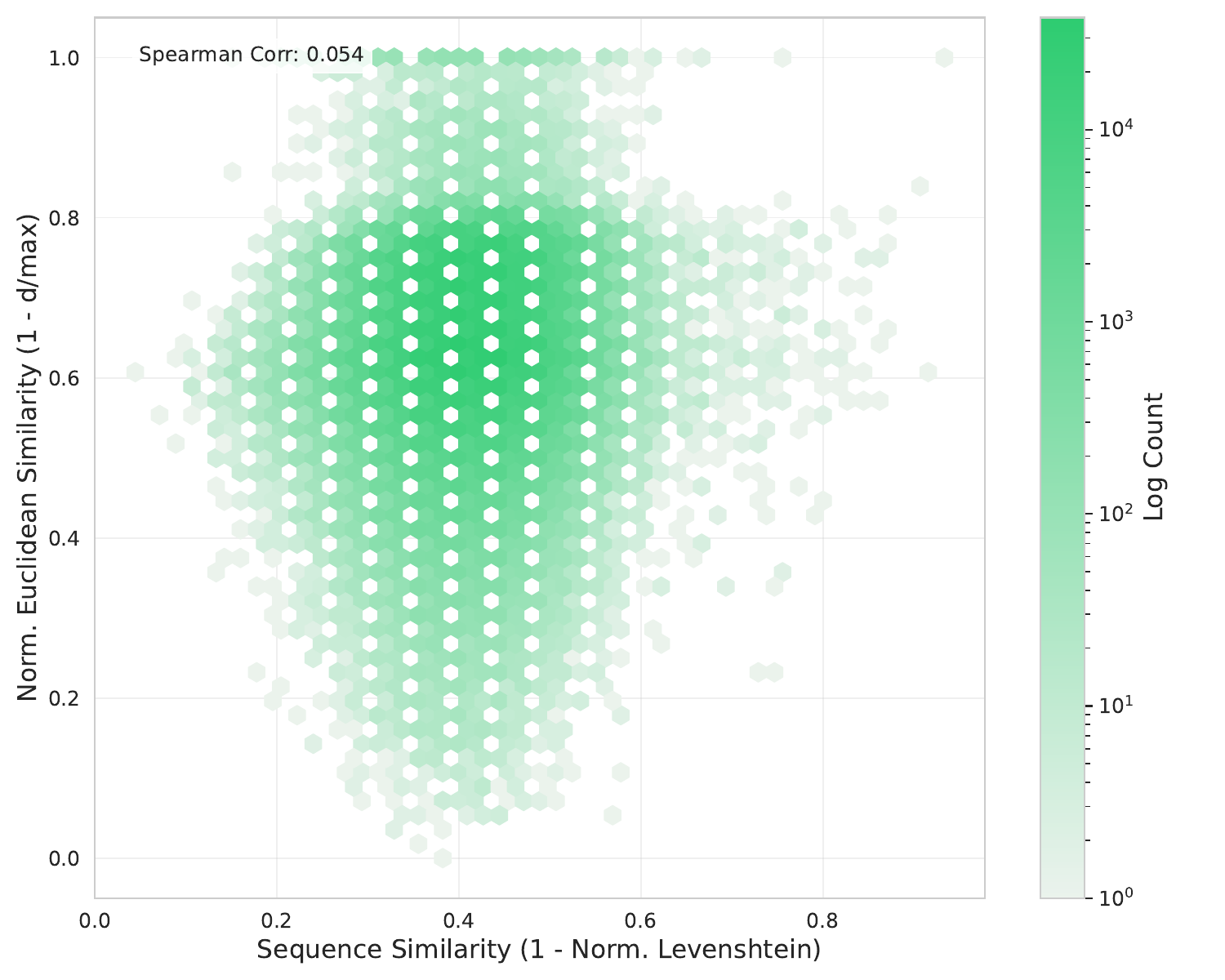}%
            \par\smallskip{\small $l = 45$}%
        \end{minipage}%
        \\[1em]%
        \begin{minipage}[t]{0.20\linewidth}%
            \centering%
            \includegraphics[width=\linewidth]{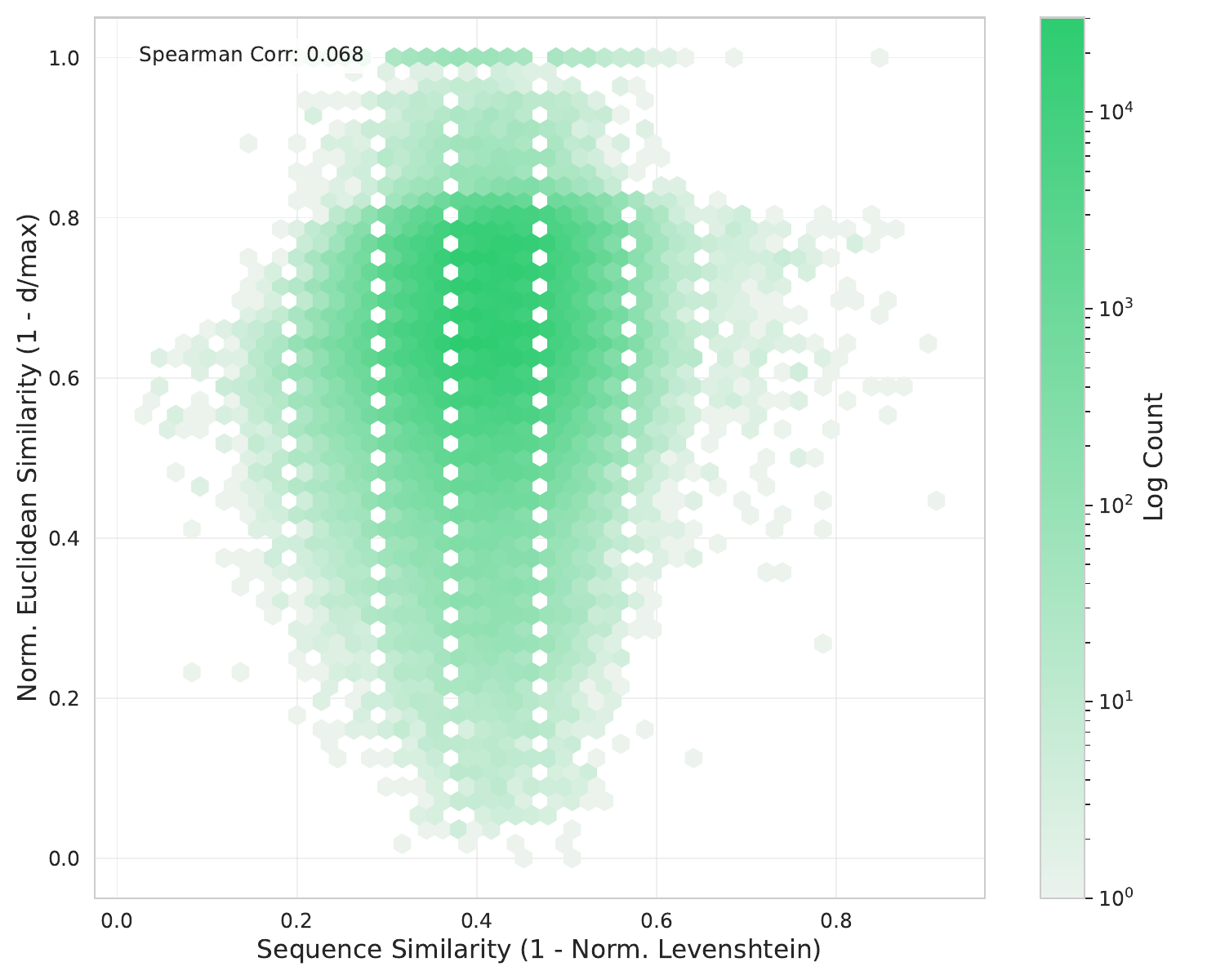}%
            \par\smallskip{\small $l = 50$}%
        \end{minipage}%
        \hspace{0.01\linewidth}%
        \begin{minipage}[t]{0.20\linewidth}%
            \centering%
            \includegraphics[width=\linewidth]{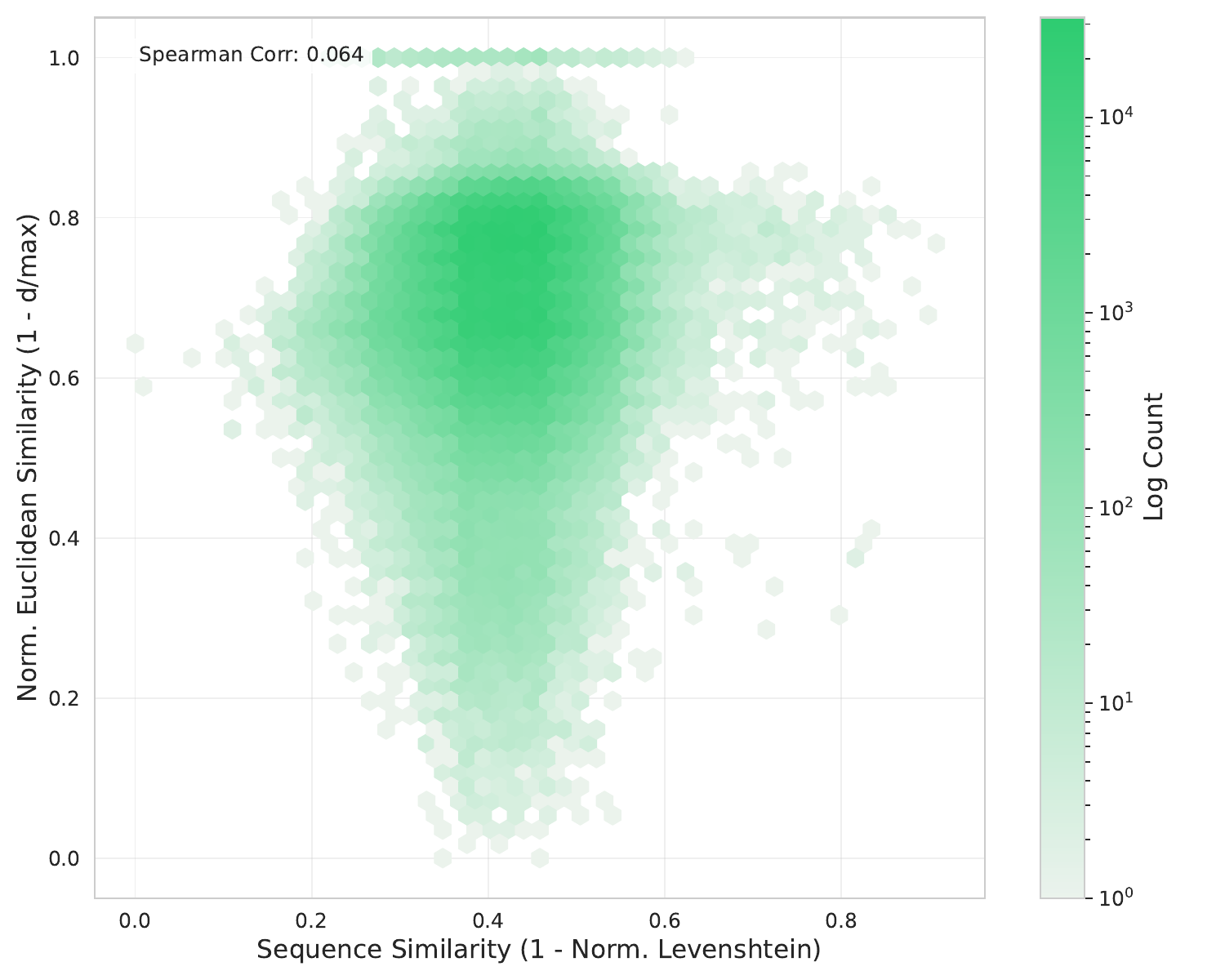}%
            \par\smallskip{\small $l = 60$}%
        \end{minipage}%
        \hspace{0.01\linewidth}%
        \begin{minipage}[t]{0.20\linewidth}%
            \centering%
            \includegraphics[width=\linewidth]{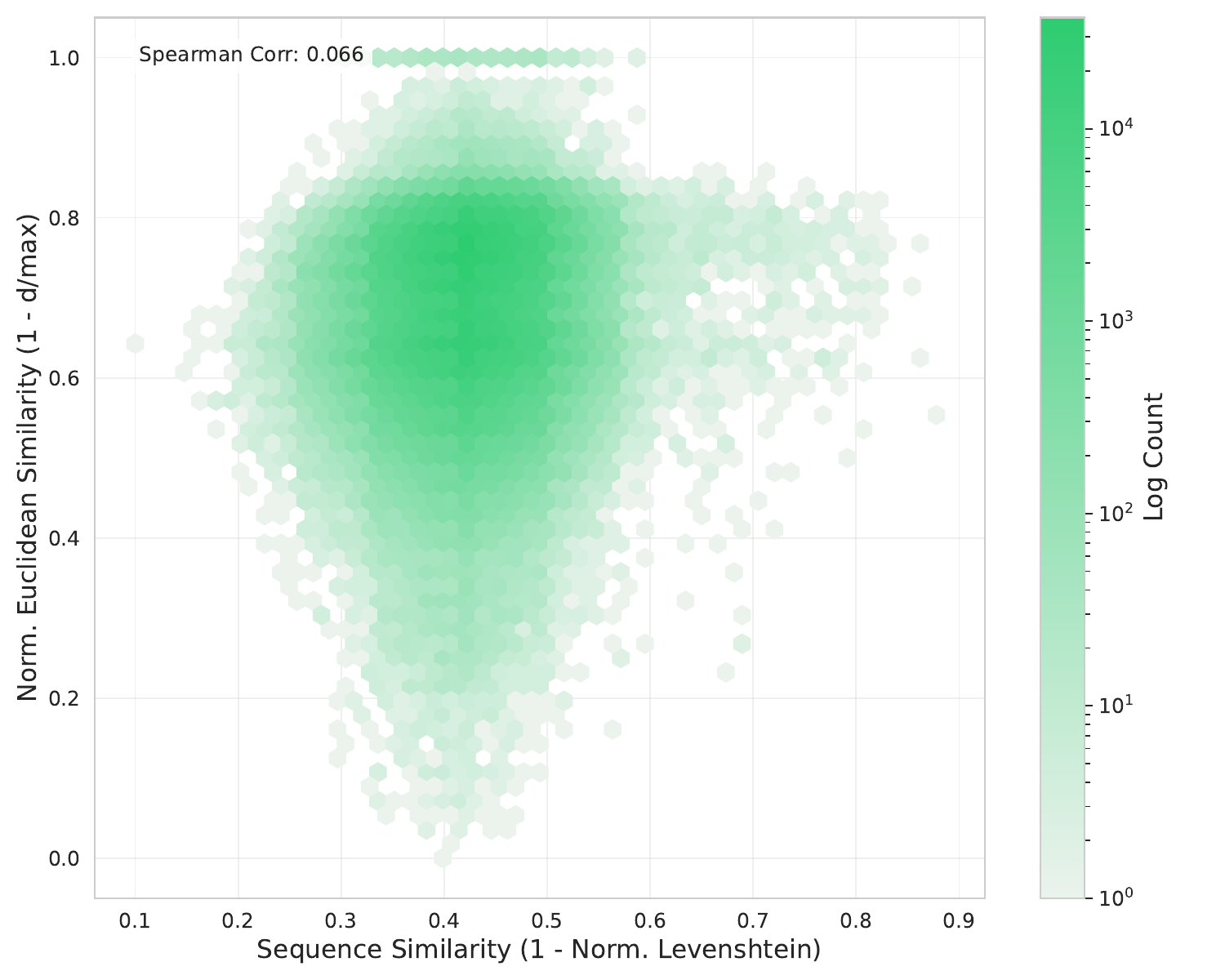}%
            \par\smallskip{\small $l = 70$}%
        \end{minipage}%
        \hspace{0.01\linewidth}%
        \begin{minipage}[t]{0.20\linewidth}%
            \centering%
            \includegraphics[width=\linewidth]{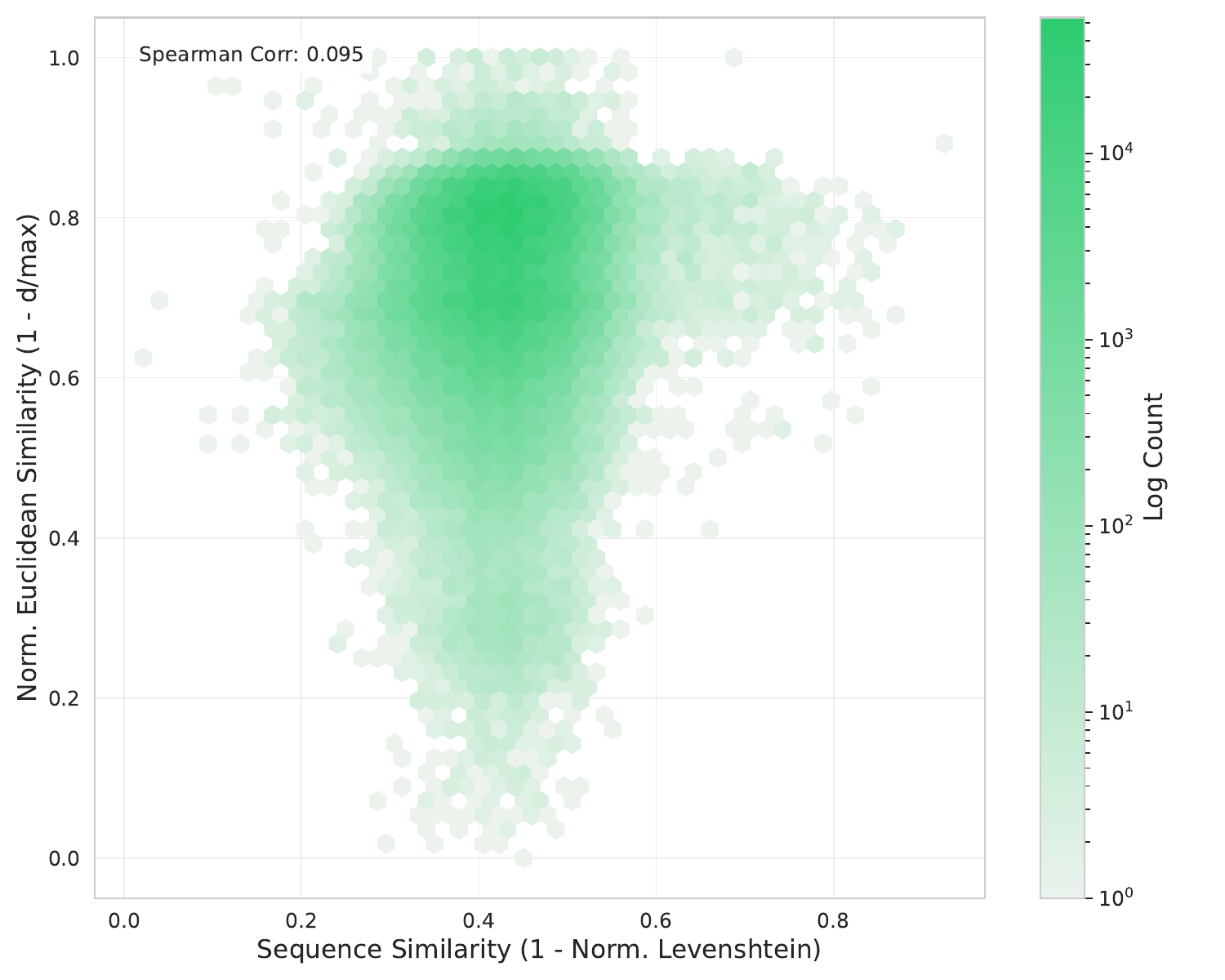}%
            \par\smallskip{\small $l = 80$}%
        \end{minipage}%
        \\[1em]%
        \begin{minipage}[t]{0.20\linewidth}%
            \centering%
            \includegraphics[width=\linewidth]{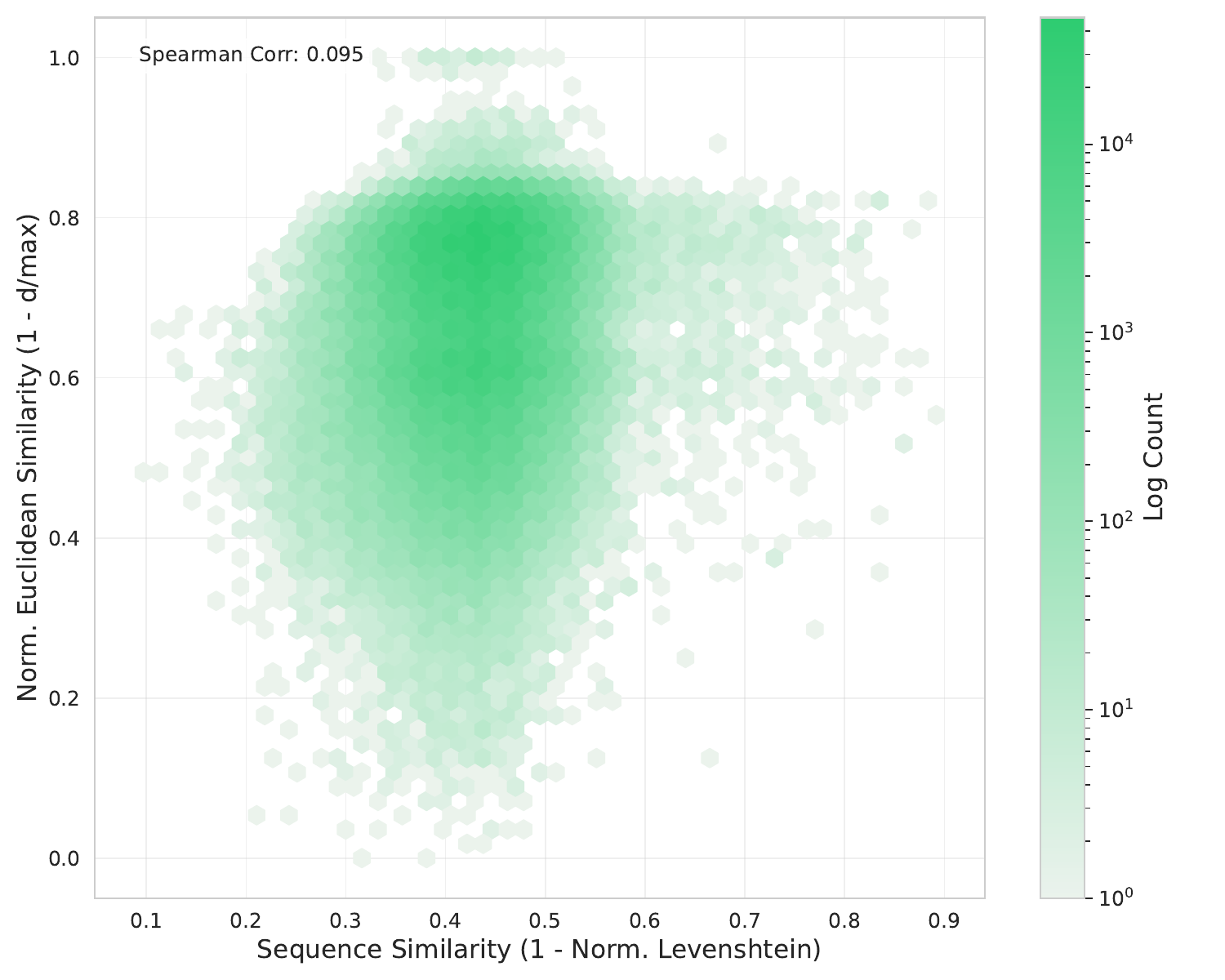}%
            \par\smallskip{\small $l = 90$}%
        \end{minipage}%
        \hspace{0.01\linewidth}%
        \begin{minipage}[t]{0.20\linewidth}%
            \centering%
            \includegraphics[width=\linewidth]{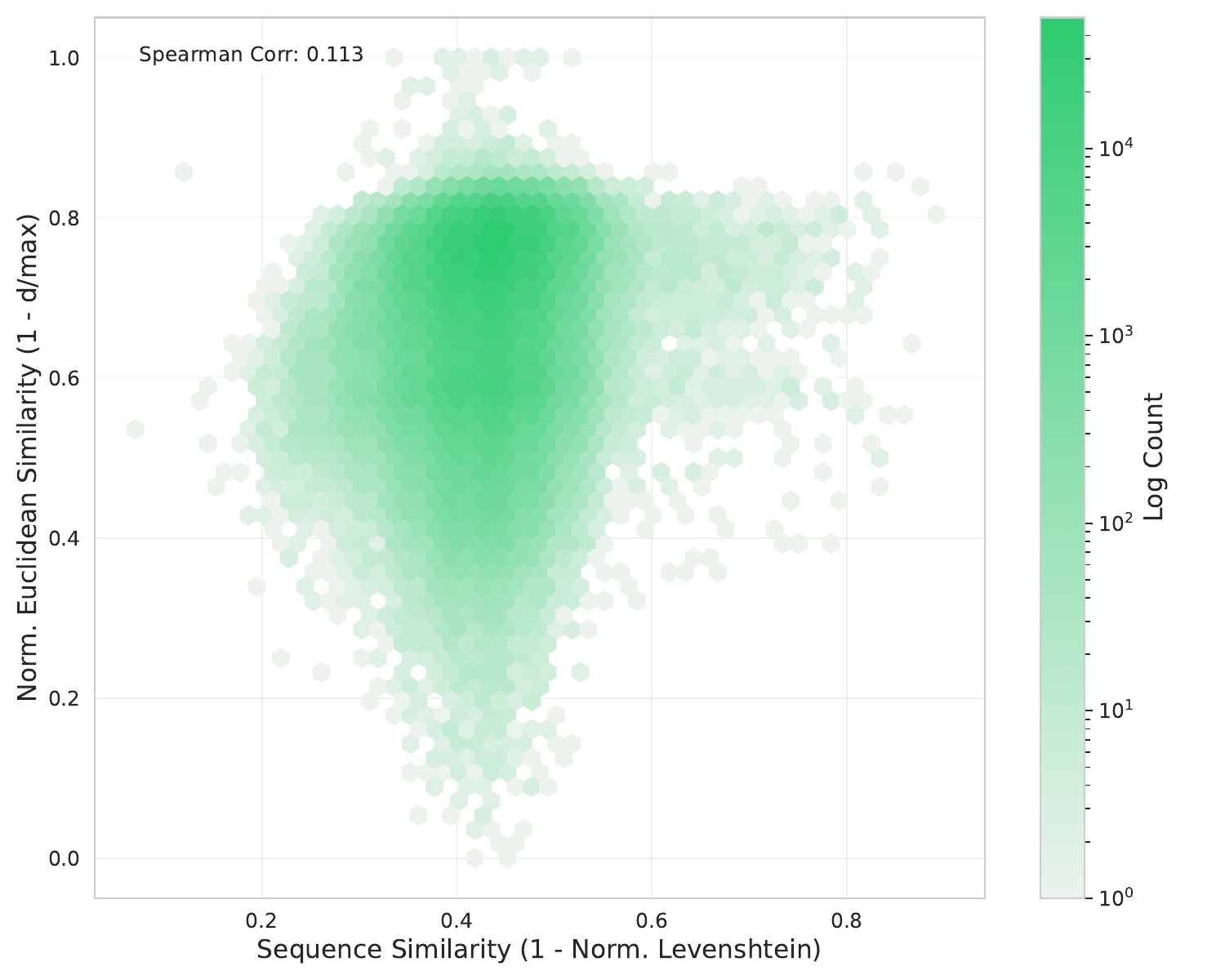}%
            \par\smallskip{\small $l = 100$}%
        \end{minipage}%
    \caption{Normalised Euclidean embedding similarity vs.\ sequence similarity for DNABERT-2 across all sequence lengths.}
    \label{fig:euclidean_dnabert2}
\end{figure}

\begin{figure}[H]
    \centering
        \begin{minipage}[t]{0.24\linewidth}%
            \centering%
            \includegraphics[width=\linewidth]{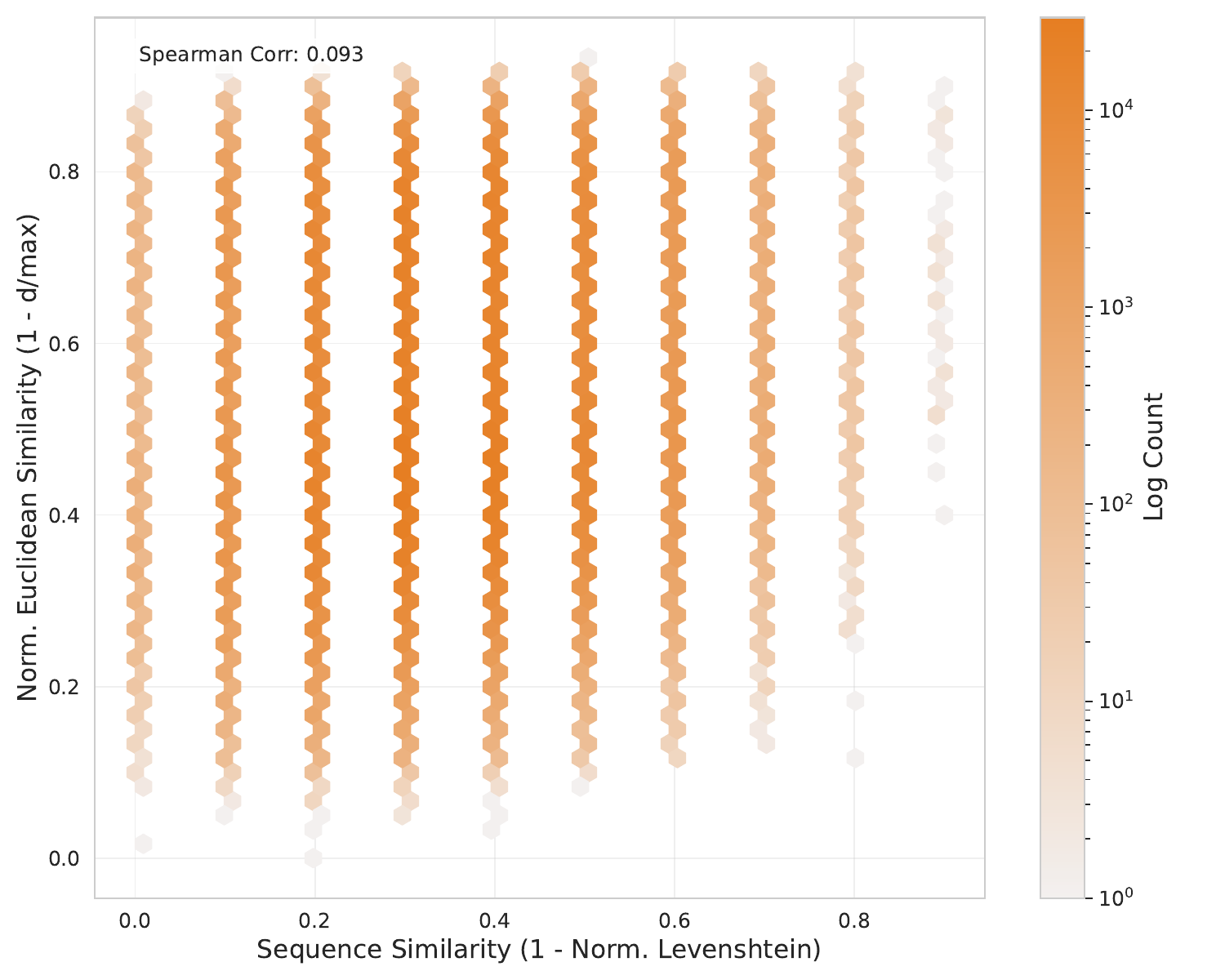}%
            \par\smallskip{\small $l = 10$}%
        \end{minipage}%
        \hspace{0.01\linewidth}%
        \begin{minipage}[t]{0.24\linewidth}%
            \centering%
            \includegraphics[width=\linewidth]{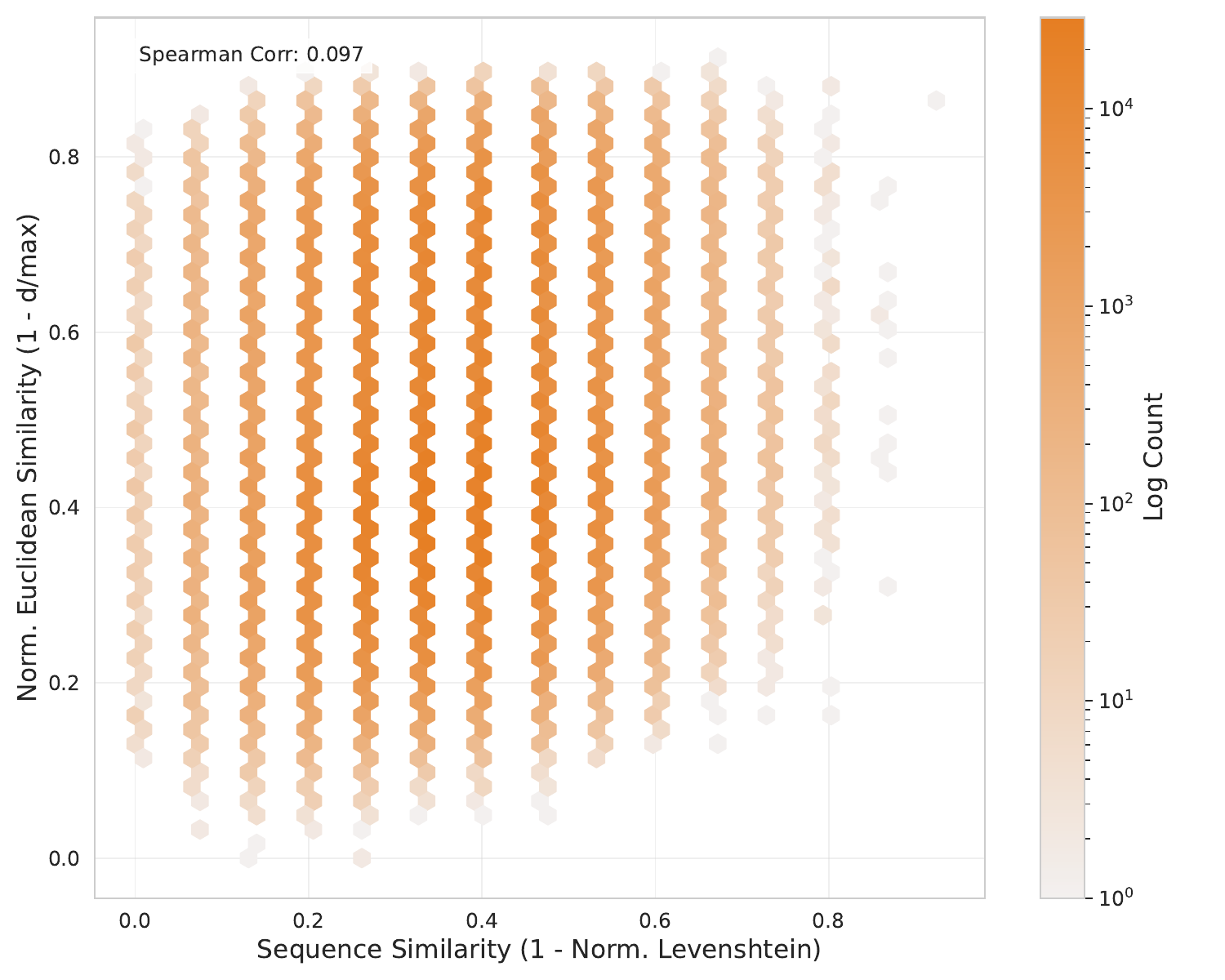}%
            \par\smallskip{\small $l = 15$}%
        \end{minipage}%
        \hspace{0.01\linewidth}%
        \begin{minipage}[t]{0.24\linewidth}%
            \centering%
            \includegraphics[width=\linewidth]{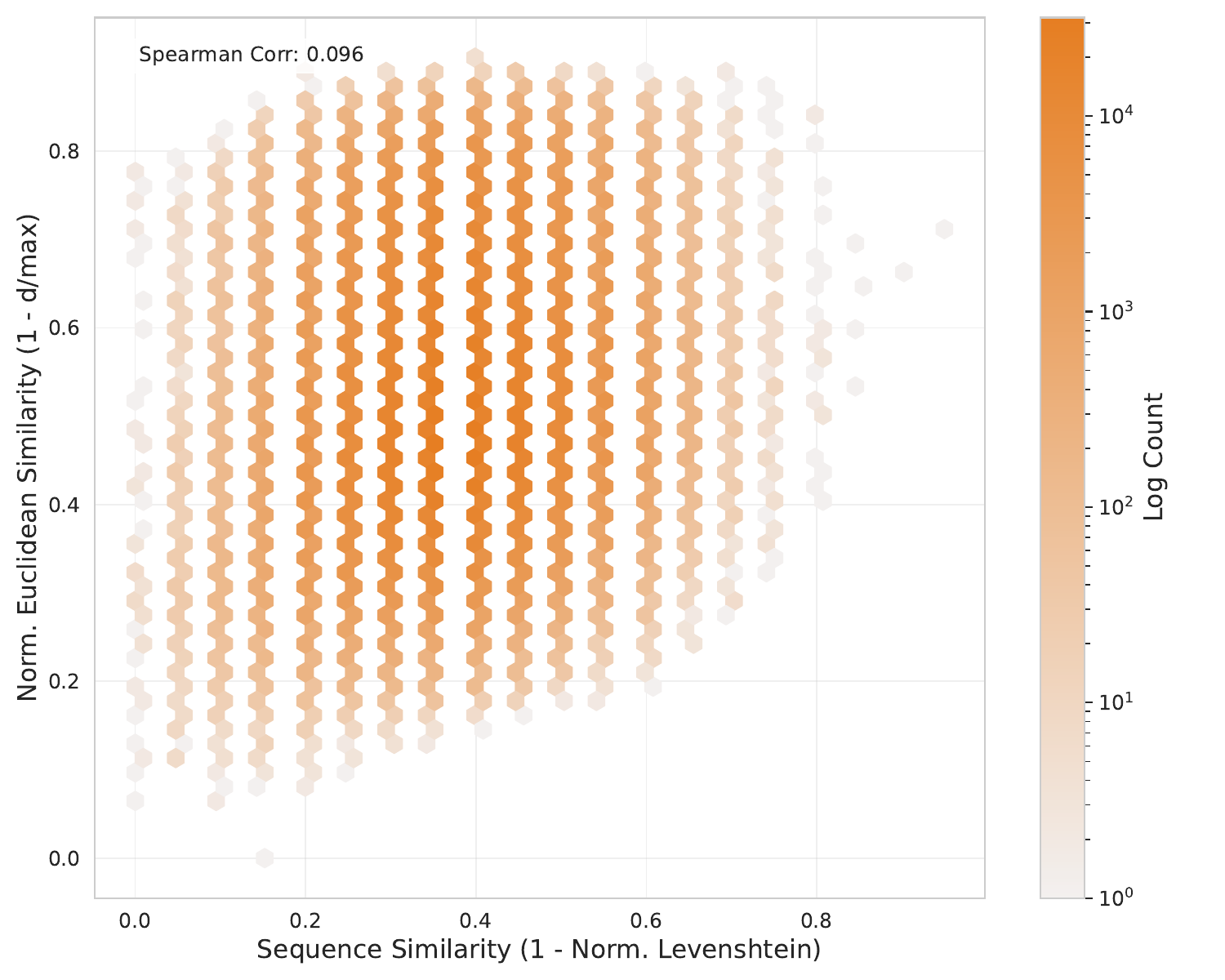}%
            \par\smallskip{\small $l = 20$}%
        \end{minipage}%
        \hspace{0.01\linewidth}%
        \begin{minipage}[t]{0.24\linewidth}%
            \centering%
            \includegraphics[width=\linewidth]{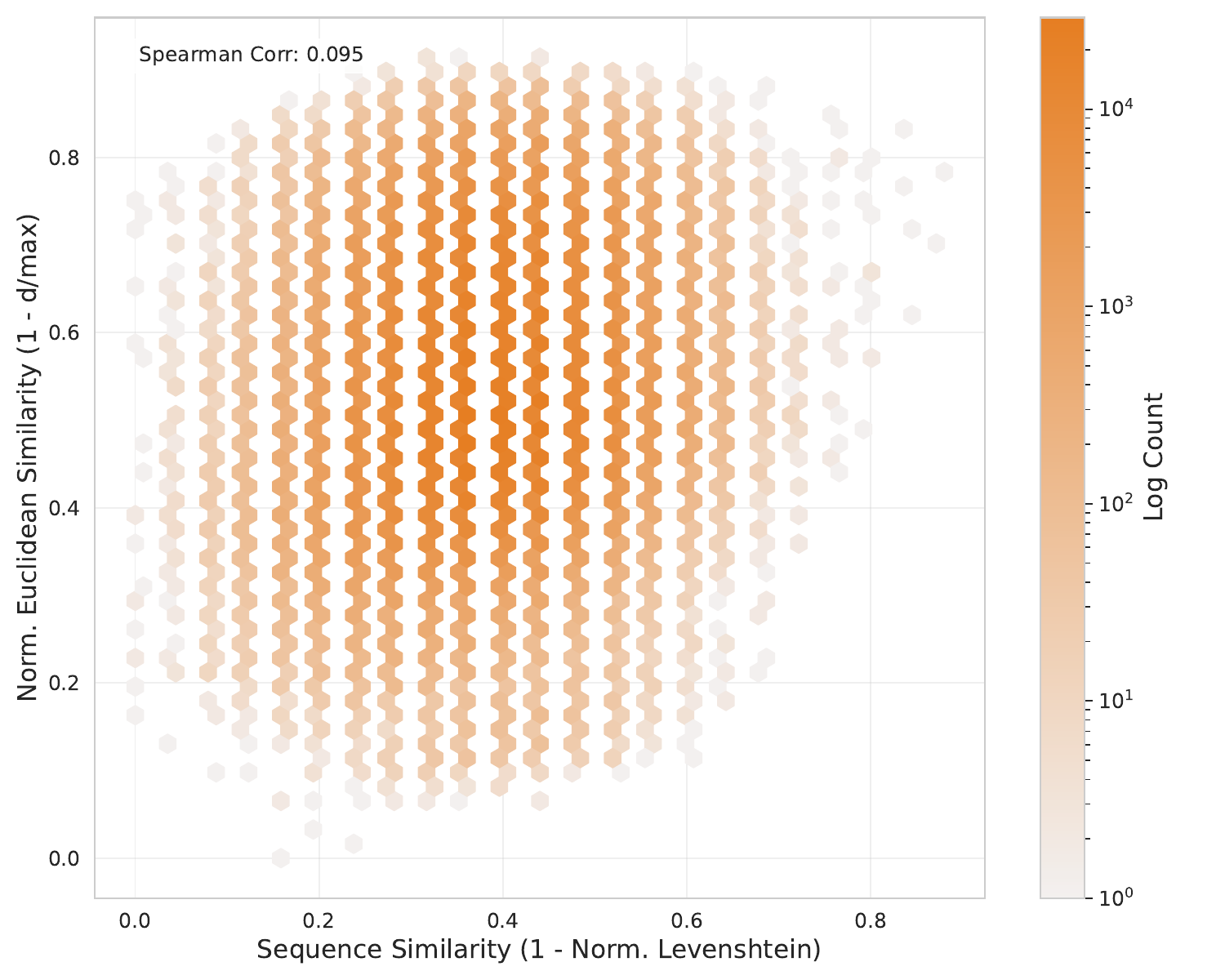}%
            \par\smallskip{\small $l = 25$}%
        \end{minipage}%
        \\[1em]%
        \begin{minipage}[t]{0.24\linewidth}%
            \centering%
            \includegraphics[width=\linewidth]{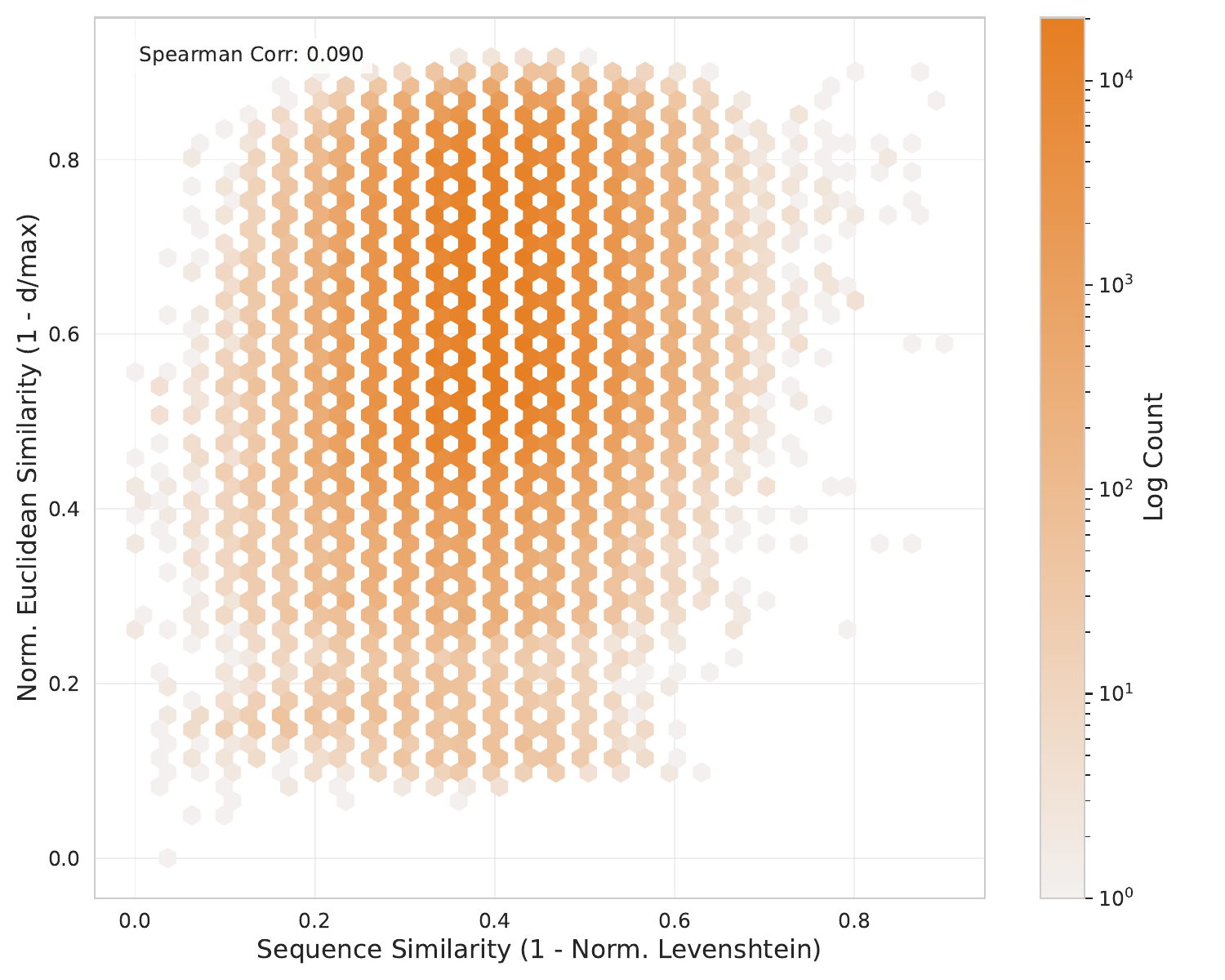}%
            \par\smallskip{\small $l = 30$}%
        \end{minipage}%
        \hspace{0.01\linewidth}%
        \begin{minipage}[t]{0.24\linewidth}%
            \centering%
            \includegraphics[width=\linewidth]{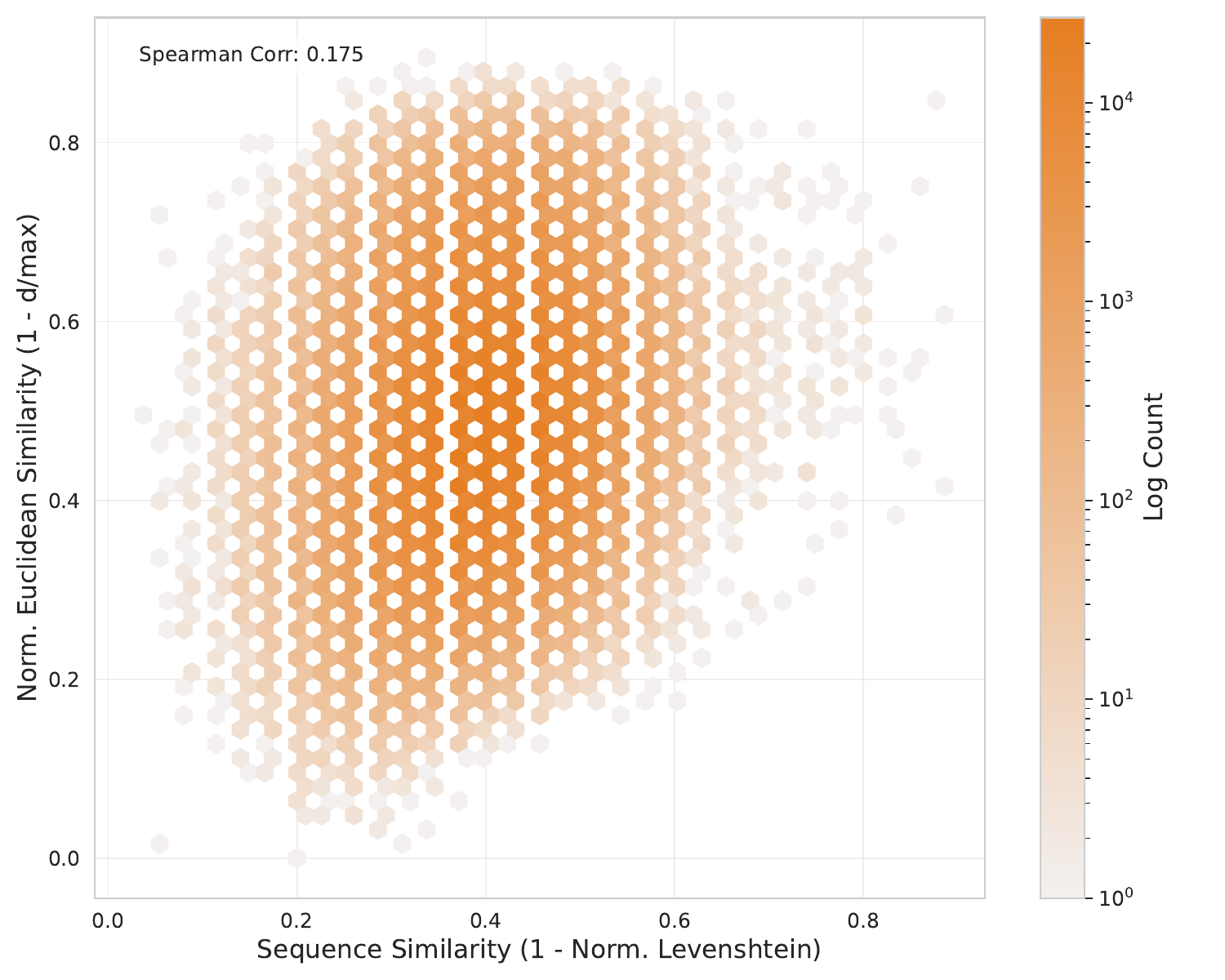}%
            \par\smallskip{\small $l = 35$}%
        \end{minipage}%
        \hspace{0.01\linewidth}%
        \begin{minipage}[t]{0.24\linewidth}%
            \centering%
            \includegraphics[width=\linewidth]{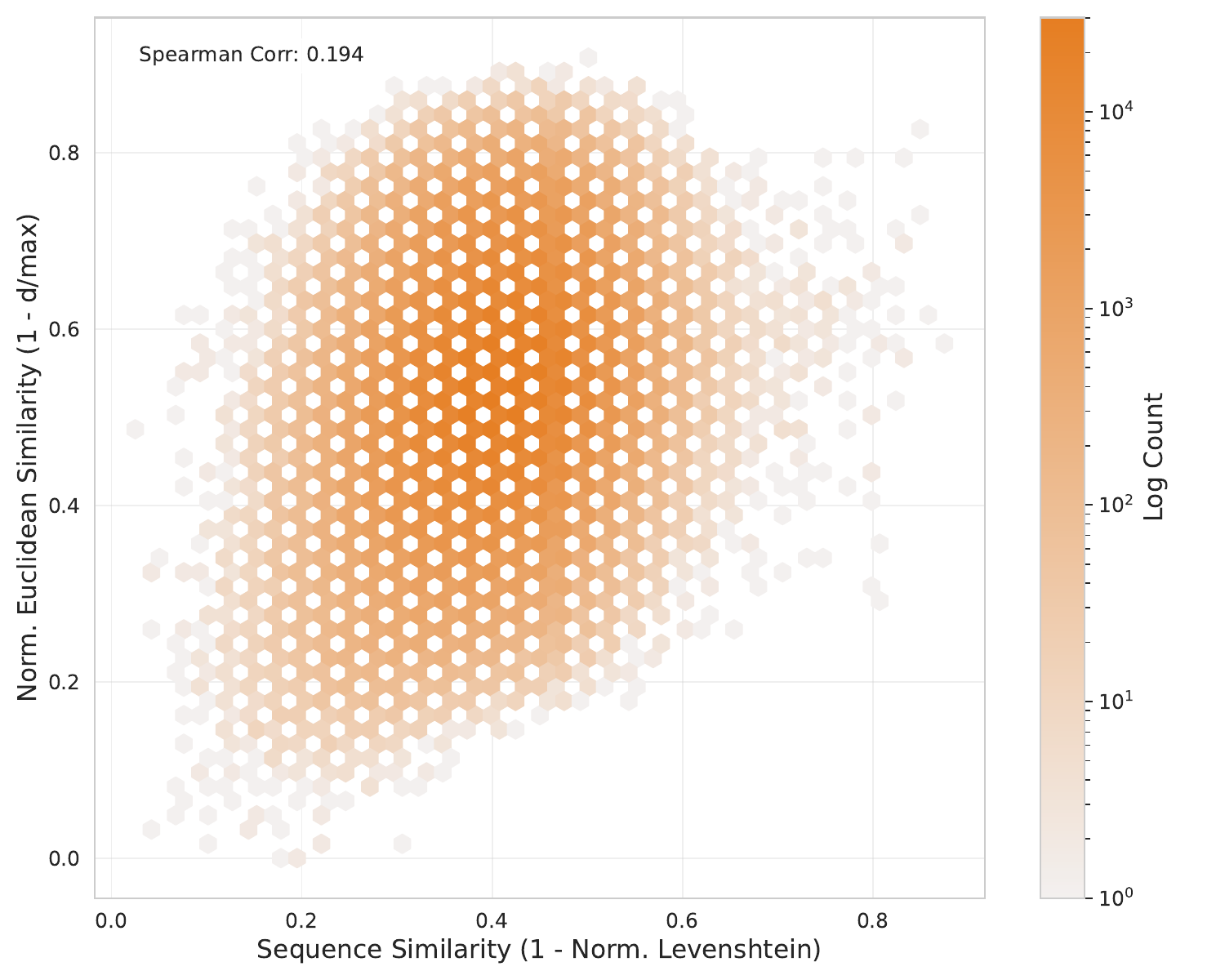}%
            \par\smallskip{\small $l = 40$}%
        \end{minipage}%
        \hspace{0.01\linewidth}%
        \begin{minipage}[t]{0.24\linewidth}%
            \centering%
            \includegraphics[width=\linewidth]{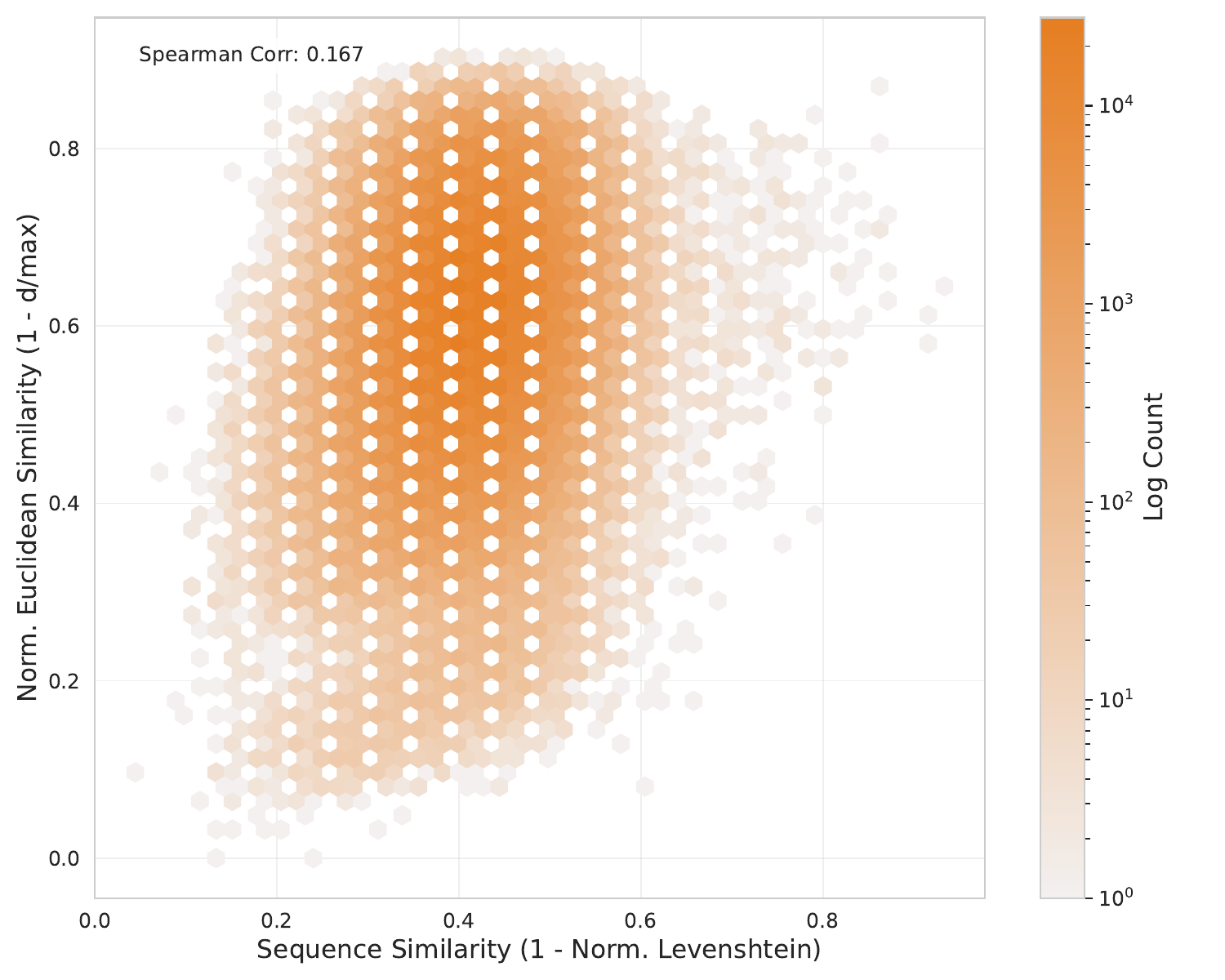}%
            \par\smallskip{\small $l = 45$}%
        \end{minipage}%
        \\[1em]%
        \begin{minipage}[t]{0.24\linewidth}%
            \centering%
            \includegraphics[width=\linewidth]{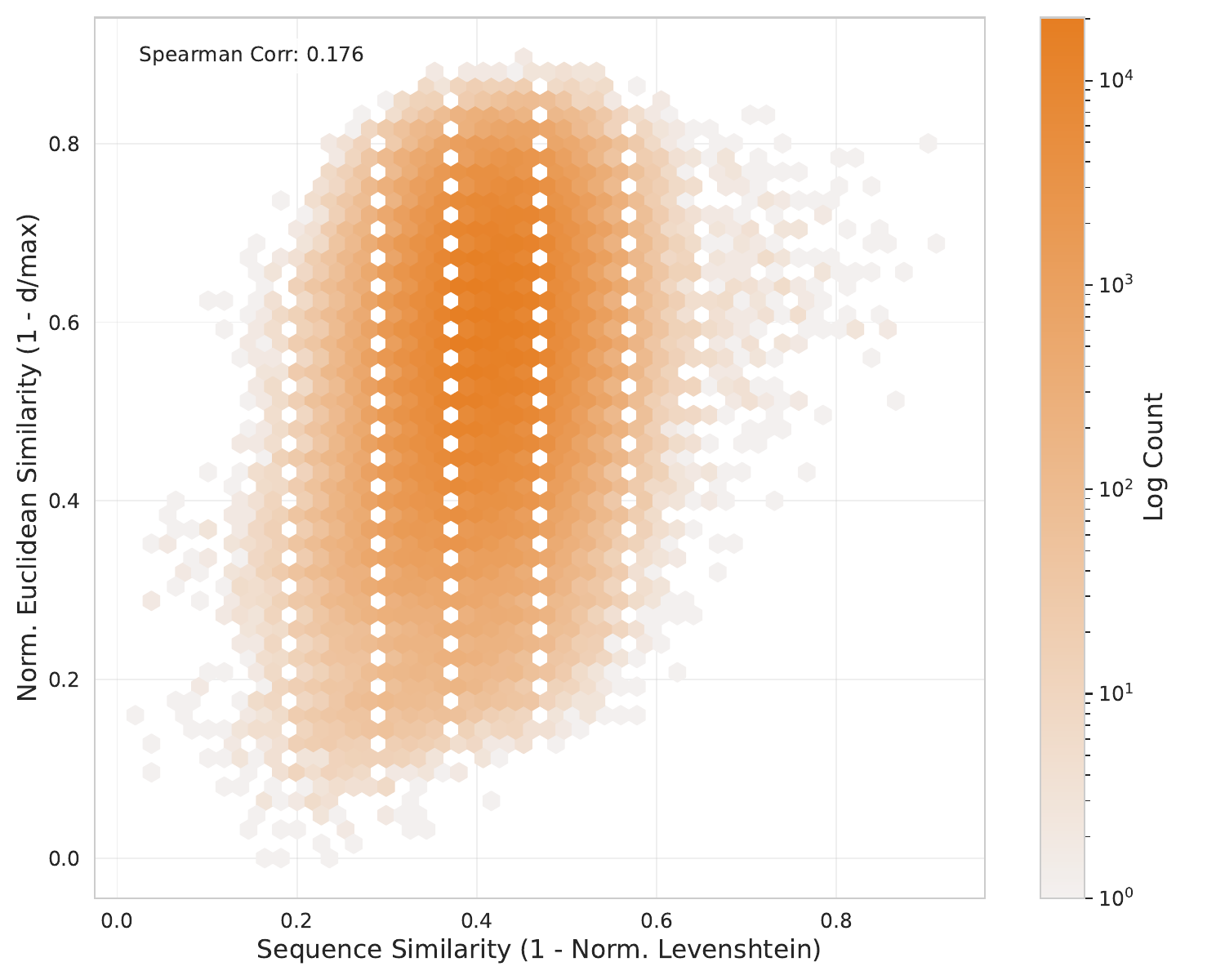}%
            \par\smallskip{\small $l = 50$}%
        \end{minipage}%
        \hspace{0.01\linewidth}%
        \begin{minipage}[t]{0.24\linewidth}%
            \centering%
            \includegraphics[width=\linewidth]{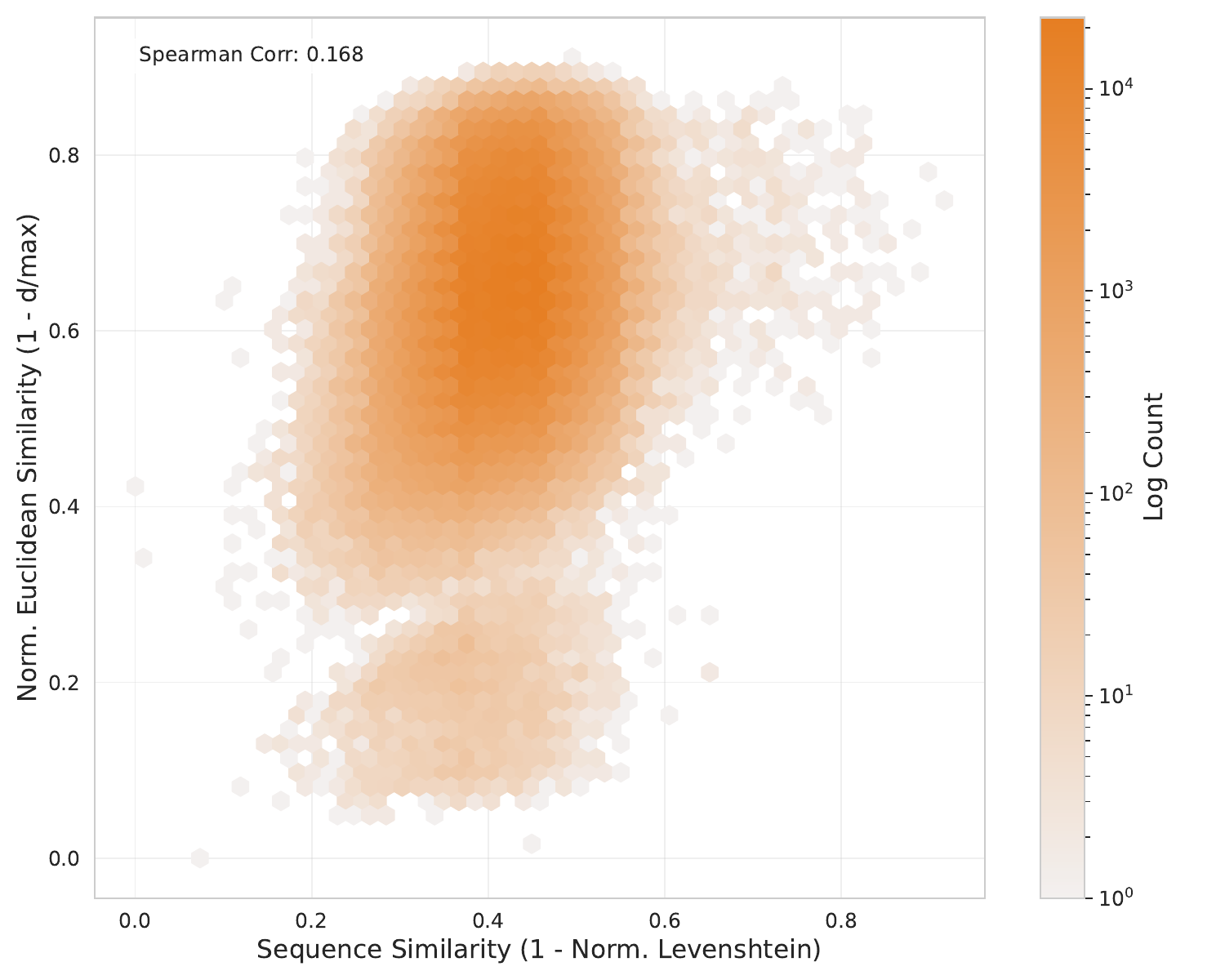}%
            \par\smallskip{\small $l = 60$}%
        \end{minipage}%
        \hspace{0.01\linewidth}%
        \begin{minipage}[t]{0.24\linewidth}%
            \centering%
            \includegraphics[width=\linewidth]{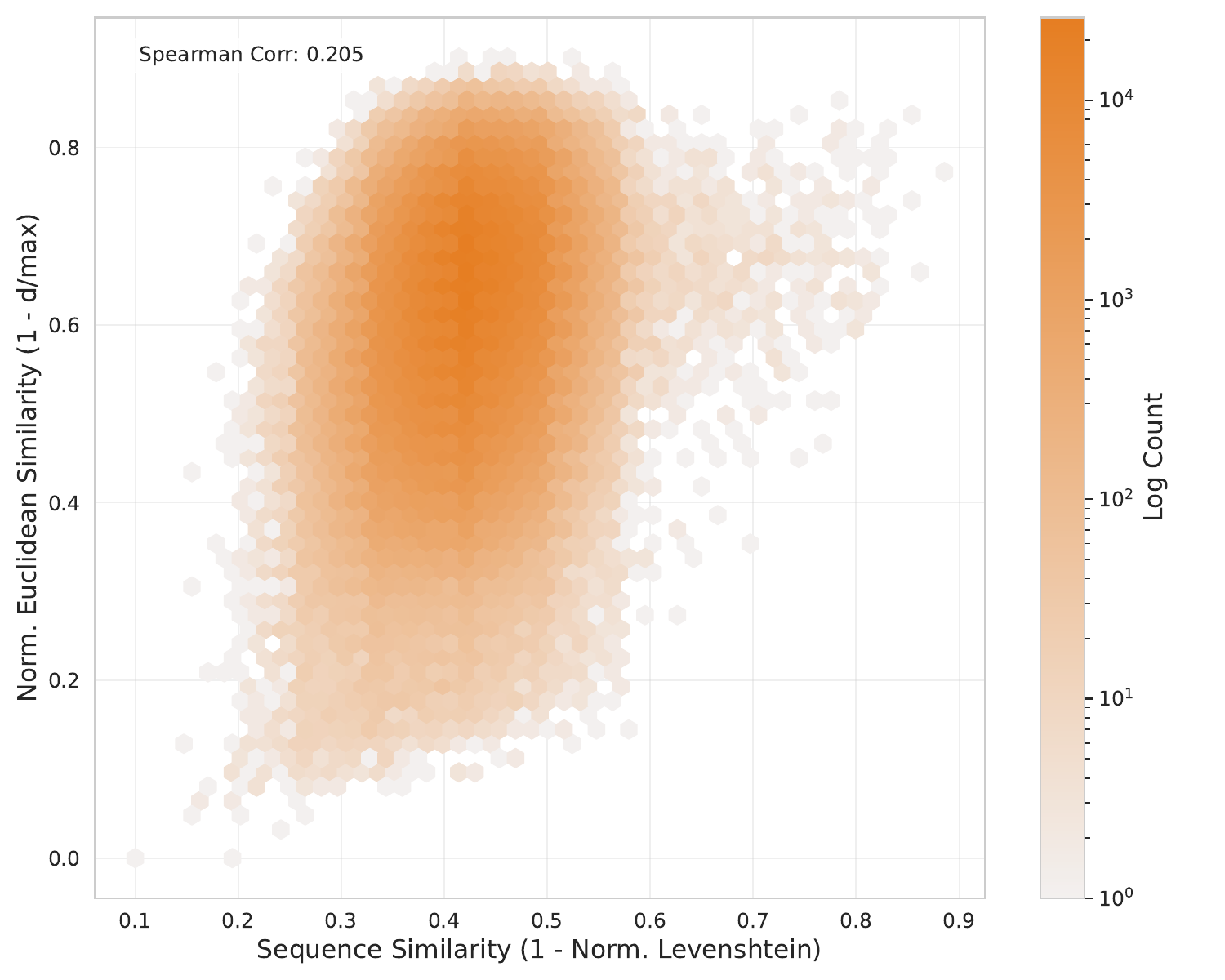}%
            \par\smallskip{\small $l = 70$}%
        \end{minipage}%
        \hspace{0.01\linewidth}%
        \begin{minipage}[t]{0.24\linewidth}%
            \centering%
            \includegraphics[width=\linewidth]{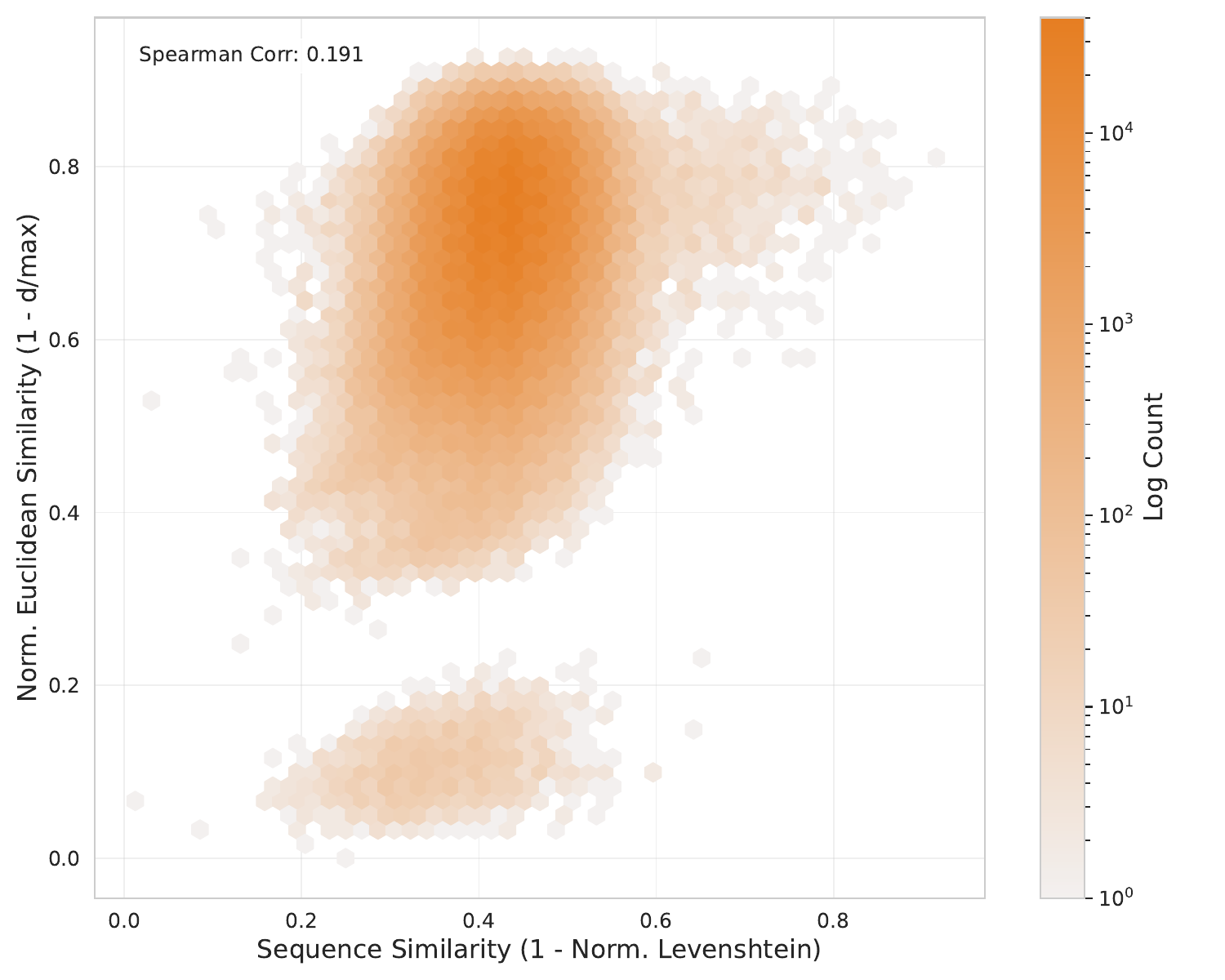}%
            \par\smallskip{\small $l = 80$}%
        \end{minipage}%
        \\[1em]%
        \begin{minipage}[t]{0.24\linewidth}%
            \centering%
            \includegraphics[width=\linewidth]{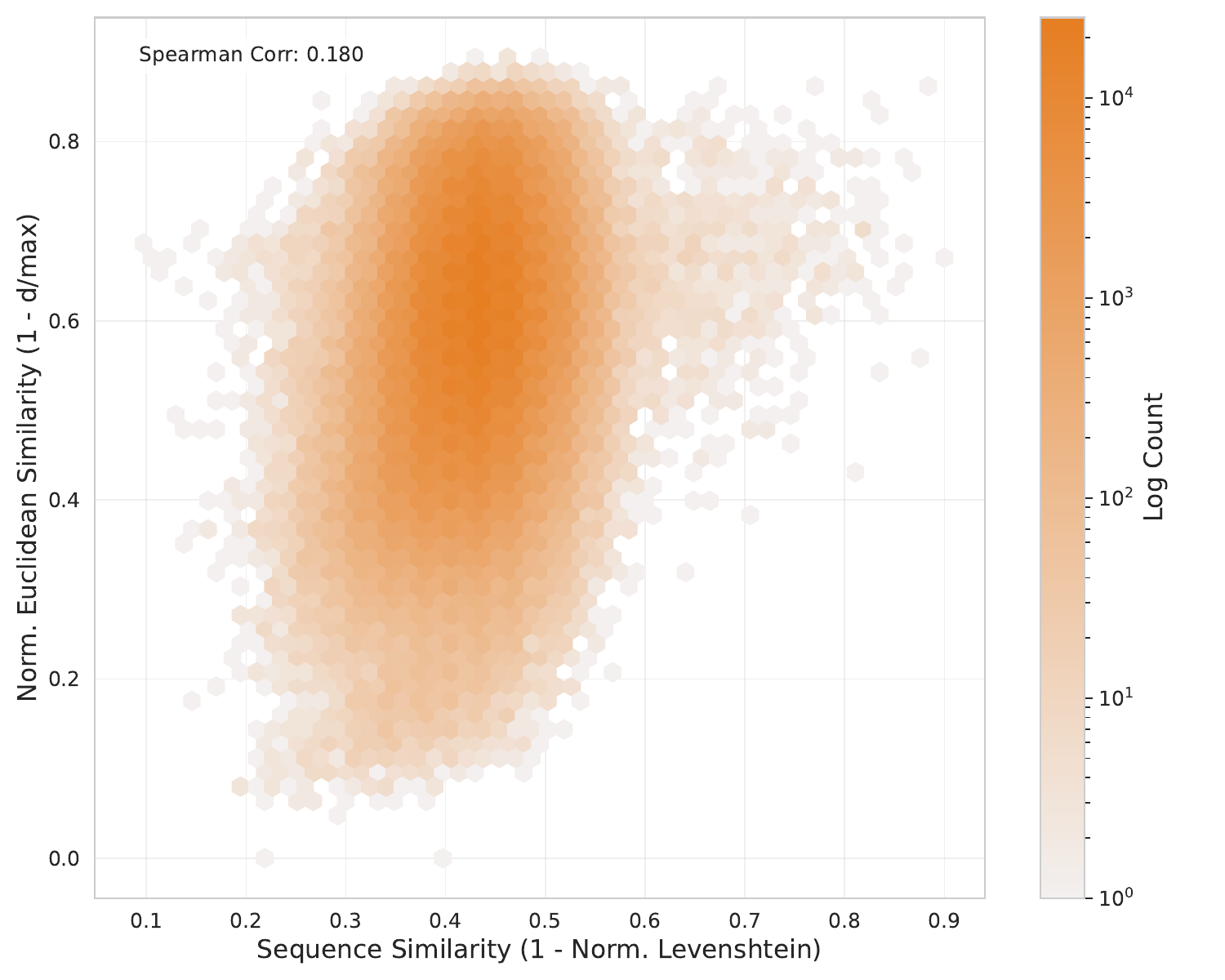}%
            \par\smallskip{\small $l = 90$}%
        \end{minipage}%
        \hspace{0.01\linewidth}%
        \begin{minipage}[t]{0.24\linewidth}%
            \centering%
            \includegraphics[width=\linewidth]{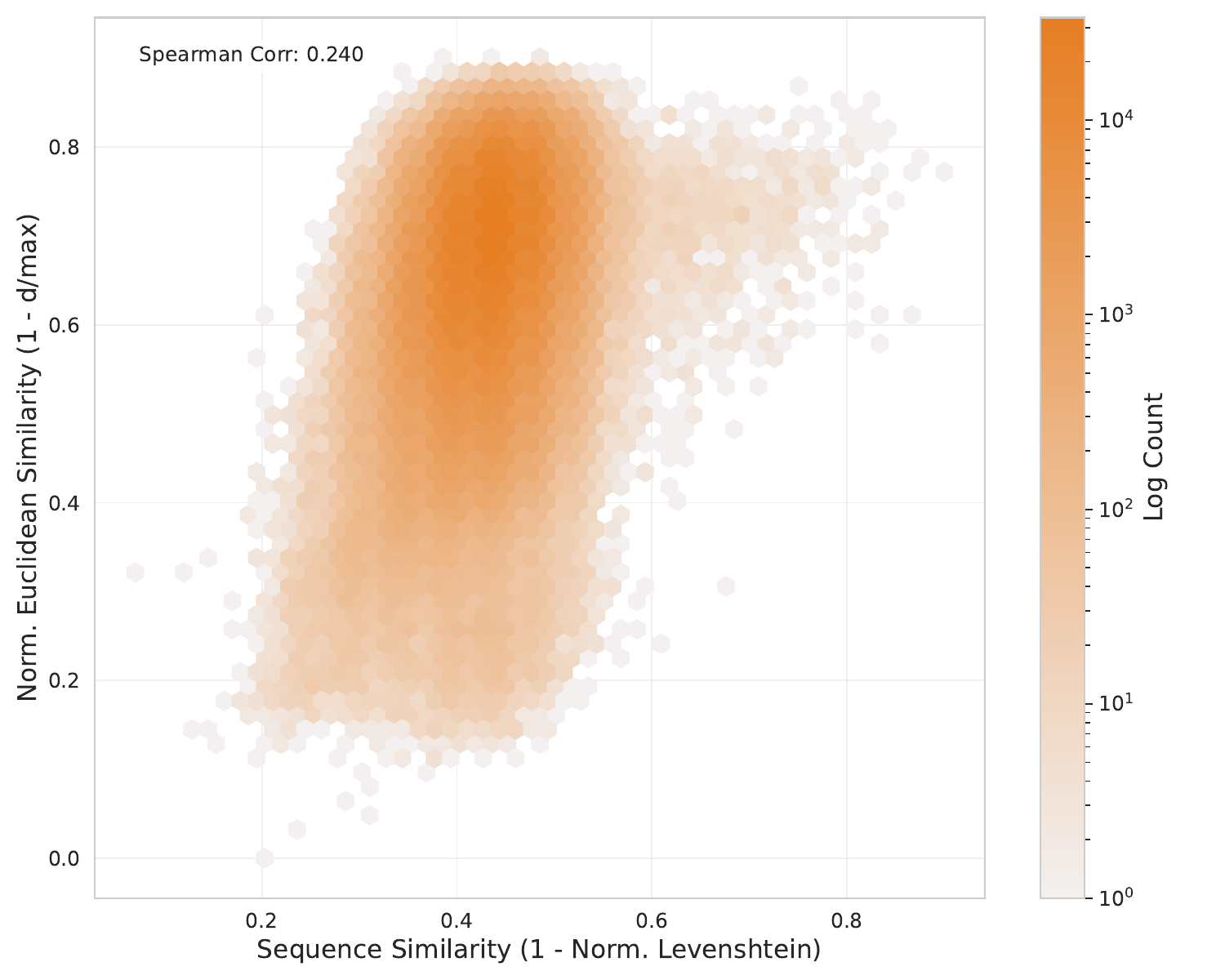}%
            \par\smallskip{\small $l = 100$}%
        \end{minipage}%
    \caption{Normalised Euclidean embedding similarity vs.\ sequence similarity for NTv2 across all sequence lengths.}
    \label{fig:euclidean_ntv2}
\end{figure}

\begin{figure}[H]
    \centering
        \begin{minipage}[t]{0.24\linewidth}%
            \centering%
            \includegraphics[width=\linewidth]{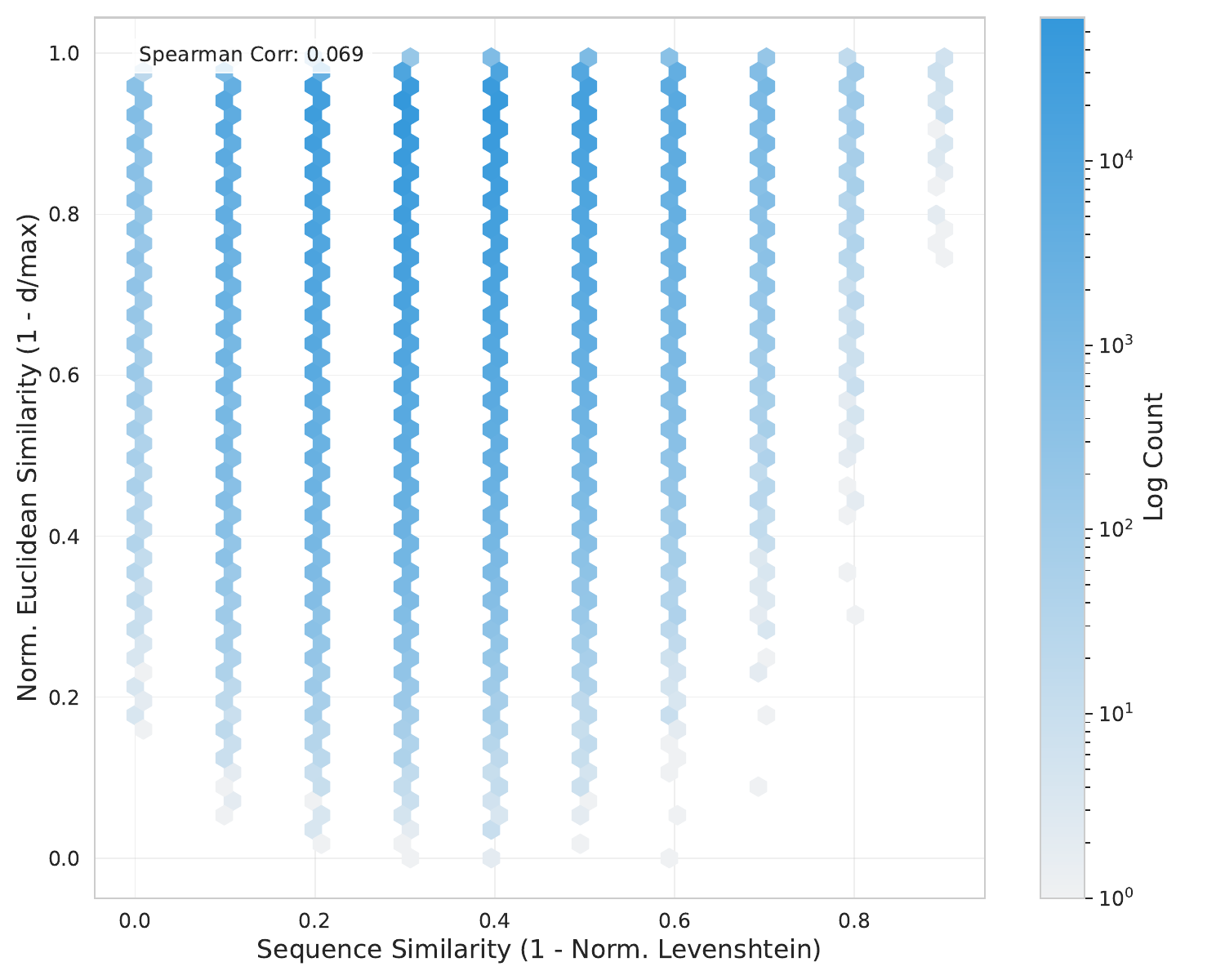}%
            \par\smallskip{\small $l = 10$}%
        \end{minipage}%
        \hspace{0.01\linewidth}%
        \begin{minipage}[t]{0.24\linewidth}%
            \centering%
            \includegraphics[width=\linewidth]{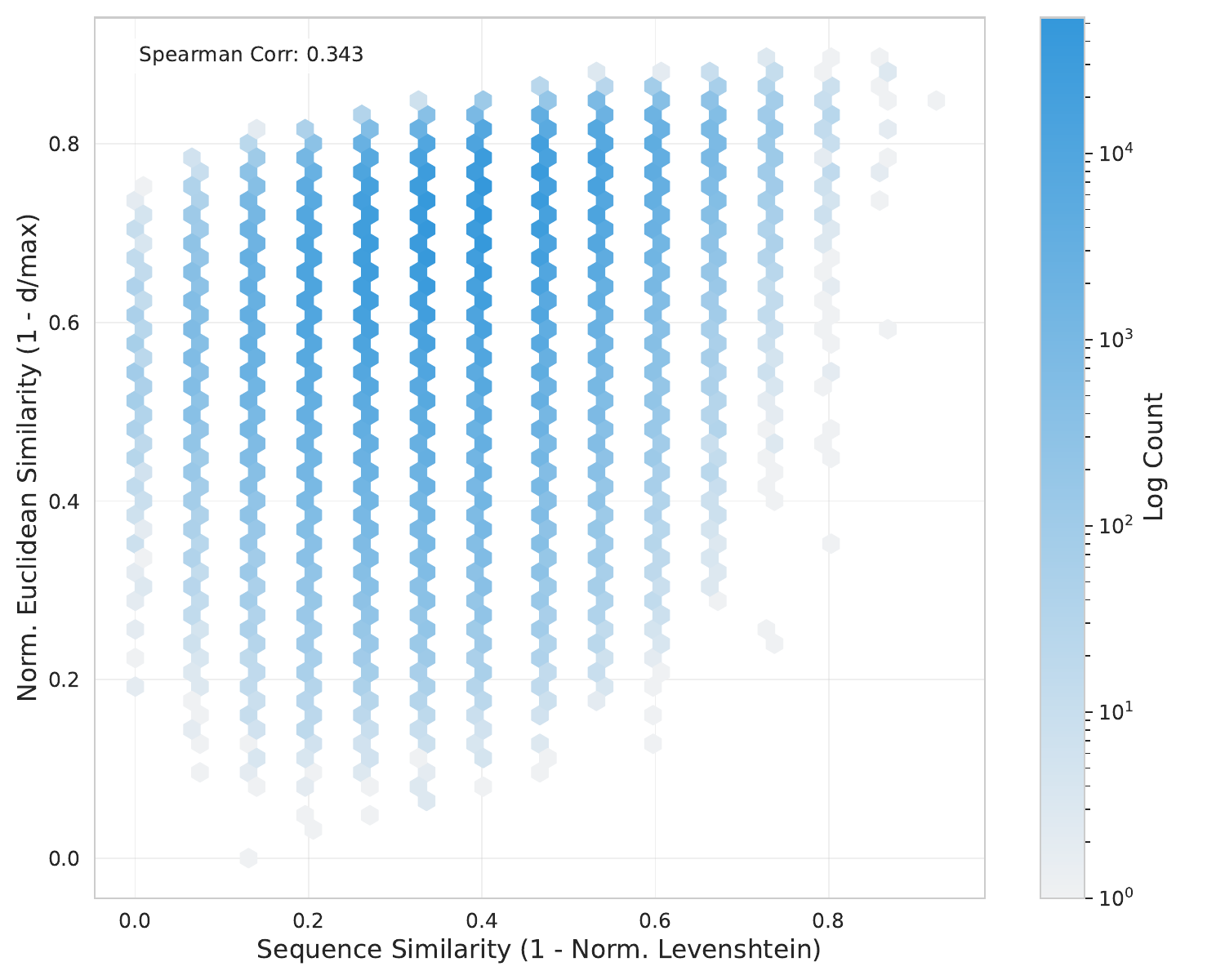}%
            \par\smallskip{\small $l = 15$}%
        \end{minipage}%
        \hspace{0.01\linewidth}%
        \begin{minipage}[t]{0.24\linewidth}%
            \centering%
            \includegraphics[width=\linewidth]{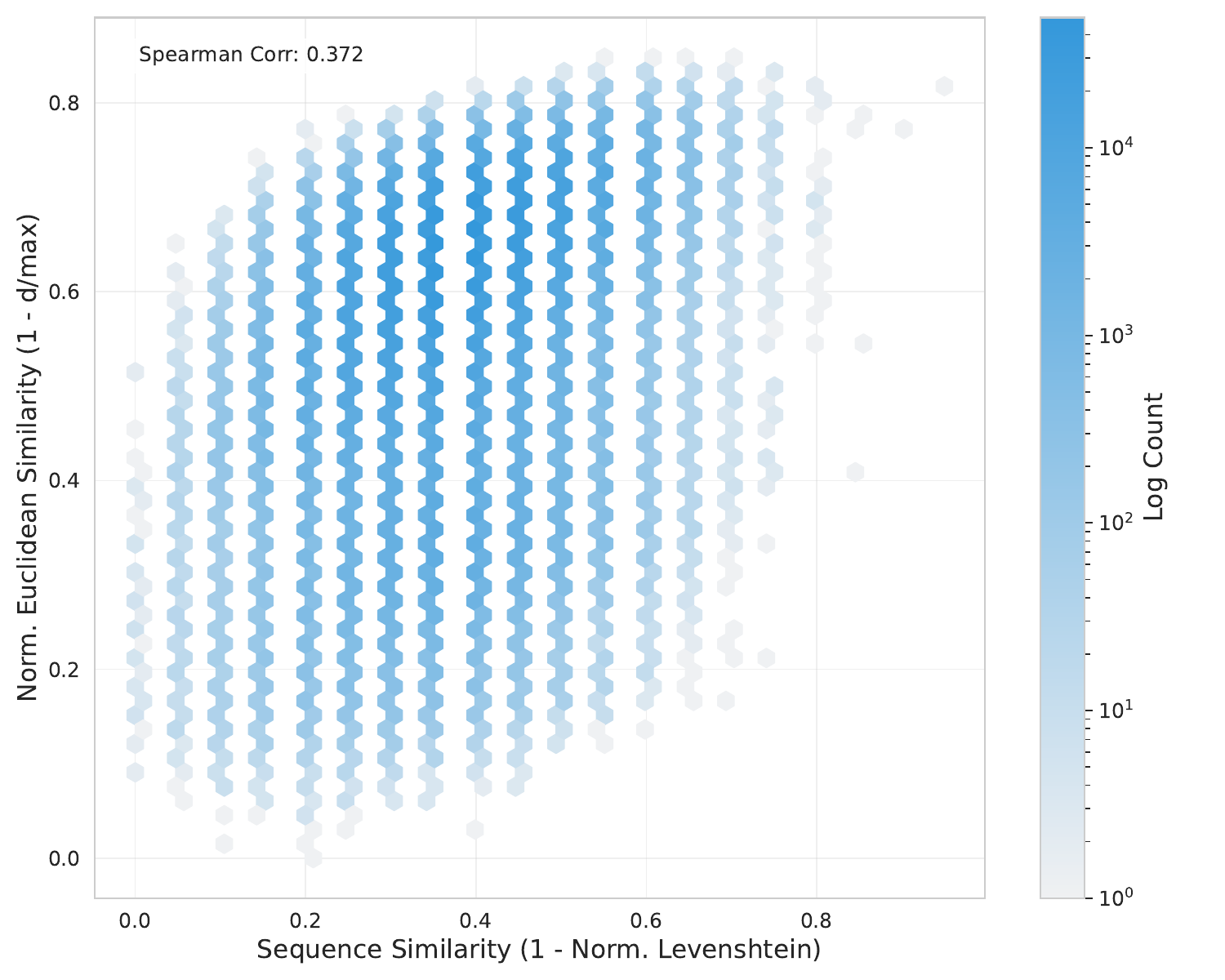}%
            \par\smallskip{\small $l = 20$}%
        \end{minipage}%
        \hspace{0.01\linewidth}%
        \begin{minipage}[t]{0.24\linewidth}%
            \centering%
            \includegraphics[width=\linewidth]{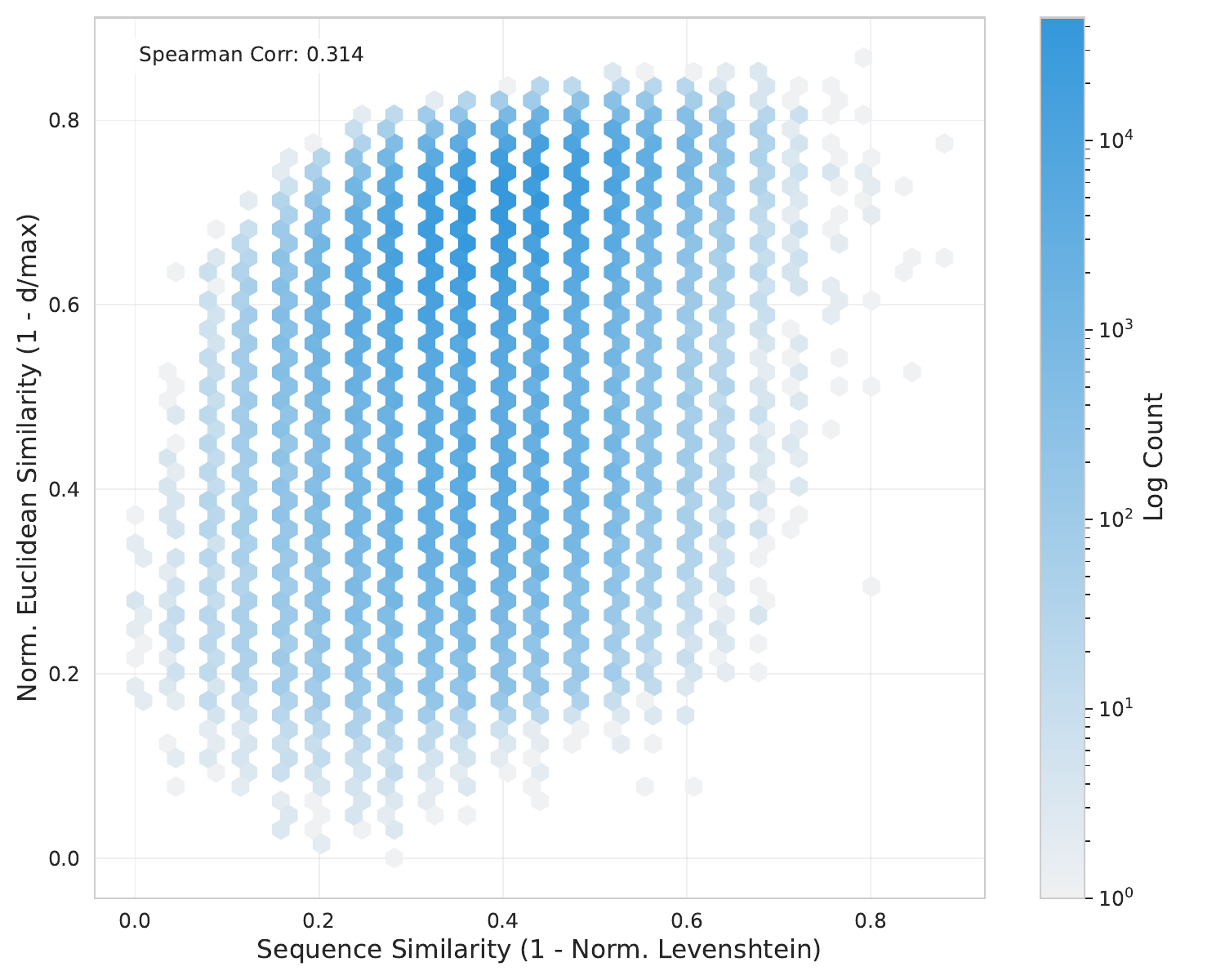}%
            \par\smallskip{\small $l = 25$}%
        \end{minipage}%
        \\[1em]%
        \begin{minipage}[t]{0.24\linewidth}%
            \centering%
            \includegraphics[width=\linewidth]{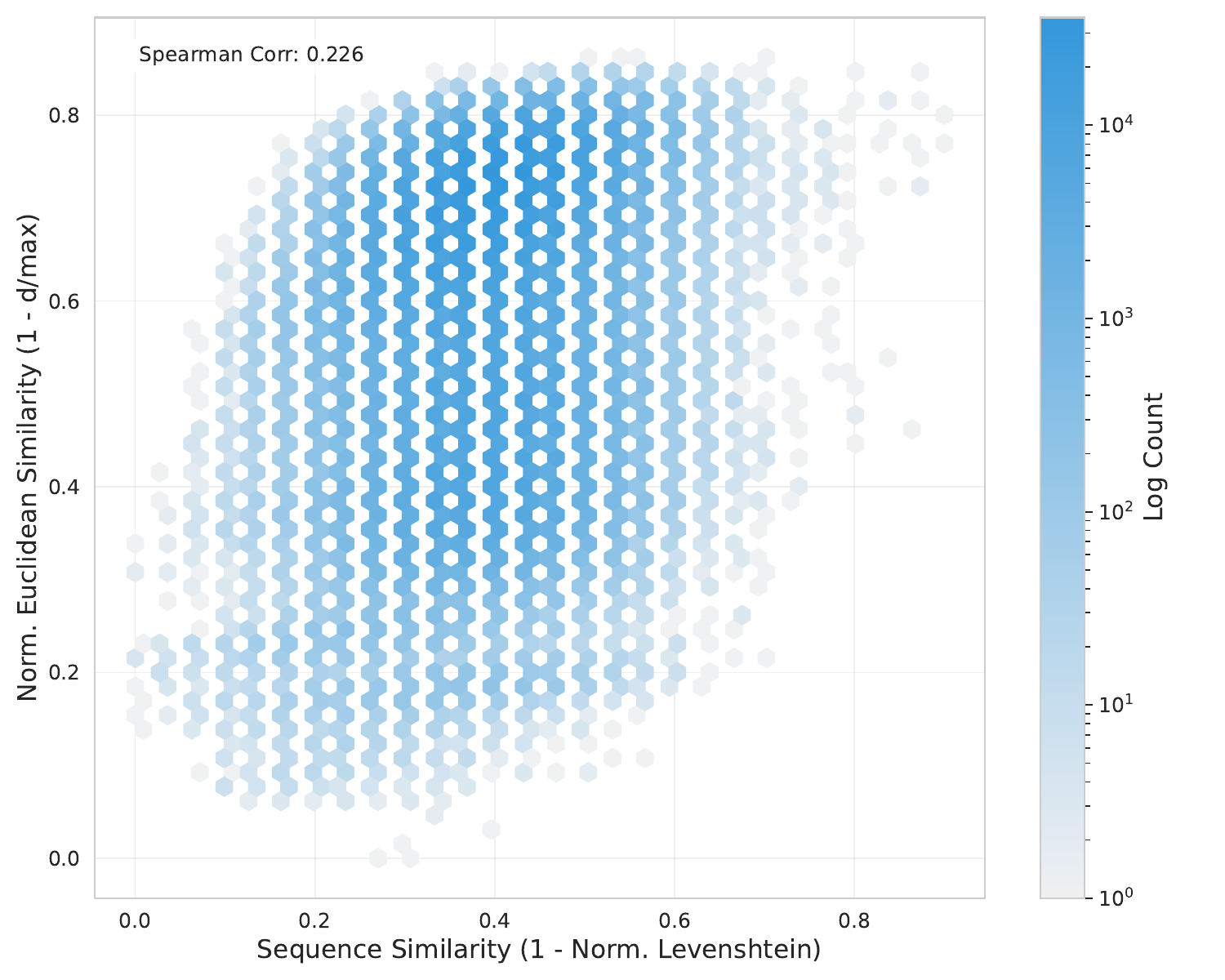}%
            \par\smallskip{\small $l = 30$}%
        \end{minipage}%
        \hspace{0.01\linewidth}%
        \begin{minipage}[t]{0.24\linewidth}%
            \centering%
            \includegraphics[width=\linewidth]{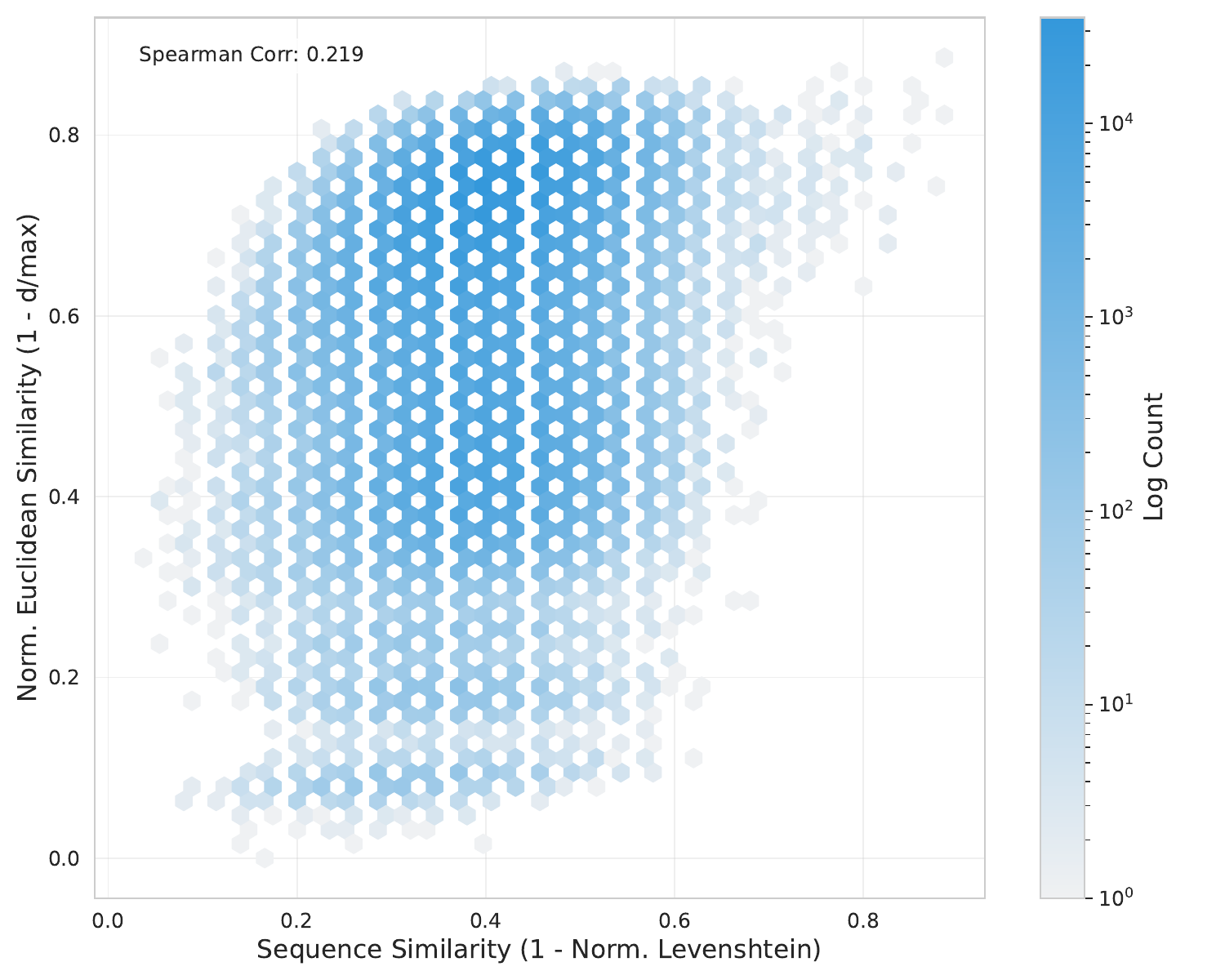}%
            \par\smallskip{\small $l = 35$}%
        \end{minipage}%
        \hspace{0.01\linewidth}%
        \begin{minipage}[t]{0.24\linewidth}%
            \centering%
            \includegraphics[width=\linewidth]{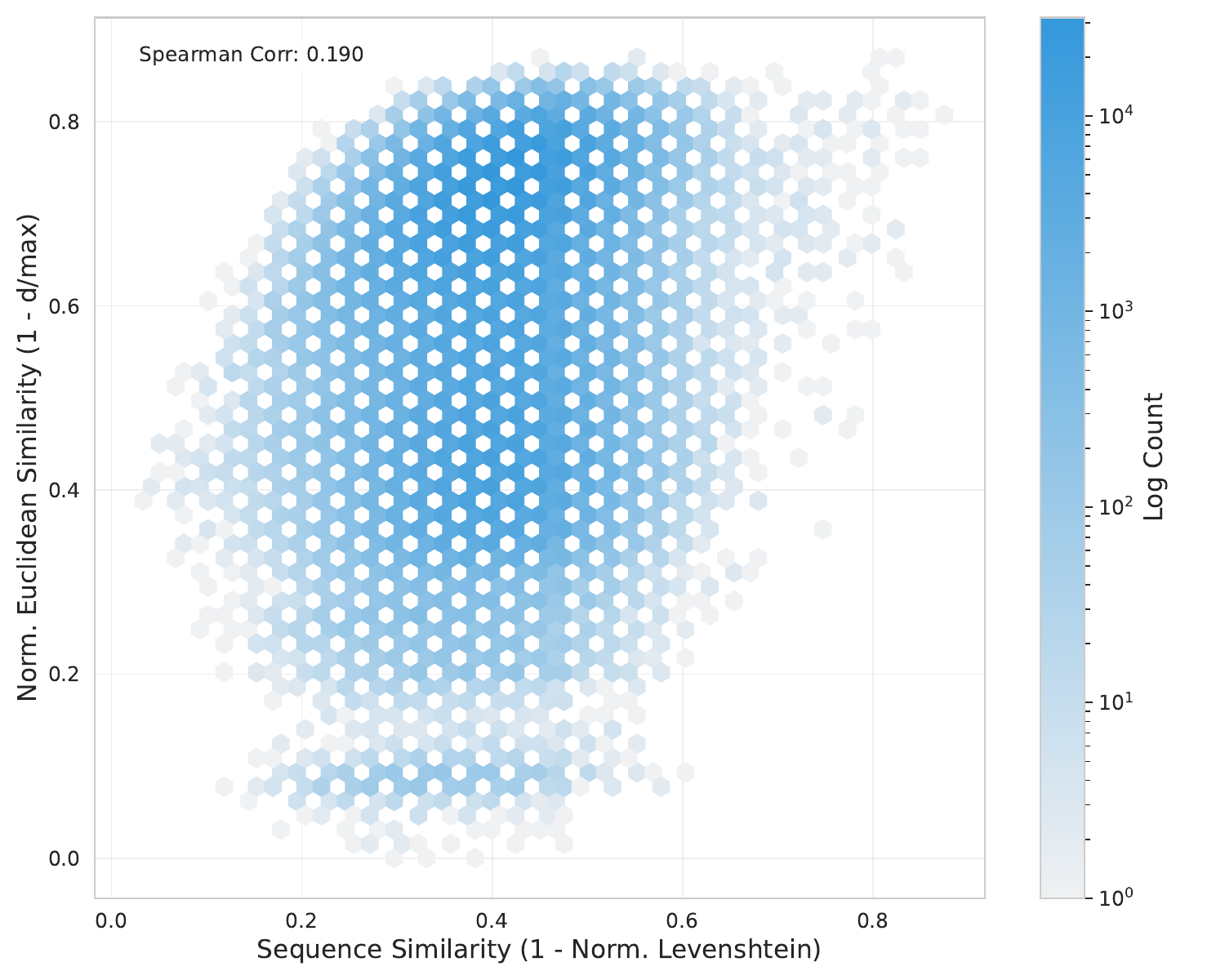}%
            \par\smallskip{\small $l = 40$}%
        \end{minipage}%
        \hspace{0.01\linewidth}%
        \begin{minipage}[t]{0.24\linewidth}%
            \centering%
            \includegraphics[width=\linewidth]{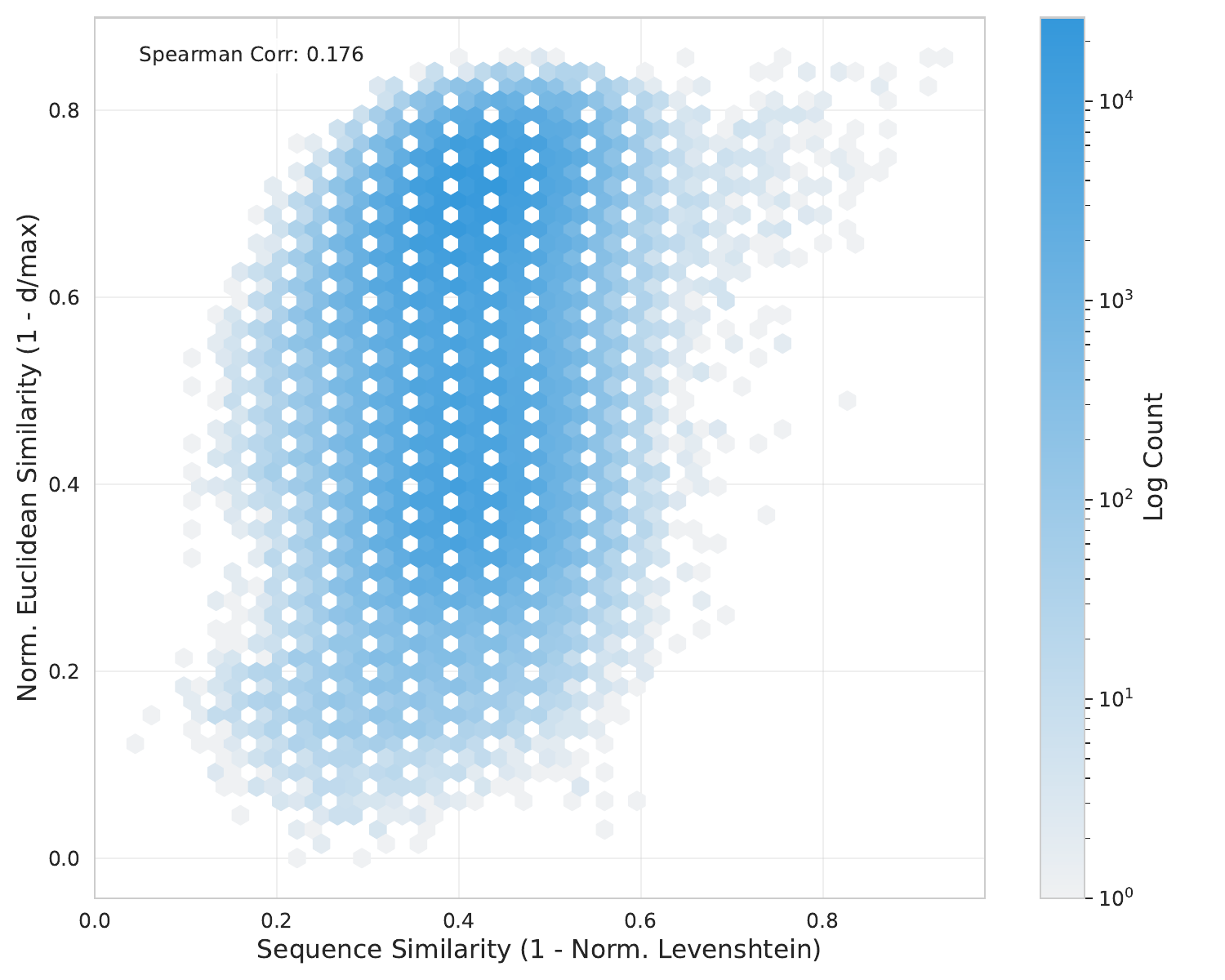}%
            \par\smallskip{\small $l = 45$}%
        \end{minipage}%
        \\[1em]%
        \begin{minipage}[t]{0.24\linewidth}%
            \centering%
            \includegraphics[width=\linewidth]{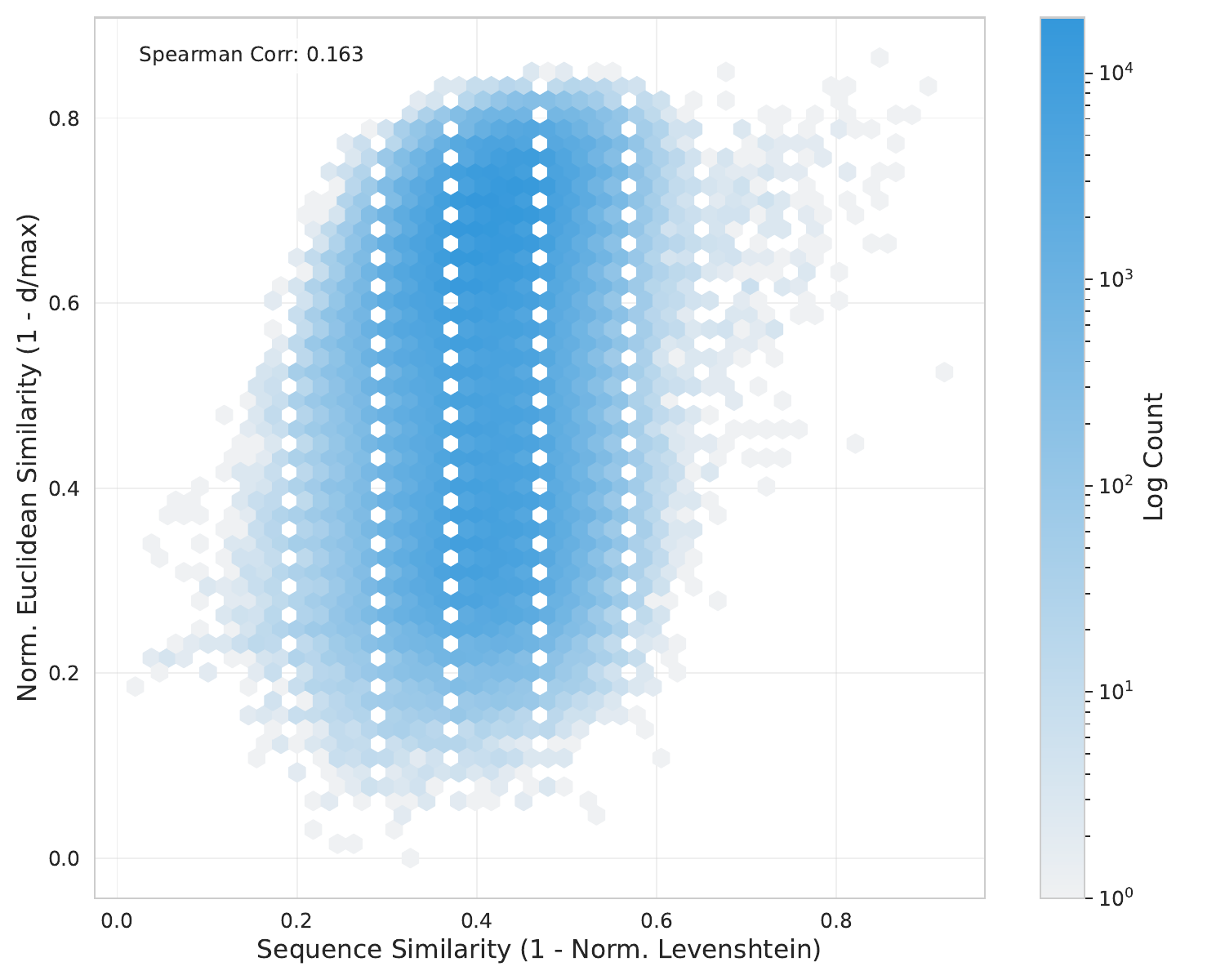}%
            \par\smallskip{\small $l = 50$}%
        \end{minipage}%
        \hspace{0.01\linewidth}%
        \begin{minipage}[t]{0.24\linewidth}%
            \centering%
            \includegraphics[width=\linewidth]{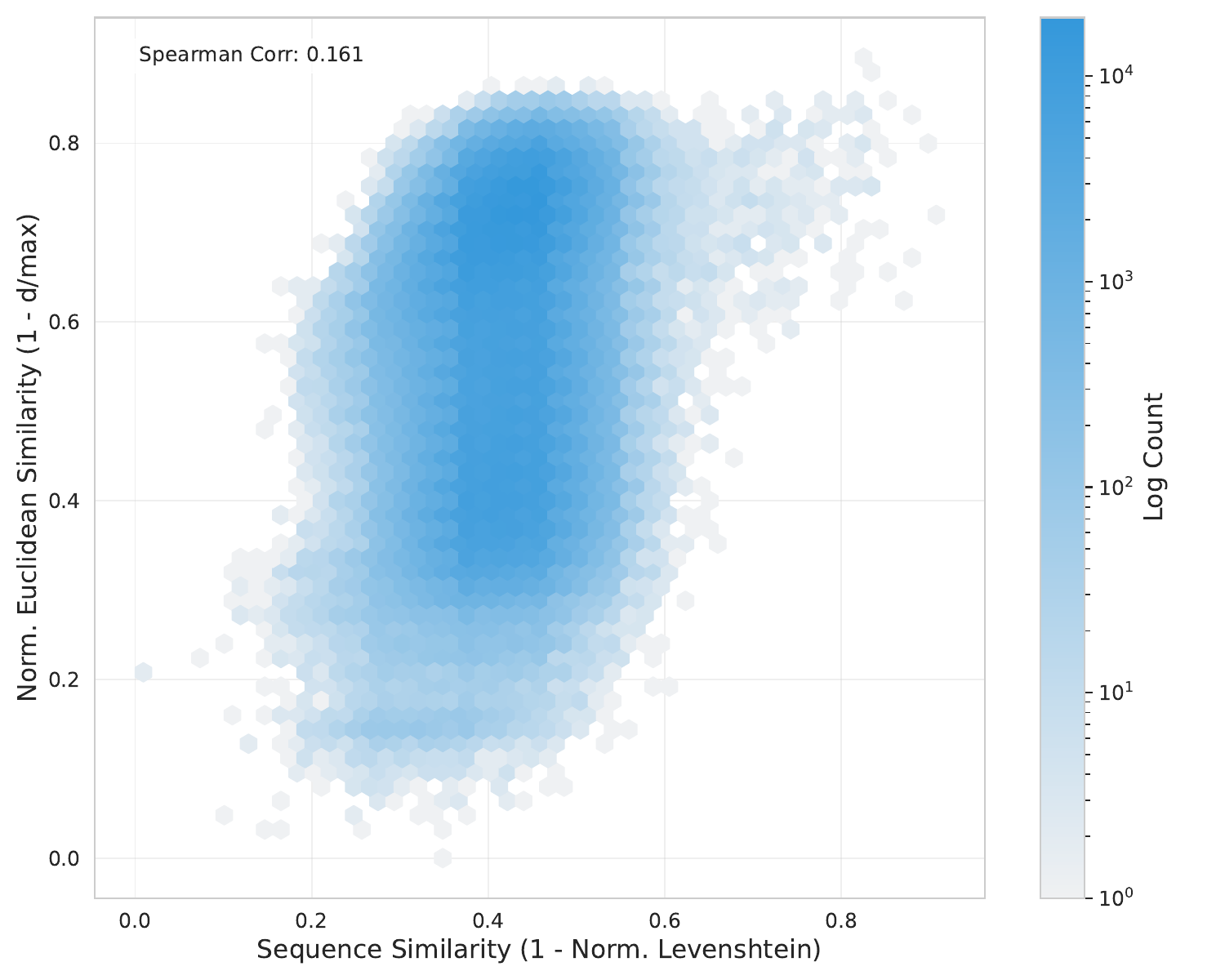}%
            \par\smallskip{\small $l = 60$}%
        \end{minipage}%
        \hspace{0.01\linewidth}%
        \begin{minipage}[t]{0.24\linewidth}%
            \centering%
            \includegraphics[width=\linewidth]{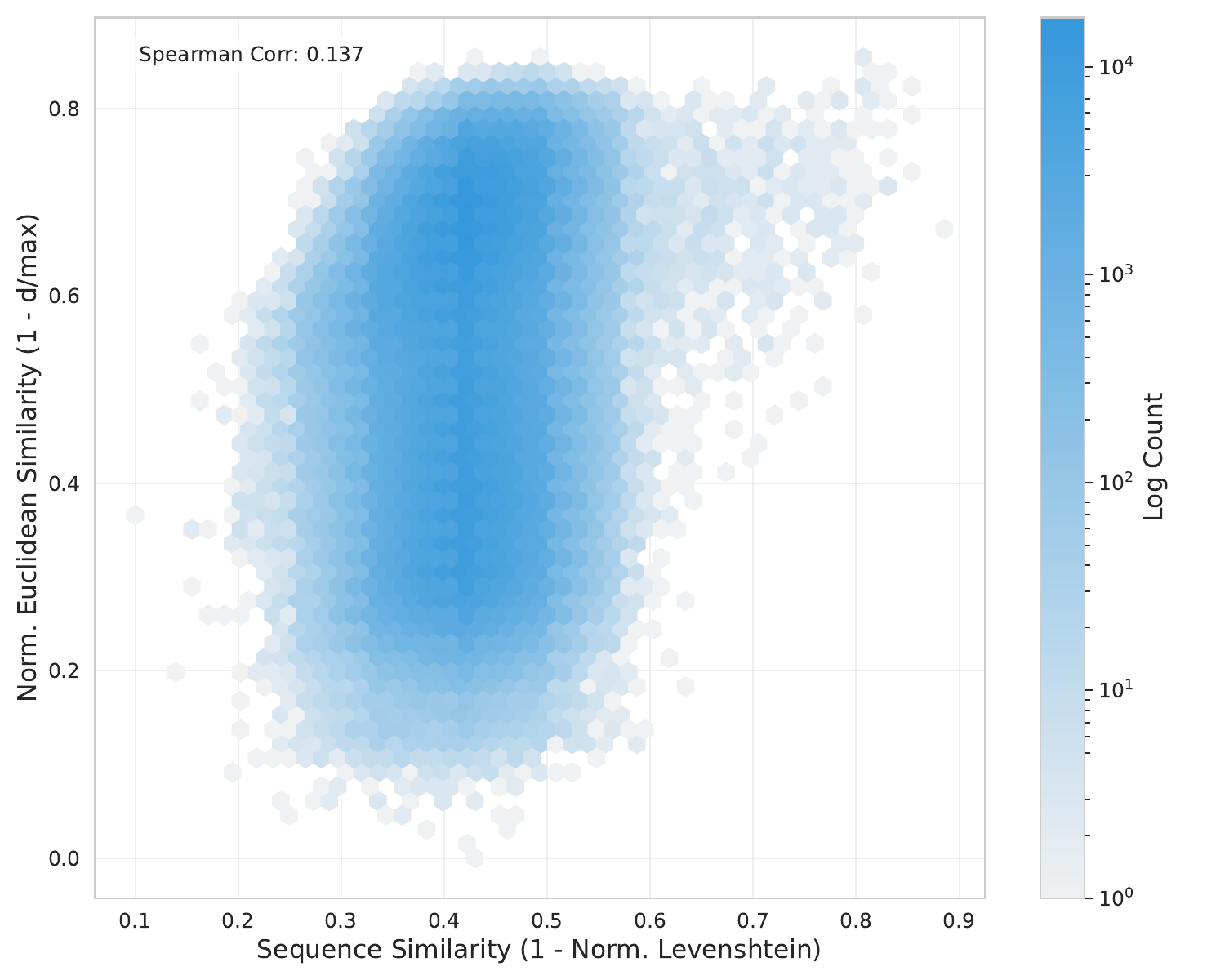}%
            \par\smallskip{\small $l = 70$}%
        \end{minipage}%
        \hspace{0.01\linewidth}%
        \begin{minipage}[t]{0.24\linewidth}%
            \centering%
            \includegraphics[width=\linewidth]{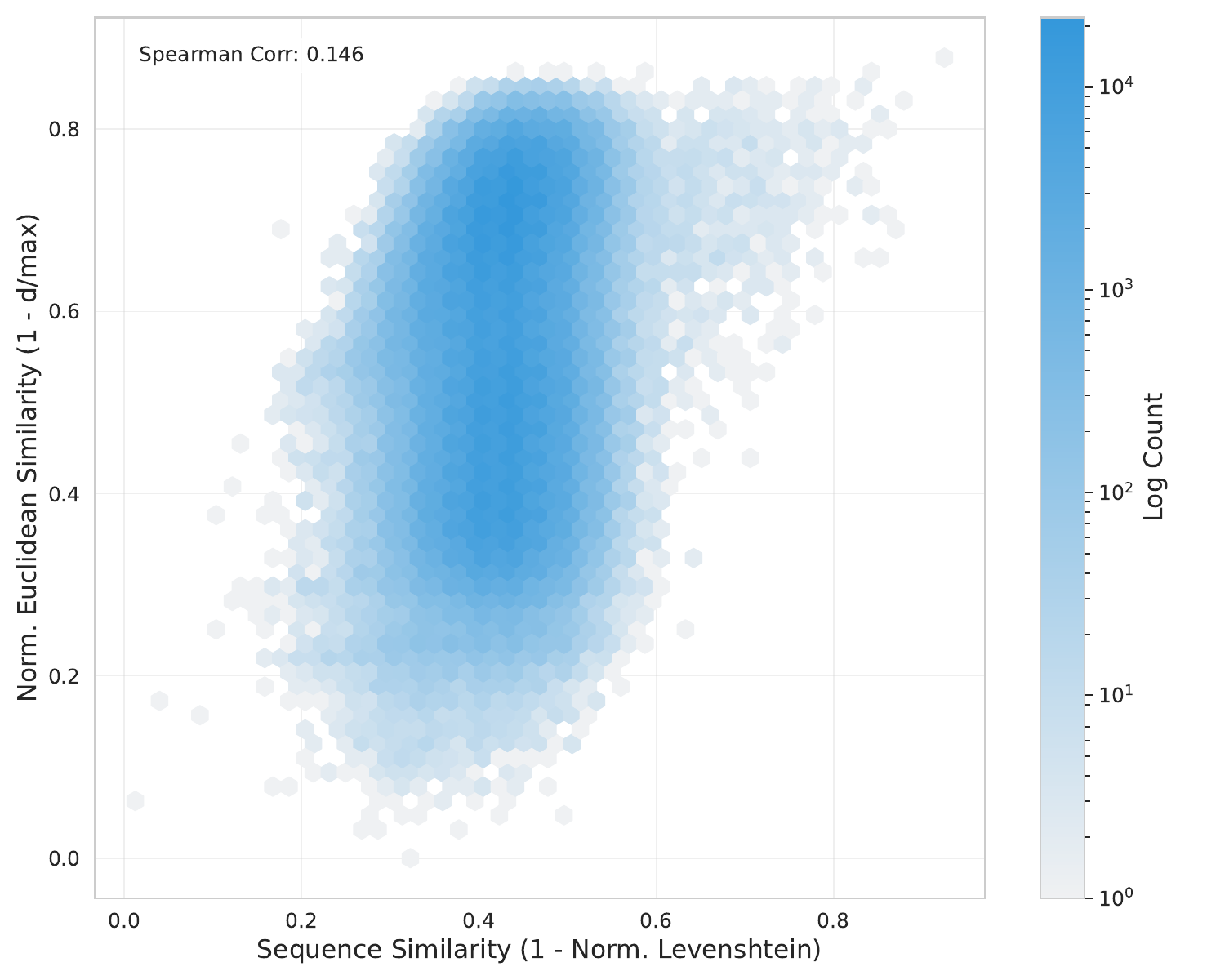}%
            \par\smallskip{\small $l = 80$}%
        \end{minipage}%
        \\[1em]%
        \begin{minipage}[t]{0.24\linewidth}%
            \centering%
            \includegraphics[width=\linewidth]{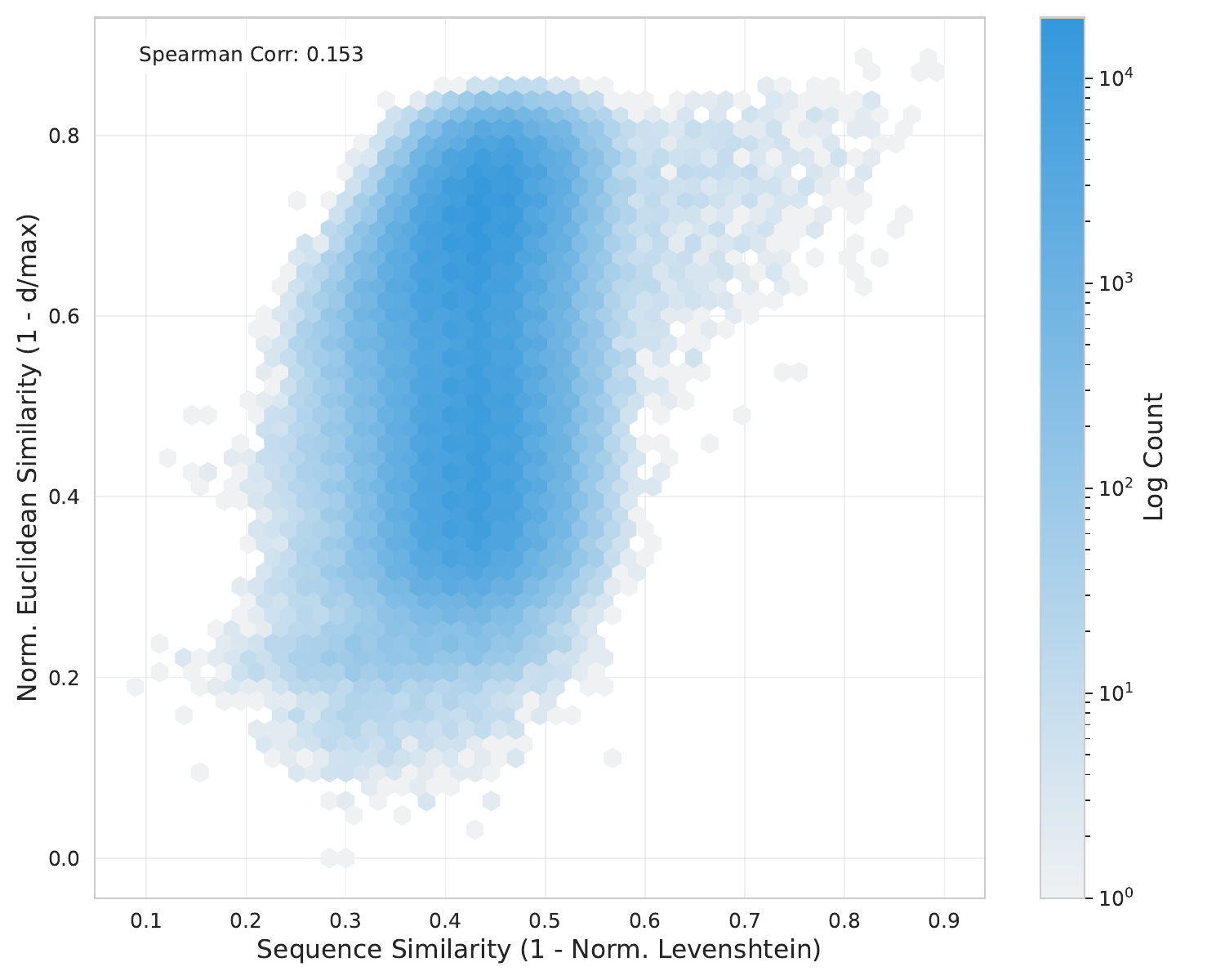}%
            \par\smallskip{\small $l = 90$}%
        \end{minipage}%
        \hspace{0.01\linewidth}%
        \begin{minipage}[t]{0.24\linewidth}%
            \centering%
            \includegraphics[width=\linewidth]{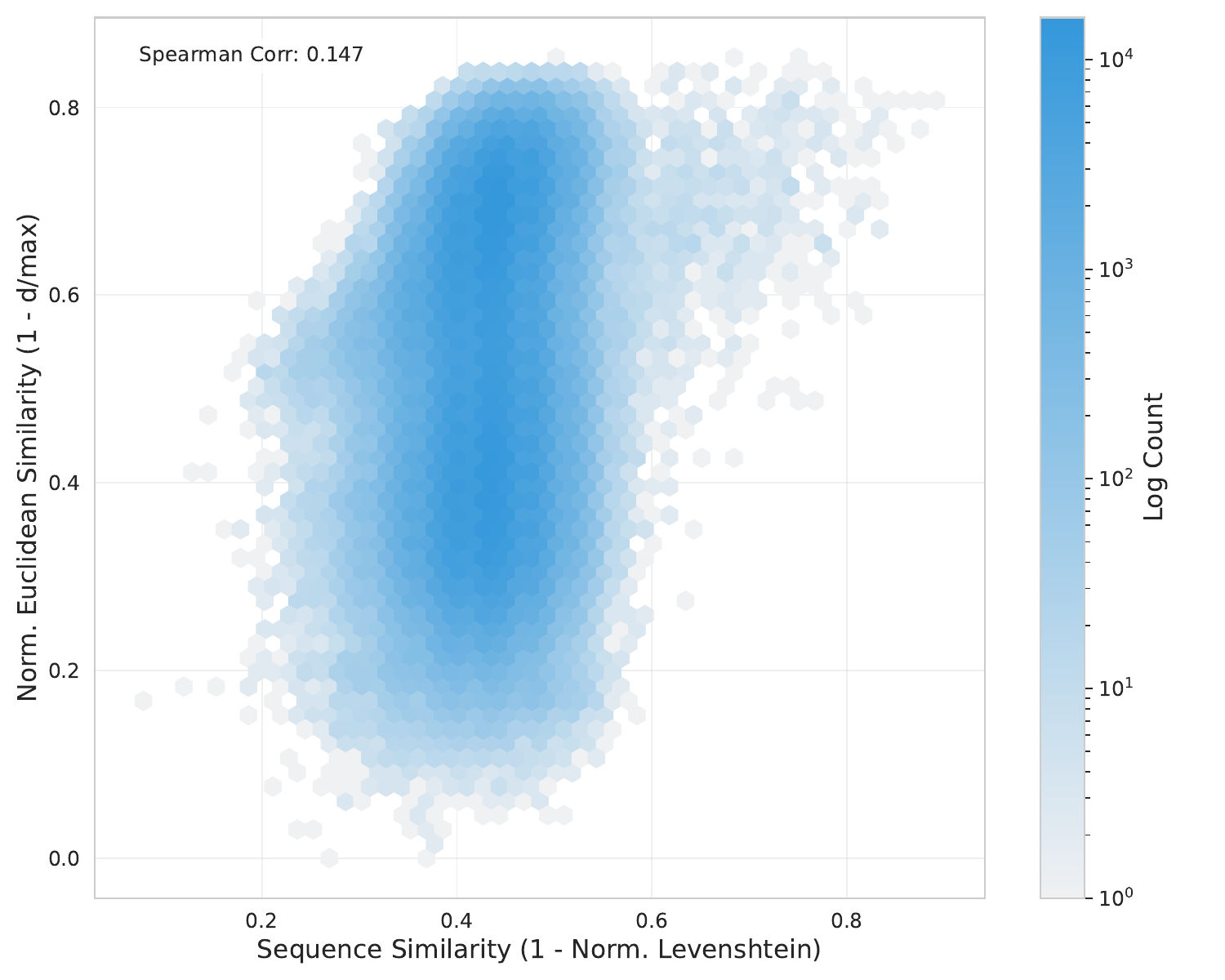}%
            \par\smallskip{\small $l = 100$}%
        \end{minipage}%
    \caption{Normalised Euclidean embedding similarity vs.\ sequence similarity for Evo~2 across all sequence lengths.}
    \label{fig:euclidean_evo2}
\end{figure}

Supplementary Figures~\ref{fig:euclidean_dnabert2}--\ref{fig:euclidean_evo2} compare the relationship between normalised Euclidean embedding similarity and sequence similarity across all models and sequence lengths.

\subsubsection*{Euclidean vs Sequence Similarity Correlation}
\label{subsec:euclidean_sim_corr}
We report the Spearman association between pairwise sequence similarity and embedding similarity, defined for the Euclidean analysis as one minus normalised distance. This diagnostic describes representation geometry but is not a validated predictor of supervised reconstruction. Inversion also depends on global identifiability, training data, decoder capacity, and optimisation. The observed associations do not rank attack performance consistently across models and lengths.

\clearpage
\clearpage

\begin{table}[H]
\centering
\resizebox{\linewidth}{!}{%
\begin{tabular}{llcccccccccccccc}
\toprule
\textbf{Model} & \textbf{Similarity} & \textbf{10} & \textbf{15} & \textbf{20} & \textbf{25} & \textbf{30} & \textbf{35} & \textbf{40} & \textbf{45} & \textbf{50} & \textbf{60} & \textbf{70} & \textbf{80} & \textbf{90} & \textbf{100} \\
\midrule
\multirow{2}{*}{DNABERT-2} & Cosine & $0.0167$ & $0.0284$ & $0.0282$ & $0.0355$ & $0.0416$ & $0.0513$ & $0.0637$ & $0.0535$ & $0.0714$ & $0.0753$ & $0.0795$ & $0.1013$ & $0.1057$ & $0.1216$ \\
 & Euclidean & $0.0290$ & $0.0360$ & $0.0351$ & $0.0377$ & $0.0336$ & $0.0470$ & $0.0557$ & $0.0539$ & $0.0678$ & $0.0640$ & $0.0662$ & $0.0950$ & $0.0949$ & $0.1133$ \\
\midrule
\multirow{2}{*}{Evo~2} & Cosine & $\mathbf{0.1495}$ & $\mathbf{0.4150}$ & $\mathbf{0.3747}$ & $\mathbf{0.3203}$ & $\mathbf{0.2384}$ & $\mathbf{0.2348}$ & $\mathbf{0.2103}$ & $\mathbf{0.1964}$ & $\mathbf{0.1883}$ & $\mathbf{0.1847}$ & $0.1648$ & $0.1749$ & $\mathbf{0.1843}$ & $0.1727$ \\
 & Euclidean & $0.0688$ & $\mathbf{0.3426}$ & $\mathbf{0.3720}$ & $\mathbf{0.3141}$ & $\mathbf{0.2261}$ & $\mathbf{0.2189}$ & $0.1898$ & $\mathbf{0.1759}$ & $0.1633$ & $0.1612$ & $0.1366$ & $0.1461$ & $0.1526$ & $0.1473$ \\
\midrule
\multirow{2}{*}{NTv2} & Cosine & $0.0912$ & $0.0940$ & $0.0900$ & $0.0849$ & $0.0836$ & $0.1558$ & $0.1757$ & $0.1548$ & $0.1619$ & $0.1500$ & $\mathbf{0.1941}$ & $\mathbf{0.1755}$ & $0.1608$ & $\mathbf{0.2331}$ \\
 & Euclidean & $\mathbf{0.0930}$ & $0.0971$ & $0.0964$ & $0.0946$ & $0.0905$ & $0.1747$ & $\mathbf{0.1945}$ & $0.1665$ & $\mathbf{0.1763}$ & $\mathbf{0.1676}$ & $\mathbf{0.2051}$ & $\mathbf{0.1908}$ & $\mathbf{0.1804}$ & $\mathbf{0.2395}$ \\
\bottomrule
\end{tabular}}
\caption{Spearman correlation between embedding similarity (cosine / Euclidean) and sequence similarity (Levenshtein) for different models and sequence lengths.}
\label{tab:spearman_correlation}
\end{table}

\clearpage
\section{Model Architecture and Size}
\label{sec:model_architecture}
Supplementary Table~\ref{tab:model_sizes} summarises the inversion model architectures and their parameter counts. The primary LCQD is trained once per foundation model across all sequence lengths, whereas the preliminary baselines are trained independently for each length. All parametric models are intentionally compact relative to the foundation models to test whether leakage can be exploited without an attacker-scale model.

Let $z\in\mathbb{R}^{d_{\mathrm{in}}}$ denote a mean-pooled embedding and let $T$ be the maximum number of output slots, including EOS. The LCQD projects $z$ into $n_{\mathrm{ctx}}$ context tokens,
\begin{equation}
C(z)=\operatorname{reshape}(W_c z+b_c)\in\mathbb{R}^{n_{\mathrm{ctx}}\times d}.
\end{equation}
A separate head produces a categorical output-content-token-count distribution $p_\ell(z)=\operatorname{softmax}(\operatorname{MLP}(z))_\ell$. Its expected length embedding conditions every learned position query,
\begin{equation}
\bar e_L(z)=\sum_{\ell=0}^{T-1}p_\ell(z)E^{\mathrm{len}}_\ell,
\qquad
q_i(z)=E^{\mathrm{pos}}_i+\bar e_L(z),
\end{equation}
and the transformer decoder computes $H=\operatorname{Dec}(q_{1:T}(z),C(z))$ followed by per-position vocabulary logits $W_oH+b_o$. Training minimises token cross-entropy plus $\lambda_{\mathrm{len}}$ times the length cross-entropy. The soft length embedding keeps the conditioning path differentiable, while inference terminates the decoded sequence at the first predicted EOS token.

\begin{table}[H]
\centering
\begin{tabular}{lll}
\toprule
\textbf{Model} & \textbf{Key Hyperparameters} & \textbf{Parameters} \\
\midrule
\multicolumn{3}{l}{\textit{Primary benchmark (multi-length)}} \\
LCQD (DNABERT-2) & input 768, 30 output slots, vocab. 4{,}096 & $\approx 21$M \\
LCQD (NTv2) & input 1{,}024, 21 output slots, vocab. 4{,}107 & $\approx 22$M \\
LCQD (Evo~2) & input 4{,}096, 101 output slots, vocab. 5 & $\approx 41$M \\
\midrule
\multicolumn{3}{l}{\textit{Baselines (per-length)}} \\
Encoder & $d=384$, $h=8$, $L=5$, $d_{ff}=1536$, drop$=0.1$ & $\approx 9$M \\
Decoder & $d=384$, $h=8$, $L=5$, $d_{ff}=1536$, drop$=0.1$ & $\approx 12$M \\
ResNet & $d=384$, 6 blocks, $k=5$ & $\approx 9$M \\
Nearest Neighbour &  & - \\
\bottomrule
\end{tabular}
\caption{Inversion model architectures and parameter counts. All LCQDs use $d=384$, $h=8$, six layers, $d_{ff}=1{,}536$, 16 context tokens, dropout 0.1, and an auxiliary length-prediction loss with weight 0.5. Their exact counts were read from the saved checkpoint state dictionaries and include the foundation-model-specific input projection, length components, and output head. Baseline counts remain approximate.}
\label{tab:model_sizes}
\end{table}

\subsection{Training Details}
All three LCQDs are trained for 20 epochs with a batch size of 32 and random seed 42. We use AdamW with weight decay $10^{-4}$ and $\epsilon=10^{-6}$, together with a OneCycle learning-rate schedule that peaks at $5\times10^{-4}$ after 30\% of the optimisation steps and then follows cosine annealing. The schedule uses a division factor of 25 and a final division factor of $10^{4}$. We clip the global gradient norm at 1.0 and disable early stopping so that every model completes the fixed learning-rate schedule. The objective is token-level cross-entropy, ignoring padded target positions, plus an auxiliary output-content-token-count cross-entropy term with weight 0.5. Model selection uses validation loss, and evaluation uses the best saved checkpoint.

Each inversion model uses the same tokeniser as the corresponding foundation model for decoding predictions back to nucleotide sequences. That is, for \textit{DNABERT-2} the inversion model predicts over the BPE vocabulary, for \textit{NTv2} over the single-nucleotide and 6-mer vocabulary, and for \textit{Evo~2} over the single-nucleotide alphabet.

\clearpage
\subsection{Baseline Models}
In addition to the primary multi-length models, we evaluate four baseline single-length architectures trained independently for each sequence length:
\begin{itemize}
    \item \textbf{Encoder-only Transformer}: The encoder projects the input embeddings into a $d_{\text{model}}$-dimensional space, applies sinusoidal positional encoding, and passes the sequence through stacked encoder layers. Mean-pooled embeddings are first projected and reshaped into a sequence of length $l$ before positional encoding. An output layer maps each position to a distribution over the tokens.
    \item \textbf{Decoder-only Transformer}: A transformer decoder with causal masking. The architecture mirrors the encoder but employs self-attention with a causal mask, preventing each position from attending to subsequent positions during reconstruction.
    \item \textbf{ResNet}: A 1D convolutional residual network consisting of stacked residual blocks, each containing two convolutional layers with batch normalisation, ReLU activation, and dropout, connected via a skip connection.
    \item \textbf{Nearest Neighbour Lookup}: A non-parametric baseline that stores all training embeddings and their corresponding sequences. At inference time, the training sequence whose embedding has the smallest Euclidean distance to the query embedding is returned as the reconstruction. We evaluated both Euclidean and cosine distances and used Euclidean distance because their correlations with sequence similarity were similar.
\end{itemize}
\newpage
\subsubsection{Tokeniser Ablation}
\label{subsec:tokeniser_ablation}
\begin{figure}[H]
    \centering
    \includegraphics[width=0.48\linewidth]{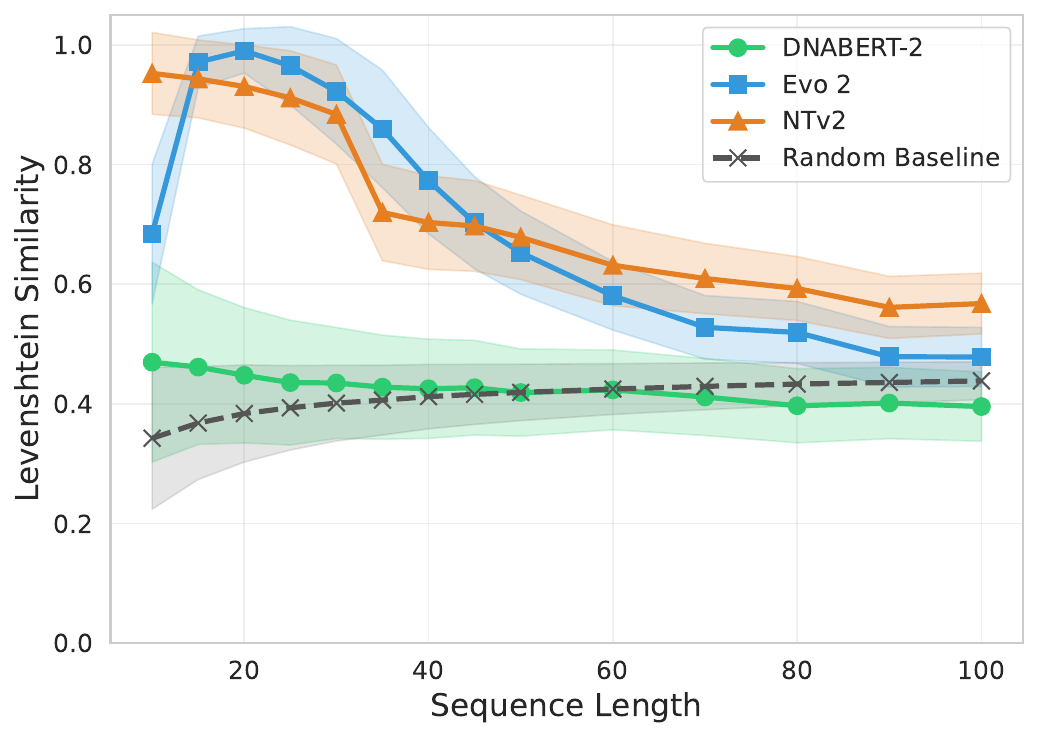}
    \caption{Levenshtein similarity for the encoder-only architecture with a fixed single-nucleotide decoder target vocabulary.}
    \label{fig:4token_levenshtein}
\end{figure}

We additionally trained inversion baselines with a fixed single-nucleotide target vocabulary, regardless of the foundation model's native tokeniser. Supplementary Figure~\ref{fig:4token_levenshtein} shows slightly lower reconstruction for DNABERT-2 and NTv2 and unchanged performance for Evo~2. This ablation changes the attack decoder's output representation; it does not manipulate the foundation model tokeniser and therefore cannot establish that foundation-model tokenisation causes the between-model privacy differences.

\vfill
\clearpage

\section{Per-Length Architecture Sweep}
\label{sec:per_length_sweep}

The main benchmark uses a single multi-length LCQD per foundation model. That choice was informed by preliminary internal tests and the sweep reported here, in which four architectures were trained \emph{independently at each sequence length}. These per-length runs use smaller training sets than the final models and are included to provide context for the architecture selection; they are not comparable to, and should not be read as, the headline values in Supplementary Table~\ref{tab:query_decoder_results}. These preliminary tests motivated the LCQD choice but do not establish superiority under matched data and compute. The LCQD additionally removes the need to train one model per release length.

\begin{figure}[H]
    \centering
    \begin{minipage}{0.47\linewidth}
        \includegraphics[width=\linewidth]{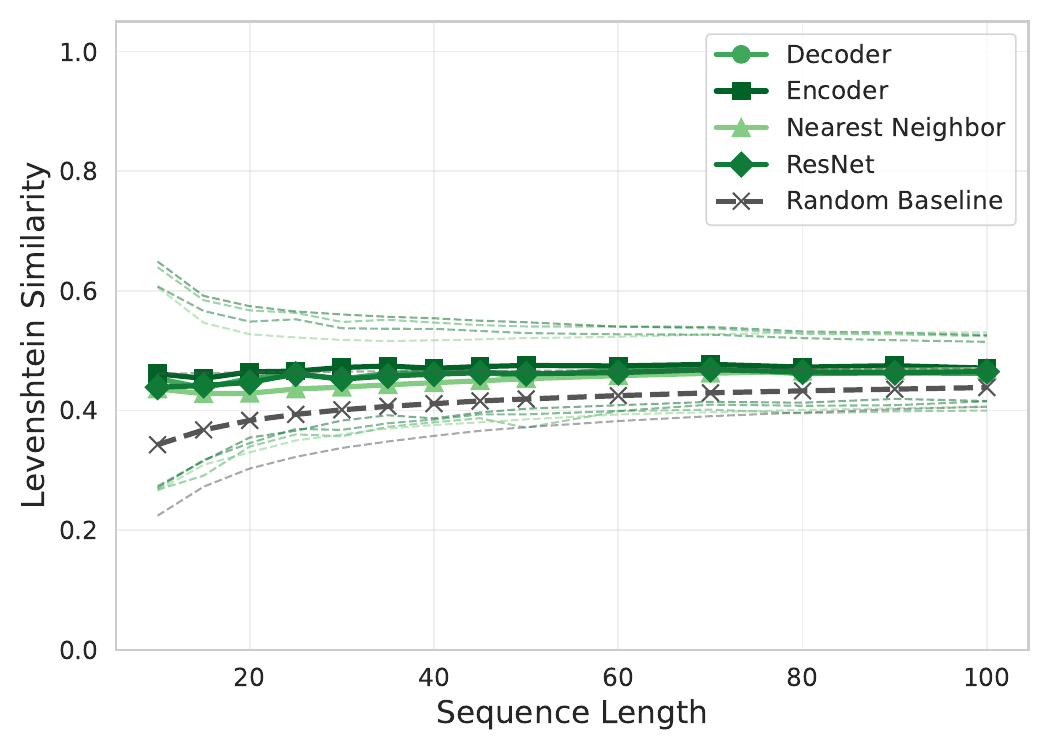}
    \end{minipage}
    \hfill
    \begin{minipage}{0.47\linewidth}
        \includegraphics[width=\linewidth]{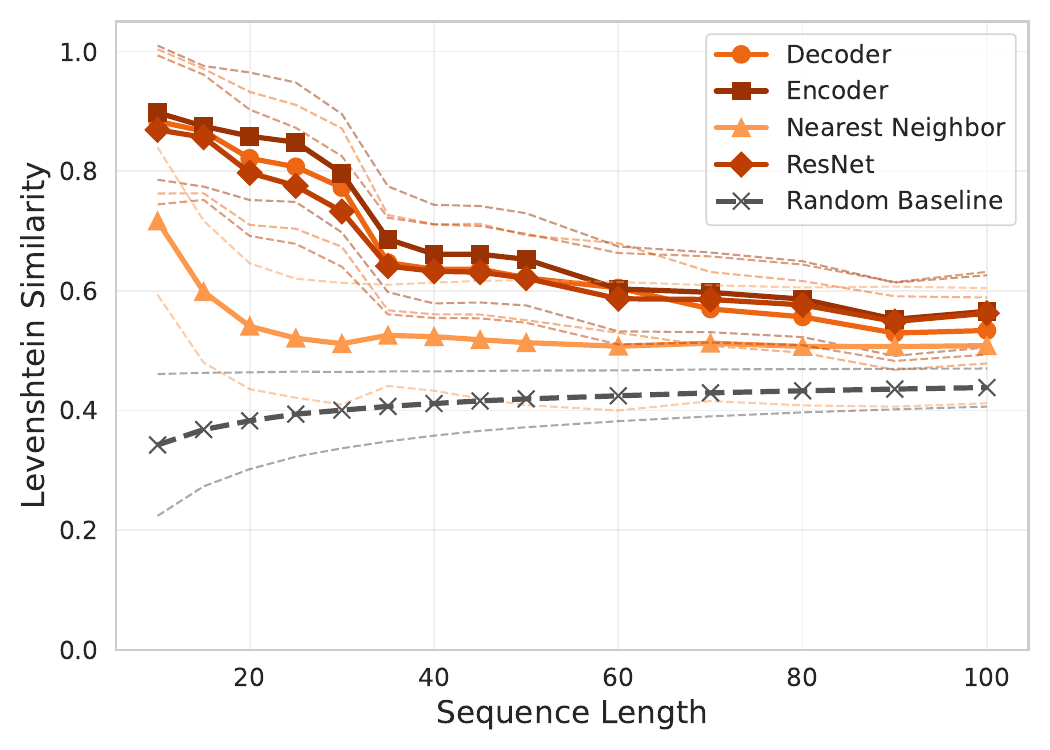}
    \end{minipage}
    \\[0.4em]
    \begin{minipage}{0.47\linewidth}
        \includegraphics[width=\linewidth]{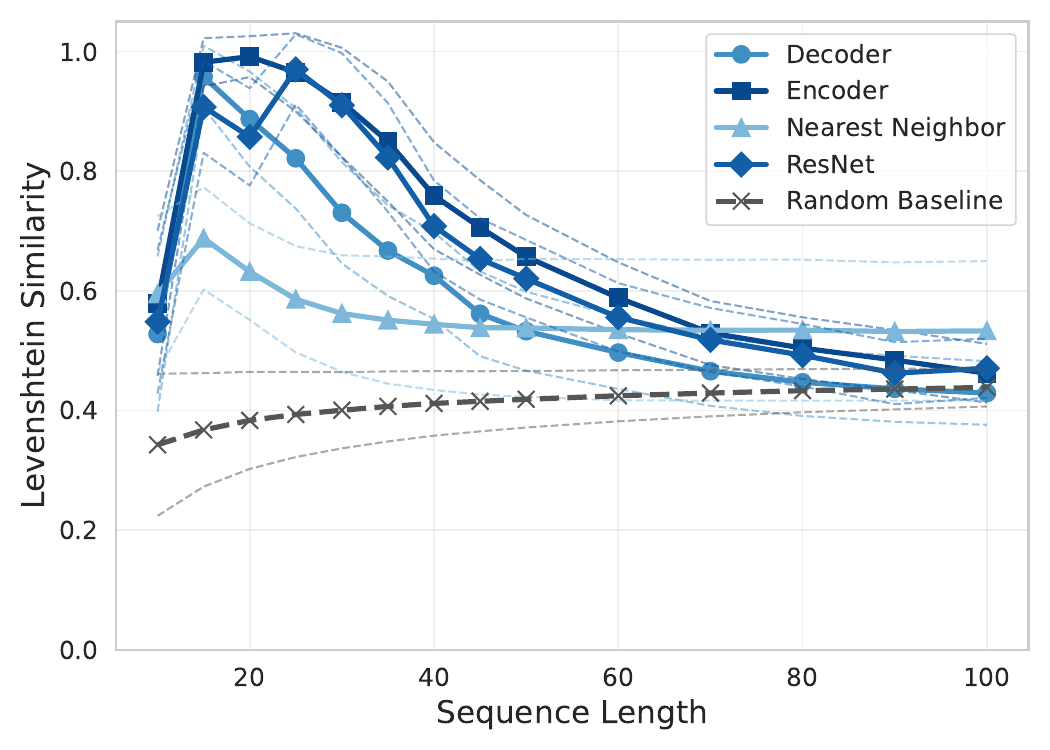}
    \end{minipage}
    \caption{Preliminary per-length architecture sweep for mean-pooled embeddings, measured by Levenshtein similarity. Top left: DNABERT-2; top right: NTv2; bottom: Evo~2. Each architecture was trained separately at every sequence length.}
    \label{fig:reconstruction_eval}
\end{figure}
\vfill
\clearpage

\section{Cross-Dataset Evaluation: 1000 Genomes Project}
\label{sec:1000g_cross_dataset}

\begin{figure}[H]
    \centering
    \includegraphics[width=0.48\linewidth]{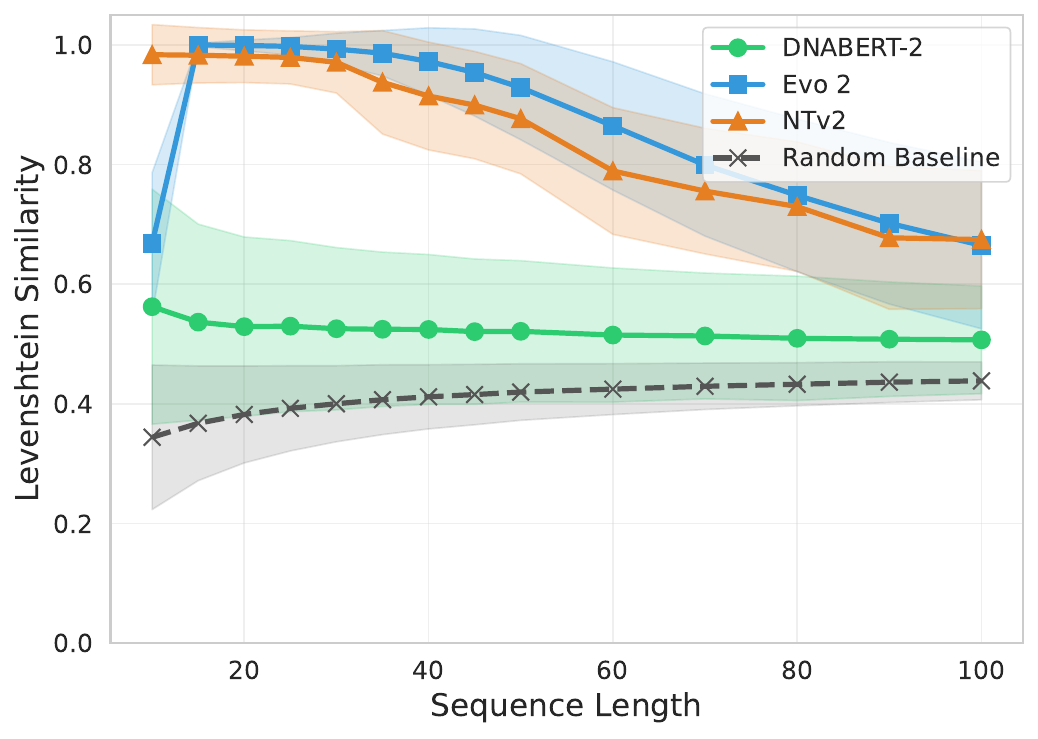}
    \caption{Levenshtein similarity across sequence lengths for the multi-length LCQD on individual-specific 1000 Genomes Project sequences.}
    \label{fig:1000g_levenshtein}
\end{figure}

To assess how well the inversion attack generalises beyond the \texttt{hg38} reference genome, we evaluate on individual-specific sequences derived from the 1000 Genomes Project. For each of the 14 benchmark lengths, we take autosomal loci from the coordinate-disjoint \texttt{hg38} test split, randomly assign each locus to a phased 1000 Genomes individual and haplotype, and apply that haplotype's biallelic SNPs to the reference window. The multi-length LCQDs trained on \texttt{hg38} are applied without retraining or fine-tuning. Supplementary Figure~\ref{fig:1000g_levenshtein} reports aggregate Levenshtein similarity.

At length 25, \num{cross.25.reference_n} of \num{cross.25.total_n} windows are reference-identical; \num{cross.25.variant_unseen_n} variant-bearing windows are absent as exact strings from decoder training. In that latter stratum, mean Levenshtein similarity is \num{cross.25.dnabert2.variant_unseen_lev} for DNABERT-2, \num{cross.25.evo2.variant_unseen_lev} for Evo~2, and \num{cross.25.ntv2.variant_unseen_lev} for NTv2.

We also count exact whole-window reconstructions in those \num{cross.25.variant_unseen_n} windows: 0 for DNABERT-2, 326 (96.4\%) for Evo~2, and 248 (73.4\%) for NTv2. An exact match necessarily recovers every alternate allele in its window, so these fractions are lower bounds on the fraction of variant-bearing windows whose alternate alleles are all recovered. An inexact window may still have its alternate alleles correct. Neither whole-window measure estimates per-site alternate-allele recall or precision, and the test covers phased biallelic SNP substitutions rather than indels or structural variants.

\vfill
\clearpage
\section{Detailed Metrics}
\label{sec:detailed_metrics}

The tables below and the measured values in the main text are generated from the evaluation outputs. Supplementary Table~\ref{tab:query_decoder_results} gives the complete multi-length LCQD reconstruction results; Supplementary Table~\ref{tab:identification_results} gives the cohort-identification thresholds for both locus-sampling modes and both threat models.

For completeness, Supplementary Table~\ref{tab:query_decoder_results} also reports the secondary positionwise match rate used by the evaluation pipeline:
\[
\operatorname{acc}_{\mathrm{pos}}(x_1,x_2)
=
\frac{\sum_{j=1}^{\min(|x_1|,|x_2|)}
\mathbbm{1}[x_1^{(j)}=x_2^{(j)}]}
{\max(|x_1|,|x_2|)}.
\]
Unmatched suffix positions therefore count as errors. Because an insertion or deletion can shift all subsequent indices, this diagnostic is retained in Supplementary Table~\ref{tab:query_decoder_results} but is not used as a headline reconstruction metric.

\begin{table}[H]
\centering
\small
\renewcommand{\arraystretch}{1.12}
\caption{Mean-pooled reconstruction by the multi-length LCQD on 50{,}000 held-out \texttt{hg38} sequences per length. Values are mean $\pm$ standard deviation over sequences; exact-match rate is the fraction reconstructed without any edit. DNABERT-2.}
\label{tab:query_decoder_results}
\begin{tabular}{rccc}
\toprule
Length (nt) & Levenshtein similarity & Positionwise match rate & Exact-match rate \\
\midrule
10 & 0.558 $\pm$ 0.198 & 0.386 $\pm$ 0.266 & 0.055 \\
15 & 0.536 $\pm$ 0.164 & 0.332 $\pm$ 0.209 & 0.015 \\
20 & 0.529 $\pm$ 0.149 & 0.313 $\pm$ 0.177 & 0.004 \\
25 & 0.528 $\pm$ 0.142 & 0.299 $\pm$ 0.158 & 0.002 \\
30 & 0.527 $\pm$ 0.134 & 0.292 $\pm$ 0.145 & 0.002 \\
35 & 0.525 $\pm$ 0.128 & 0.284 $\pm$ 0.136 & 0.002 \\
40 & 0.523 $\pm$ 0.124 & 0.278 $\pm$ 0.129 & 0.002 \\
45 & 0.521 $\pm$ 0.120 & 0.273 $\pm$ 0.121 & 0.001 \\
50 & 0.520 $\pm$ 0.118 & 0.269 $\pm$ 0.116 & 0.001 \\
60 & 0.514 $\pm$ 0.111 & 0.261 $\pm$ 0.104 & 0.001 \\
70 & 0.512 $\pm$ 0.105 & 0.256 $\pm$ 0.095 & 0.000 \\
80 & 0.510 $\pm$ 0.103 & 0.254 $\pm$ 0.089 & 0.000 \\
90 & 0.509 $\pm$ 0.095 & 0.253 $\pm$ 0.082 & 0.000 \\
100 & 0.508 $\pm$ 0.089 & 0.252 $\pm$ 0.076 & 0.000 \\
\bottomrule
\end{tabular}
\end{table}
\begin{table}[H]
\centering
\small
\renewcommand{\arraystretch}{1.12}
\ContinuedFloat
\caption{Mean-pooled LCQD reconstruction results (continued): Evo~2.}
\begin{tabular}{rccc}
\toprule
Length (nt) & Levenshtein similarity & Positionwise match rate & Exact-match rate \\
\midrule
10 & 0.669 $\pm$ 0.118 & 0.655 $\pm$ 0.128 & 0.006 \\
15 & 1.000 $\pm$ 0.003 & 1.000 $\pm$ 0.003 & 0.999 \\
20 & 0.999 $\pm$ 0.010 & 0.999 $\pm$ 0.012 & 0.987 \\
25 & 0.997 $\pm$ 0.016 & 0.997 $\pm$ 0.019 & 0.962 \\
30 & 0.993 $\pm$ 0.026 & 0.992 $\pm$ 0.033 & 0.899 \\
35 & 0.986 $\pm$ 0.039 & 0.983 $\pm$ 0.050 & 0.805 \\
40 & 0.973 $\pm$ 0.056 & 0.966 $\pm$ 0.075 & 0.667 \\
45 & 0.954 $\pm$ 0.072 & 0.942 $\pm$ 0.099 & 0.498 \\
50 & 0.929 $\pm$ 0.087 & 0.908 $\pm$ 0.119 & 0.318 \\
60 & 0.864 $\pm$ 0.107 & 0.822 $\pm$ 0.143 & 0.085 \\
70 & 0.799 $\pm$ 0.119 & 0.741 $\pm$ 0.154 & 0.046 \\
80 & 0.748 $\pm$ 0.128 & 0.682 $\pm$ 0.161 & 0.036 \\
90 & 0.701 $\pm$ 0.135 & 0.633 $\pm$ 0.164 & 0.024 \\
100 & 0.665 $\pm$ 0.140 & 0.597 $\pm$ 0.166 & 0.016 \\
\bottomrule
\end{tabular}
\end{table}
\begin{table}[H]
\centering
\small
\renewcommand{\arraystretch}{1.12}
\ContinuedFloat
\caption{Mean-pooled LCQD reconstruction results (continued): NTv2.}
\begin{tabular}{rccc}
\toprule
Length (nt) & Levenshtein similarity & Positionwise match rate & Exact-match rate \\
\midrule
10 & 0.983 $\pm$ 0.051 & 0.983 $\pm$ 0.055 & 0.880 \\
15 & 0.982 $\pm$ 0.048 & 0.981 $\pm$ 0.054 & 0.826 \\
20 & 0.981 $\pm$ 0.045 & 0.980 $\pm$ 0.049 & 0.768 \\
25 & 0.979 $\pm$ 0.043 & 0.979 $\pm$ 0.046 & 0.711 \\
30 & 0.971 $\pm$ 0.051 & 0.970 $\pm$ 0.055 & 0.580 \\
35 & 0.936 $\pm$ 0.086 & 0.929 $\pm$ 0.101 & 0.477 \\
40 & 0.914 $\pm$ 0.090 & 0.903 $\pm$ 0.107 & 0.347 \\
45 & 0.900 $\pm$ 0.089 & 0.886 $\pm$ 0.108 & 0.269 \\
50 & 0.875 $\pm$ 0.092 & 0.856 $\pm$ 0.114 & 0.175 \\
60 & 0.789 $\pm$ 0.105 & 0.743 $\pm$ 0.145 & 0.081 \\
70 & 0.755 $\pm$ 0.104 & 0.697 $\pm$ 0.140 & 0.070 \\
80 & 0.729 $\pm$ 0.109 & 0.663 $\pm$ 0.142 & 0.071 \\
90 & 0.677 $\pm$ 0.119 & 0.590 $\pm$ 0.155 & 0.050 \\
100 & 0.675 $\pm$ 0.116 & 0.585 $\pm$ 0.152 & 0.042 \\
\bottomrule
\end{tabular}
\end{table}

\begin{table}[H]
\centering
\small
\setlength{\tabcolsep}{4pt}
\renewcommand{\arraystretch}{1.15}
\begin{tabular}{lllrrrrr}
\toprule
\textbf{Model} & \textbf{Locus sampling} & \textbf{Coordinate access} & $k_{0.5}$ & $k_{0.9}$ & $k_{0.99}$ & $k_{\max}$ & \textbf{Top-1 at $k_{\max}$} \\
\midrule
DNABERT-2 & random & known (Oracle) & 1{,}024 & 4{,}096 & 4{,}096 & 4,096 & 0.997 \\
DNABERT-2 & random & unknown (alignment) & -- & -- & -- & 4,096 & 0.001 \\
DNABERT-2 & snp-enriched & known (Oracle) & 64 & 256 & 256 & 2,048 & 1.000 \\
DNABERT-2 & snp-enriched & unknown (alignment) & -- & -- & -- & 2,048 & 0.002 \\
Evo~2 & random & known (Oracle) & 128 & 256 & 512 & 4,096 & 1.000 \\
Evo~2 & random & unknown (alignment) & 512 & 1{,}024 & 2{,}048 & 4,096 & 1.000 \\
Evo~2 & snp-enriched & known (Oracle) & 16 & 16 & 32 & 2,048 & 1.000 \\
Evo~2 & snp-enriched & unknown (alignment) & 32 & 64 & 64 & 2,048 & 1.000 \\
NTv2 & random & known (Oracle) & 128 & 256 & 512 & 4,096 & 1.000 \\
NTv2 & random & unknown (alignment) & 512 & 1{,}024 & 2{,}048 & 4,096 & 1.000 \\
NTv2 & snp-enriched & known (Oracle) & 16 & 16 & 32 & 2,048 & 1.000 \\
NTv2 & snp-enriched & unknown (alignment) & 32 & 64 & 128 & 2,048 & 1.000 \\
\bottomrule
\end{tabular}
\caption{Cohort identification for 1{,}000 individuals using length-25 reconstructions. Each point averages top-1 identification accuracy over 50 seeded random locus permutations. $k_p$ is the smallest evaluated locus budget reaching top-1 accuracy $p$; ``--'' indicates the threshold was not reached within the evaluated grid. $k_{\max}$ is the largest budget evaluated for that sampling mode.}
\label{tab:identification_results}
\end{table}

\vfill
\clearpage

\begin{figure}[H]
    \centering
    \begin{minipage}{0.32\linewidth}
        \centering
        \includegraphics[width=\linewidth,trim={0 0 3.6cm 0},clip]{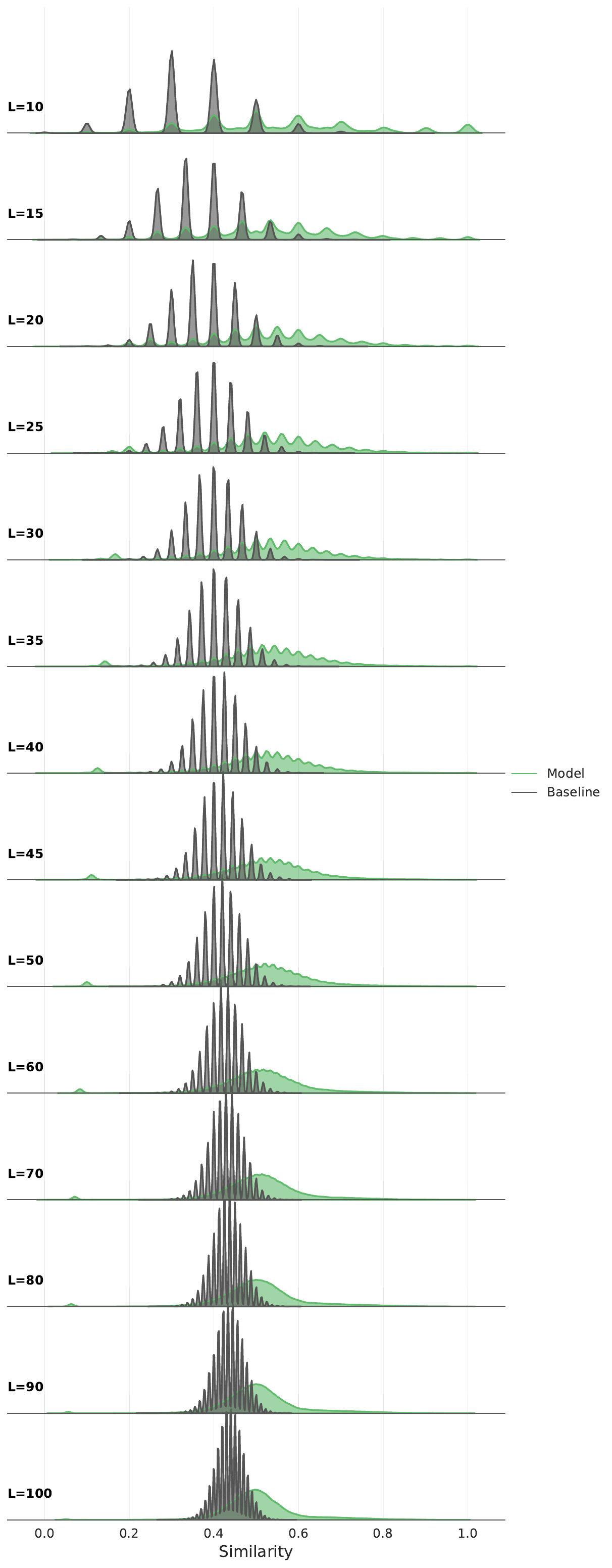}
        \\ \small (a) DNABERT-2
    \end{minipage}
    \hfill
    \begin{minipage}{0.32\linewidth}
        \centering
        \includegraphics[width=\linewidth,trim={0 0 3.6cm 0},clip]{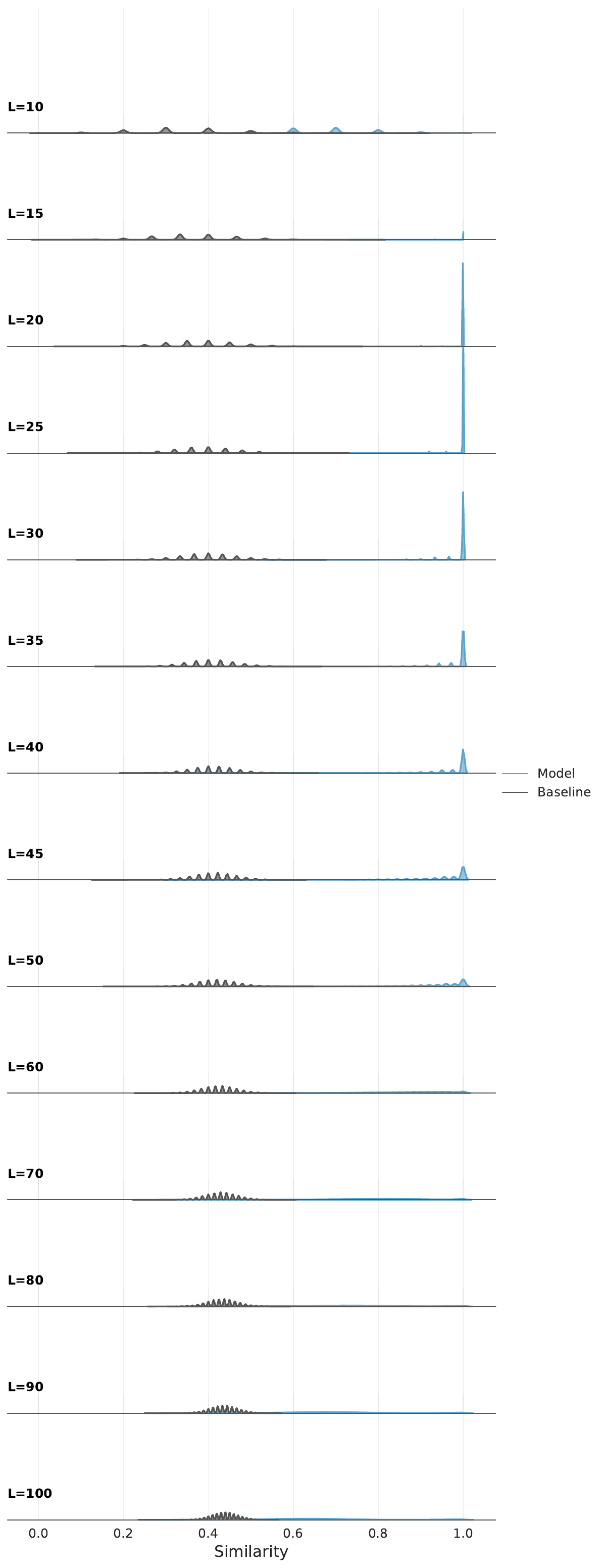}
        \\ \small (b) Evo~2
    \end{minipage}
    \hfill
    \begin{minipage}{0.32\linewidth}
        \centering
        \includegraphics[width=\linewidth,trim={0 0 3.6cm 0},clip]{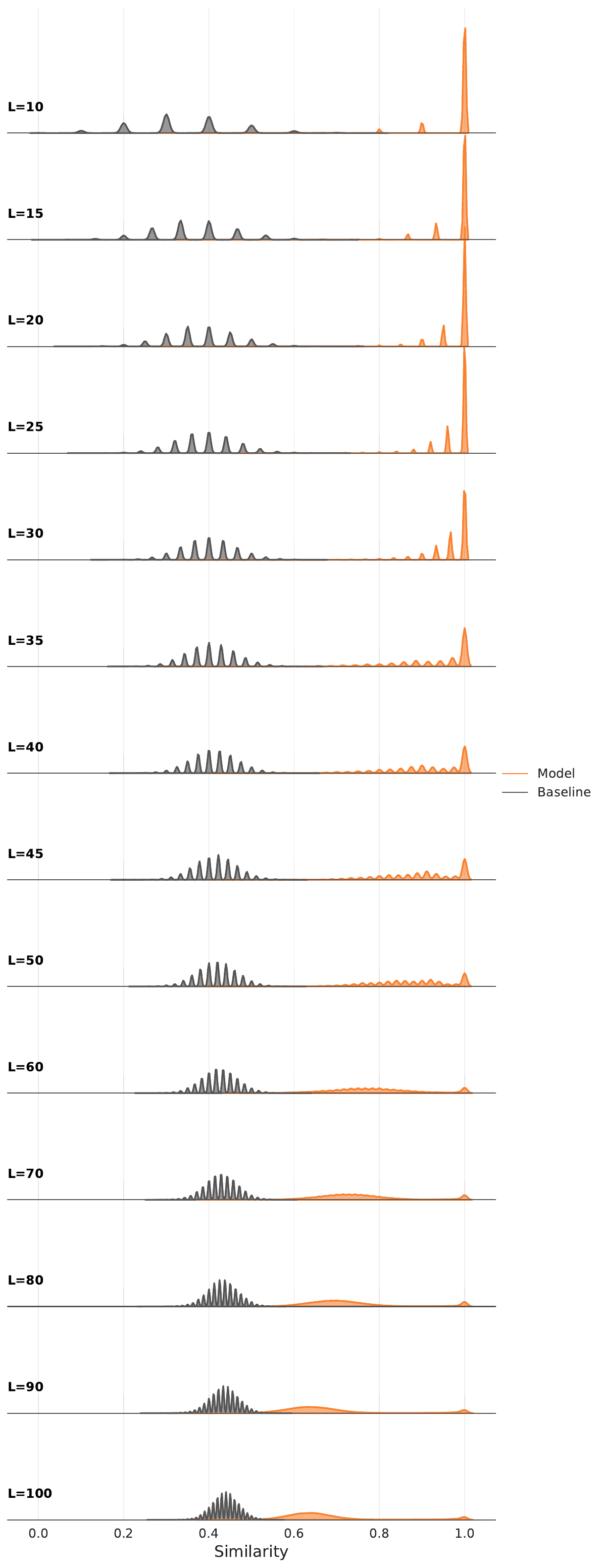}
        \\ \small (c) NTv2
    \end{minipage}
    \caption{Full distribution of per-sequence Levenshtein similarity for the multi-length LCQD at each sequence length $l = 10$ to $100$, for (a) DNABERT-2, (b) Evo~2 and (c) NTv2, against the random baseline (grey). These are the distributions summarised by the means in Supplementary Table~\ref{tab:query_decoder_results}; the spread matters because a low mean can still hide a substantial fraction of near-perfect reconstructions.}
    \label{fig:ridge_plots_mean}
\end{figure}

\end{document}